\documentclass[review]{elsarticle}
\usepackage{hyperref}
\usepackage{pdflscape}
\usepackage{scrextend}
\usepackage{amsfonts}
\usepackage{booktabs}
\usepackage{amsmath}
\usepackage{enumitem}
\usepackage{graphicx}
\usepackage{caption}
\usepackage{subcaption}
\usepackage{multirow}
\journal{Computers \& Fluids}
\usepackage{geometry}
\begin{document}

\begin{frontmatter}

\title{Fifth order finite volume WENO in general orthogonally$-$curvilinear coordinates}


\author[mymainaddress]{Mohammad Afzal Shadab \fnref{myfootnote}\corref{mycorrespondingauthor}}
\ead{mashadab@connect.ust.hk}

\author[balsaraaddress]{Dinshaw Balsara \fnref{myfootnote2}}
\ead{dbalsara@nd.edu}

\author[mymainaddress]{Wei Shyy \fnref{myfootnote3}}
\ead{weishyy@ust.hk}

\author[mymainaddress,mysecondaryaddress]{Kun Xu \fnref{myfootnote4}}
\ead{makxu@ust.hk}
\cortext[mycorrespondingauthor]{Corresponding author}

\address[mymainaddress]{Department of Mechanical and Aerospace Engineering, Hong Kong University of Science and Technology, Clear Water Bay, Peninsula, Hong Kong}
\address[balsaraaddress]{Department of Physics, University of Notre Dame, Indiana, United States IN 46556-5670}
\address[mysecondaryaddress]{Department of Mathematics, Hong Kong University of Science and Technology, Clear Water Bay, Peninsula, Hong Kong}

\fntext[myfootnote]{Research Postgraduate Student, Department of Mechanical and Aerospace Engineering, Hong Kong University of Science and Technology, Clear Water Bay, Hong Kong}
\fntext[myfootnote2]{Professor, Department of Physics, University of Notre Dame, Indiana, United States IN 46556-5670}
\fntext[myfootnote3]{President and Chair Professor, Department of Mechanical and Aerospace Engineering, Hong Kong University of Science and Technology, Clear Water Bay, Hong Kong}
\fntext[myfootnote4]{Chair Professor, Department of Mechanical and Aerospace Engineering and Department of Mathematics, Hong Kong University of Science and Technology, Clear Water Bay, Hong Kong}

\begin{abstract}
High order reconstruction in the finite volume (FV) approach is achieved by a more fundamental form of the fifth order WENO reconstruction in the framework of orthogonally$-$curvilinear coordinates, for solving hyperbolic conservation equations. The derivation employs a piecewise parabolic polynomial approximation to the zone averaged values $(\bar{Q}_{i})$ to reconstruct the right (${{q}_{i}^+}$), middle (${{q}_{i}^M}$), and left (${{q}_{i}^-}$) interface values. The grid dependent linear weights of the WENO are recovered by inverting a Vandermonde$-$like linear system of equations with spatially varying coefficients. A scheme for calculating the linear weights, optimal weights, and smoothness indicator on a regularly$-$/irregularly$-$spaced grid in orthogonally$-$curvilinear coordinates is proposed. A grid independent relation for evaluating the smoothness indicator is derived from the basic definition. Finally, a computationally efficient extension to multi-dimensions is proposed along with the procedures for flux and source term integrations. Analytical values of the linear weights, optimal weights, and weights for flux and source term integrations are provided for a regularly$-$spaced grid in Cartesian, cylindrical, and spherical coordinates. Conventional fifth order WENO$-$JS can be fully recovered in the case of limiting curvature $(R\to\infty)$. The fifth order finite volume WENO$-$C (orthogonally$-$curvilinear version of WENO) reconstruction scheme is tested for several 1D and 2D benchmark tests involving smooth and discontinuous flows in cylindrical and spherical coordinates.     
\end{abstract}


\end{frontmatter}

\section{Introduction} \label{intro}

Finite volume weighted essentially non$-$oscillatory (WENO) reconstruction scheme represents the state of art numerical methods in one$-$ and two$-$dimensional hyperbolic conservation laws \cite{titarev2005weno,Mignone-2014,Titarev-2004,Jiang-1996,Liu-1994,balsara2000monotonicity}. Finite volume methods deal with the volume averages, which changes only when there is an imbalance of the fluxes across the control volume \cite{Mignone-2014}. Flux evaluation at an interface requires an important task of reconstructing the cell averaged value at the interface \cite{Mignone-2014}. High order reconstruction is preferred for the cases of complex flow phenomena including discontinuous flows \cite{dumbser2013efficient,shu2009high}, smooth flows with turbulence \cite{dumbser2013high} \cite{Shu-2003}, aeroacoustics \cite{Shu-2003}, sediment transport \cite{vcrnjaric2004extension} and magnetohydrodynamics (MHD) \cite{Jiang-1999,balsara2009divergence,balsara2009efficient}. In a plethora of reconstruction techniques including $p^{th}$ order accurate essentially non$-$oscillatory (ENO) scheme \cite{Casper-1993}, second order total variation diminishing (TVD) methods \cite{Mignone-2014}, discontinuous Galerkin methods \cite{Shu-2003}, and modified piecewise parabolic method (PPM) \cite{Mignone-2014,colella1984piecewise,colella2008limiter,mccorquodale2011high}, WENO stands a chance by its virtue of attaining a convexly combined $(2p-1)^{th}$ order of convergence for smooth flows aided with a novel ENO strategy for maintaining high order accuracy even for the discontinuous flows \cite{Mignone-2014,Casper-1993}.

The conventional WENO scheme is specifically designed for the reconstruction in Cartesian coordinates on uniform grids \cite{Jiang-1996, Liu-1994}. For an arbitrary curvilinear mesh, the procedure of using a Jacobian, in order to map a general curvilinear mesh to a uniform Cartesian mesh, is employed \cite{Casper-1993}. However, the employment of Cartesian-based reconstruction scheme on a curvilinear grid suffers from a number of drawbacks, e.g., in the original PPM paper \cite{colella1984piecewise}, reconstruction was performed in volume coordinates (than the linear ones) so that algorithm for a Cartesian mesh can be used on a cylindrical/spherical mesh. However, the resulting interface states became first order accurate even for smooth flows \cite{colella1984piecewise}. Another example can be the volume average assignment to the geometrical cell center of finite volume than the centroid \cite{monchmeyer1989conservative, doi:10.1093/mnras/250.3.581, ziegler2011semi}. The reconstruction in general coordinates can be performed with the aid of two techniques: genuine multi$-$dimensional reconstruction and dimension$-$by$-$dimension reconstruction \cite{Casper-1993}. Genuine multi$-$dimensional reconstruction is computationally expensive and highly complicated since it considers all of the finite volumes while constructing the polynomial \cite{Casper-1993}. A better approach is to perform a dimension$-$by$-$dimension reconstruction since it consists of less expensive one$-$dimensional sweeps in every dimension and most of the problems of engineering interests are considered in orthogonally$-$curvilinear coordinates like Cartesian, cylindrical, and spherical coordinates with regularly$-$spaced and irregularly$-$spaced grids. A breakthrough in the field of high order reconstruction in these coordinates is the application of the Vandermonde$-$like linear systems of equations with spatially varying coefficients \cite{Mignone-2014}. It is reintroduced in the present work to build a basis for the derivation of the high order WENO schemes. Mignone \cite{Mignone-2014} restricted the work to the usage of the third order WENO approach with the weight functions provided by Yamaleev and Carpenter \cite{Yamaleev-2009} and did not extend it to multi$-$dimensions (2D and 3D). In Mignone's paper \cite{Mignone-2014}, modified piecewise parabolic method (PPM$_5$) of order $\sim 2-3$ gave better results when compared with the modified third order WENO. However, the latter reconstruction scheme gave consistent values for all the numerical tests performed. Also, there is a drop of accuracy in the modified third order WENO scheme for discontinuous flow cases \cite{Mignone-2014} when the standard weights derived by Jiang and Shu \cite{Jiang-1996} are used, as they are specifically restricted to the Cartesian grids.

The motivation for the present work is to develop a fifth order finite volume WENO$-$C reconstruction scheme in orthogonally$-$curvilinear coordinates for regularly$-$spaced and irregularly$-$spaced grids. It is based on the concepts of linear weights by Mignone \cite{Mignone-2014} and optimal weights, smoothness indicators by Jiang and Shu \cite{Jiang-1996}. Also, the present work provides a computationally efficient extension of this scheme to multi$-$dimensions and deals with the source terms straightforwardly.

The present work is divided into four sections. Section \ref{WENO} includes the fifth order finite volume WENO$-$C reconstruction procedure for a regularly$-$/irregularly$-$spaced grid in orthogonally$-$curvilinear coordinates. It is followed by Section \ref{tests} in which 1D and 2D numerical benchmark tests involving smooth and discontinuous flows in cylindrical and spherical coordinates are presented. Finally, Section \ref{conclusions} concludes the paper. Appendix at the end is divided into two sections. The first section includes the analytical values of the weights required for WENO$-$C reconstruction and flux/source term integration for standard uniform grids, whereas the second section includes linear stability analysis of the proposed scheme.

\section{Fifth order finite volume WENO$-$C reconstruction} \label{WENO} 
\subsection{Finite volume discretization in curvilinear coordinates} 
The scalar conservation law in an orthogonal system of coordinates $(x_1,x_2,x_3)$ having the scale factors $h_1,h_2,h_3$ and unit vectors $(\bf{\hat{e}_1,\hat{e}_2,\hat{e}_3})$ in the respective directions, is given in Eq. (\ref{eq:1}). 

\begin{equation} \label{eq:1}
\frac{\partial{Q}}{\partial{t}}+\nabla{\bf{.F}} =S
\end{equation}

where $Q$ is the conserved quantity of the fluid, ${\bf{F}}=(F_1,F_2,F_3)$ is the corresponding flux vector, and $S$ is the source term. The divergence operator is further expressed in the form of Eq. (\ref{eq:2}).

\begin{equation} \label{eq:2}
\nabla{\bf{.F}}=\frac{1}{h_1h_2h_3}\bigg[{\frac{\partial}{\partial{x_1}}(h_2h_3F_1)+\frac{\partial}{\partial{x_2}}(h_1h_3F_2)+\frac{\partial}{\partial{x_3}}(h_1h_2F_3)}\bigg]
\end{equation}

Eq. (\ref{eq:1}) is discretized over a computational domain comprising $N_1 \times N_2 \times N_3$ cells in the corresponding directions with the grid sizes given in Eq. (\ref{eq:3}).
\begin{equation} \label{eq:3}
\Delta{x_{1,i}}=x_{1,i+\frac{1}{2}}-x_{1,i-\frac{1}{2}},\quad \Delta{x_{2,j}}=x_{2,j+\frac{1}{2}}-x_{2,j-\frac{1}{2}},\quad \Delta{x_{3,k}}=x_{3,k+\frac{1}{2}}-x_{3,k-\frac{1}{2}}
\end{equation}

For the sake of simplicity, the notation $(i,j,k)$ is mentioned as $\bf{i}$ where $\bf{i} \in \mathbb{Z}^3$; and $\bf\mathbb{Z}^3$ is a vector of coordinate index in the computational domain with $1\le i \le N_1$, $1\le j \le N_2$, and $1\le k \le N_3$. Also, the position of a cell interface orthogonal to any direction $(d)$ is given by $\bf\hat{e}_d$ and it is denoted by $\bf{i}\pm\frac{1}{2}\bf{\hat{e}_d}$. For example, $\bf{i}\pm\frac{1}{2}\bf{\hat{e}_1}$ refers to the ${i\pm\frac{1}{2}}$ interfaces of the cell $\bf i$ in $\bf{\hat{e}_1}$ direction. The cell volume is given in Eq. (\ref{eq:4}). 

\begin{equation}\label{eq:4}
\Delta{\mathcal{V}_{i,j,k}=\int_{x_{3,k-\frac{1}{2}}}^{x_{3,k+\frac{1}{2}}}\int_{x_{2,j-\frac{1}{2}}}^{x_{2,j+\frac{1}{2}}}\int_{x_{1,i-\frac{1}{2}}}^{x_{1,i+\frac{1}{2}}}h_1h_2h_3dx_1dx_2dx_3}
\end{equation}
The flux $F_d$ is averaged over the surface$-$area $A_d$ of the interface $\bf{i}+\frac{1}{2}\bf{\hat{e}_1}$, as given in Eq. (\ref{eq:5}).

\begin{equation}\label{eq:5}
 \tilde{F}_{1,{\bf{i}+\frac{1}{2}\bf{\hat{e}_1}}}=\frac{1}{A_{1,{\bf{i}+\frac{1}{2}\bf{\hat{e}_1}}}}\int_{x_{3,k-\frac{1}{2}}}^{x_{3,k+\frac{1}{2}}}\int_{x_{2,j-\frac{1}{2}}}^{x_{2,j+\frac{1}{2}}}F_1 h_2 h_3 dx_2 dx_3
\end{equation}
 where the cross$-$sectional area ${A_{1,{\bf{i}+\frac{1}{2}\bf{\hat{e}_1}}}}$ is provided in Eq. (\ref{eq:6}). Here the scale factors $h_2,h_3$ are the functions of the position vector at the interface ${\bf{i}+\frac{1}{2}\bf{\hat{e}_1}}$.
\begin{equation}\label{eq:6}
 A_{1,{\bf{i}+\frac{1}{2}\bf{\hat{e}_1}}}=\int_{x_{3,k-\frac{1}{2}}}^{x_{3,k+\frac{1}{2}}}\int_{x_{2,j-\frac{1}{2}}}^{x_{2,j+\frac{1}{2}}}h_2 h_3 dx_2 dx_3
\end{equation}

Similarly, the expressions for the other directions ($d=2,3$) can be obtained by cyclic permutations. The final form of the discretized conservation law can be derived by integrating Eq. (\ref{eq:1}) over the cell volume and applying the Gauss theorem to the flux term yielding Eq. (\ref{eq:7}), where $\bar{Q}_{\bf{i}}$ and $\bar{S}_{\bf{i}}$ are respectively the conservative variable and the source term averaged over the finite volume $\bf{i}$.

\begin{equation}\label{eq:7}
 \frac{\partial}{\partial{t}} \bar{Q}_{\textbf{i}}+ \frac{1}{\Delta \mathcal{V}_\textbf{i}}\sum\limits_{d}\bigg[(A_d\tilde{F}_d)_{\textbf{i}+\frac{1}{2}\bf{\hat{e}_d}}-(A_d\tilde{F}_d)_{\textbf{i}-\frac{1}{2}\bf{\hat{e}_d}}\bigg]=\bar{S}_{\textbf{i}}
\end{equation}

In cylindrical coordinates, ($x_1,x_2,x_3$)$\equiv$($R,\theta,z$), ($h_1,h_2,h_3$)$\equiv$($1,R,1$), and Eq. (\ref{eq:7}) transforms into Eq. (\ref{eq:8}).

\begin{equation}\label{eq:8}
\begin{split}
 \frac{\partial}{\partial{t}}\bar{Q}_{\bf{i}}=-{\frac{(\tilde{F}_RR)_{\bf{i}+\frac{1}{2}\bf{\hat{e}_r}}-(\tilde{F}_RR)_{\bf{i}-\frac{1}{2}\bf{\hat{e}_r}}}{\Delta{\mathcal{V}_{R,i}}}}-\frac{(\tilde{F}_\theta)_{\bf{i}+\frac{1}{2}\bf{\hat{e}_{\theta}}}-(\tilde{F}_\theta)_{\bf{i}-\frac{1}{2}\bf{\hat{e}_{\theta}}}}{R_i\Delta{\theta_j}} \\
 -\frac{(\tilde{F}_z)_{\bf{i}+\frac{1}{2}\bf{\hat{e}_z}}-(\tilde{F}_z)_{\bf{i}-\frac{1}{2}\bf{\hat{e}_z}}}{\Delta{z_k}}+\bar{S}_{\bf{i}}
 \end{split}
\end{equation}

where ($\tilde{F}_R,\tilde{F_{\theta}},\tilde{F}_z$) are the surface averaged flux vector ($\bf{F}$) components in ($R,\theta,z$) directions and $\Delta{\mathcal{V}_{R,i}}=(R_{i+\frac{1}{2}}^2-R_{i-\frac{1}{2}}^2)/2$ is the cell radial volume.

In spherical coordinates, ($x_1,x_2,x_3$)$\equiv$($r,\theta,\phi$), ($h_1,h_2,h_3$)$\equiv$($1,r,rsin\theta$), and Eq. (\ref{eq:7}) transforms into Eq. (\ref{eq:9}).

\begin{equation}\label{eq:9}
\begin{split}
 \frac{\partial}{\partial{t}}\bar{Q}_{\bf{i}}=-{\frac{(\tilde{F}_rr^2)_{\bf{i}+\frac{1}{2}\bf{\hat{e}_r}}-(\tilde{F}_rr^2)_{\bf{i}-\frac{1}{2}\bf{\hat{e}_r}}}{\Delta{\mathcal{V}_{r,i}}}}-\frac{(\tilde{F}_{\theta}sin\theta)_{\bf{i}+\frac{1}{2}\bf{\hat{e}_{\theta}}}-(\tilde{F}_{\theta}sin\theta)_{\bf{i}-\frac{1}{2}\bf{\hat{e}_{\theta}}}}{\tilde{r}_i\Delta{\mu_j}} \\
 -{\frac{\Delta{\theta_j}}{\Delta{\mu_j}}}\frac{(\tilde{F}_\phi)_{\bf{i}+\frac{1}{2}\bf{\hat{e}_{\phi}}}-(\tilde{F}_\phi)_{\bf{i}-\frac{1}{2}\bf{\hat{e}_{\phi}}}}{\tilde{r}_i\Delta{\phi_k}}+\bar{S}_{\bf{i}}
 \end{split}
\end{equation}

where ($\tilde{F}_r,\tilde{F_{\theta}},\tilde{F}_{\phi}$) are the surface averaged flux vector components in ($r,\theta,\phi$) directions and the remaining geometrical factors are provided in Eq. (\ref{eq:10}).

\begin{equation} \label{eq:10}
\Delta{\mathcal{V}_{r,i}}=\frac{(r_{i+\frac{1}{2}}^3-r_{i-\frac{1}{2}}^3)}{3};\quad \tilde{r}_i=\frac{2}{3}\frac{(r_{i+\frac{1}{2}}^3-r_{i-\frac{1}{2}}^3)}{(r_{i+\frac{1}{2}}^2-r_{i-\frac{1}{2}}^2)}; \quad \Delta{\mu_j}=cos{\theta_{j-\frac{1}{2}}}-cos{\theta_{j+\frac{1}{2}}}
\end{equation}

\subsection{Evaluation of the linear weights} \label{linearweights}

A non$-$uniform grid spacing with zone width $\Delta{\xi}_{i}={\xi}_{i+\frac{1}{2}}-{\xi}_{i-\frac{1}{2}}$ is considered having $\xi \in (x_1,x_2,x_3)$ as the coordinate along the reconstruction direction and ${\xi}_{i+\frac{1}{2}}$ denoting the location of the cell interface between zones $i$ and $i+1$. Let $\bar{Q}_{i}$ be the cell average of conserved quantity $Q$ inside zone $i$ at some given time, which can be expressed in form of Eq. (\ref{eq:11}).

\begin{equation} \label{eq:11}
\bar{Q}_{i} = \frac{1}{{\Delta\mathcal{V}_{i}}}{\int_{{\xi}_{i-\frac{1}{2}}}^{{\xi}_{i+\frac{1}{2}}}Q_i(\xi)\frac{\partial{\mathcal{V}}}{\partial\xi}d\xi} 
\end{equation}

where the local cell volume $\Delta{\mathcal{V}}_i$ of $i^{th}$ cell in the direction of reconstruction given in Eq. (\ref{eq:12})
\begin{equation} \label{eq:12}
\Delta{\mathcal{V}}_i={\int_{{\xi}_{i-\frac{1}{2}}}^{{\xi}_{i+\frac{1}{2}}}\frac{\partial{\mathcal{V}}}{\partial\xi}d\xi}
\end{equation}

$\frac{\partial{\mathcal{V}}}{\partial\xi}$ is a one$-$dimensional Jacobian whose values for volumetric operations are summarized in Table \ref{tab:1} for structured grids in standard coordinates.

\begin{table}[h!] 
\centering
    \caption{One$-$dimensional Jacobian $\big(\frac{\partial{\mathcal{V}}}{\partial\xi}\big)$ values for the regularly$-$spaced grids for volumetric operations}
    \begin{tabular}{ |c | c | c |}
    \hline
    \mbox{$Coordinates$} & {$Direction(s)$} & {$\frac{\partial{\mathcal{V}}}{\partial\xi}$}\\
    \hline
    Cartesian & $x, y, z$ & $\xi^0$\\
    \hline
    \multirow{2}{*}{Cylindrical} & $R$ & $\xi^1$\\
    \cline{2-3}
              & $\theta,z$ & $\xi^0$\\
    \hline
    \multirow{3}{*}{Spherical} & $r$ & $\xi^2$\\
    \cline{2-3}
              & $\theta$ & $sin\xi$\\
    \cline{2-3}
              & $\phi$ & $\xi^0$\\
    \hline
    \end{tabular}
\label{tab:1}
\end{table}

Now, our aim is to find a $p^{th}$ order accurate approximation to the actual solution by constructing a $(p-1)^{th}$ order polynomial distribution, as given in Eq. (\ref{eq:13}).

\begin{equation} \label{eq:13}
Q_i(\xi) = a_{i,0} +a_{i,1}({\xi}-{\xi_i^c})+a_{i,2}({\xi}-{\xi_i^c})^2 +...+a_{i,p-1}({\xi}-{\xi_i^c})^{p-1}
\end{equation}

where ${a_{i,n}}$ corresponds to a vector of the coefficients which to be determined and ${\xi_i^c}$ can be taken as the cell centroid. However, the final values at the interface are independent of the particular choice of ${\xi_i^c}$ and one may as well set ${\xi_i^c}=0$ \cite{Mignone-2014}. Unlike the cell center, the centroid is not equidistant from the cell interfaces in the case of curvilinear coordinates, and the cell averaged values are assigned at the centroid \cite{Mignone-2014}. Further, the method has to be locally conservative, i.e., the polynomial $Q_i(\xi)$ must fit the neighboring cell averages, satisfying Eq. (\ref{eq:14}).

\begin{equation} \label{eq:14}
 {\int_{{\xi}_{i+s-\frac{1}{2}}}^{{\xi}_{i+s+\frac{1}{2}}}Q_i(\xi)\frac{\partial{\mathcal{V}}}{\partial\xi}d\xi} = {{\Delta\mathcal{V}_{i+s}}}\bar{Q}_{i+s}\quad\quad\textrm{for}\quad-i_L\le s \le i_R
\end{equation}

where the stencil includes $i_L$ cells to the left and $i_R$ cells to the right of the $i^{th}$ zone such that $i_L+i_R+1 = p$. Implementing Eqs. (\ref{eq:12}) and (\ref{eq:13}) in Eq. (\ref{eq:14}) along with a simplification leads to a $p\times p$ linear system (\ref{eq:15}) in the coefficients \{${a_{i,n}}$\}.

\begin{equation} \label{eq:15}
\begin{pmatrix}
    \beta_{i-i_L,0}       & \dots  & \beta_{i-i_L,p-1} \\
    \vdots       & \ddots & \vdots \\
    \beta_{i+i_R,0}       & \dots & \beta_{i+i_R,p-1}
\end{pmatrix}
\begin{pmatrix}
    a_{i,0} \\
    \vdots \\
    a_{i,p-1} 
\end{pmatrix}
=
\begin{pmatrix}
    \bar{Q}_{i-i_L} \\
    \vdots \\
    \bar{Q}_{i+i_R} 
\end{pmatrix}
\end{equation}

where 
\begin{equation} \label{eq:16}
\beta_{i+s,n}=\frac{1}{\Delta{\mathcal{V}}_{i+s}}{\int_{{\xi}_{i+s-\frac{1}{2}}}^{{\xi}_{i+s+\frac{1}{2}}}({\xi-\xi_i^c})^{n}\frac{\partial{\mathcal{V}}}{\partial\xi}d\xi}
\end{equation}

Eq. (\ref{eq:15}) can be written in the short notation using a $p\times p$ matrix $\bf{B}$ with the rows ranging from $s=-i_L,...,i_R$ and columns ranging from $n=0,...,p-1$. 
\begin{equation} \label{eq:17}
\sum \limits_{n=0}^{p-1}{\bf{B}}_{sn}a_{i,n}=\bar{Q}_{i+s}
\end{equation}
However, evaluation of the weights $a_{i,k}$ in Eqs. (\ref{eq:15}) and (\ref{eq:17}) requires zone averaged values $\bar{Q}_{i}$, thus, increasing the computational cost of the whole process as it needs to be evaluated at every time step. The coefficients $\{a_{i,n}\}$ extracted from Eq. (\ref{eq:15}) will also satisfy condition (\ref{eq:18}).
\begin{equation} \label{eq:18}
q_i^+=\lim_{\xi \to \xi_{i+\frac{1}{2}}^{(-)}}Q_i(\xi)=\sum \limits_{n=0}^{p-1}a_{i,n}(\xi_{i+\frac{1}{2}}-\xi_i^c)^n; \quad q_i^-=\lim_{\xi \to \xi_{i-\frac{1}{2}}^{(+)}}Q_i(\xi)=\sum \limits_{n=0}^{p-1}a_{i,n}(\xi_{i-\frac{1}{2}}-\xi_i^c)^n
\end{equation}

A more efficient approach for evaluating left and right interface values is using a linear combination of the adjacent cell averaged values \cite{Mignone-2014}, as given in Eq. (\ref{eq:19}).

 \begin{equation} \label{eq:19}
 q_i^{\pm} = \sum\limits_{s=-i_L}^{i_R}{w_{i,s}^{\pm}}\bar{Q}_{i+s}
 \end{equation}
 
 From Eq. (\ref{eq:17}), after inverting the matrix ${\bf{B}}$, we get relation (\ref{eq:20}).
 \begin{equation} \label{eq:20}
a_{i,n}=\sum \limits_{s=-i_L}^{i_R}{\bf{C}}_{ns}\bar{Q}_{i+s}
\end{equation}
 where ${\bf{C}}={\bf{B}}^{-1}$ corresponds to the inverse of matrix ${\bf{B}}$, which will exist only if matrix ${\bf{B}}$ exists and is nonsingular.
 
 After combining Eqs. (\ref{eq:18}) and (\ref{eq:20}), we get 
\begin{equation} \label{eq:21}
q_i^\pm=\sum \limits_{n=0}^{p-1}\bigg(\sum \limits_{s=-i_L}^{i_R}{\bf{C}}_{ns}\bar{Q}_{i+s} \bigg)(\xi_{i\pm \frac{1}{2}}-\xi_i^c)^n=\sum \limits_{s=-i_L}^{i_R}\bar{Q}_{i+s}\bigg(\sum \limits_{n=0}^{p-1}{\bf{C}}_{ns}(\xi_{i\pm \frac{1}{2}}-\xi_i^c)^n\bigg) 
\end{equation}

By comparing Eqs. (\ref{eq:19}) and (\ref{eq:21}), we can extract the matrix of weights $w_{i,s}^\pm$.
\begin{equation} \label{eq:22}
w_{i,s}^\pm=\sum \limits_{n=0}^{p-1}{\bf{C}}_{ns}(\xi_{i\pm \frac{1}{2}}-\xi_i^c)^n 
\end{equation}

Since, ${\bf{C}}_{ns}={(\bf{C}^T)_{sn}}=((\bf{B}^{T})^{-1})_{sn}$, Eq. (\ref{eq:22}) can be finally written in the form of Eq. (\ref{eq:23}).
\begin{equation} \label{eq:23}
{\sum \limits_{s=-i_L}^{i_R}{{(\bf{B}}^{T})_{ns}w_{i,s}^\pm=(\xi_{i\pm \frac{1}{2}}-\xi_i^c)^n}} 
\end{equation}

Therefore, it is evident that the weights $w_{i,s}^{\pm}$ are shown to satisfy Eq. (\ref{eq:24}) \cite{Mignone-2014}, which is the fundamental equation for reconstruction in orthogonally$-$curvilinear coordinates.

\begin{equation} \label{eq:24}
\begin{pmatrix}
    \beta_{i-i_L,0}       & \dots  & \beta_{i-i_L,p-1} \\
    \vdots                & \ddots & \vdots \\
    \beta_{i+i_R,0}       & \dots  & \beta_{i+i_R,p-1}
\end{pmatrix}
^T
\begin{pmatrix}
    w_{i,-i_L}^{\pm} \\
    \vdots \\
    w_{i,i_R}^{\pm} 
\end{pmatrix}
=
\begin{pmatrix}
    1 \\
    \vdots \\
    (\xi_{i\pm\frac{1}{2}}-\xi_i^c)^{p-1} 
\end{pmatrix}
\end{equation}

Also, the grid dependent linear weights ($w_{i,s}^\pm$) satisfy the normalization condition (\ref{eq:25})\cite{Mignone-2014}.

\begin{equation} \label{eq:25}
{\sum \limits_{s=-i_L}^{i_R}w_{i,s}^\pm=1}
\end{equation}

Some important remarks on the linear weights in the proposed scheme are as follows:
\begin{enumerate}
   \item Eq. (\ref{eq:24}) is capable of evaluating the grid generated linear weights for any regularly$-$/irregularly$-$spaced mesh in orthogonally$-$curvilinear coordinates. 
   It is observed that these weights are independent of the mesh size for standard regularly$-$spaced grid cases, but depend on the grid type. Also, they can be evaluated and stored (at a nominal cost) independently before the actual computation, after the grid type is finalized.
   \item For fifth order WENO, three sets of third order ($p=3$) stencils ($S_k$) are chosen namely 
   \begin{itemize}
       \item    $S_0 (i-2,i-1,i)$  ::  $-i_L=2,i_R=0$ \item    $S_1  (i-1,i,i+1)$  ::  $-i_L=1,i_R=1$ \item    $S_2 (i,i+1,i+2)$  ::  $-i_L=0,i_R=2$.
   \end{itemize}
   In addition to this, another symmetric stencil $S_5$ :: $(i-2,i-1,i,i+1,i+2)$ is used to extract the values of the optimal weights in the subsection \ref{optimalweights}.
   \item The final interface values (\ref{eq:19}) and the linear weights depend only on the order of the reconstruction polynomial and not on $\xi_i^c$, which can be set to zero \cite{Mignone-2014}.
   \item The values are simplified when the Jacobian is a simple power of $\xi$ i.e. $\frac{\partial{\mathcal{V}}}{\partial\xi}=\xi^m$. Then, $\beta_{i+s,n}$ of Eq. (\ref{eq:16}) can be written in the simplified form (\ref{eq:26}).
  
  \begin{equation} \label{eq:26}
\beta_{i+s,n}=\frac{m+1}{n+m+1}\frac{\xi_{i+s+\frac{1}{2}}^{n+m+1}-\xi_{i+s-\frac{1}{2}}^{n+m+1}}{\xi_{i+s+\frac{1}{2}}^{m+1}-\xi_{i+s-\frac{1}{2}}^{m+1}}
 \end{equation} 
  \item For the spherical$-$meridional coordinate, $\beta_{i+s,n}$ of Eq. (\ref{eq:16}) becomes highly complex as ($\frac{\partial{\mathcal{V}}}{\partial\xi}=sin\xi$). The value of $\beta_{i+s,n}$ can be computed from Eq. (\ref{eq:27}) and needs to be solved numerically e.g. by using LU decomposition method. 
   \begin{equation} \label{eq:27}
\beta_{i+s,n}=\frac{1}{cos\xi_{i_{s-}}-cos\xi_{i_{s+}}}\sum_{k=0}^{n}k!\begin{pmatrix}
    n\\
    k
\end{pmatrix}
\bigg[\xi_{i_{s-}}^{n-k}cos\bigg(\xi_{i_{s-}}+\frac{k\pi}{2}\bigg)- \xi_{i_{s+}}^{n-k}cos\bigg(\xi_{i_{s+}}+\frac{k\pi}{2}\bigg) \bigg]
 \end{equation} 
 where $i_{s\pm}$ refers to $i+s\pm \frac{1}{2}$. 
 \item For non$-$standard grids, $\frac{\partial{\mathcal{V}}}{\partial{\xi}}$ is not a simple function, which makes the direct integration highly complex and time consuming. Therefore, such cases are tackled using numerical integration of the Eq. (\ref{eq:16}) and then matrix inversion of the Eq. (\ref{eq:24}). 
 \item Eq. (\ref{eq:24}) can also be used to compute the point$-$values of $Q(\xi)$ at any other points than the interfaces e.g. the cell center ($q_i^M$). The value at the cell center is obtained by setting the right hand side of the matrix (\ref{eq:24}) as $(1,0,0,...,0)^T$ with $\xi_i^c=0$, which is important in the case of nonlinear systems of equations where the reconstruction of the primitive variables is done instead of the conserved variables \cite{Mignone-2014}.
 \item The linear positive ($w_i^+$), middle ($w_i^M$) and negative ($w_i^-$) weights for the WENO reconstruction for the standard cases of regularly$-$spaced grid in Cartesian, cylindrical, and spherical coordinates are summarized in the \ref{Cartesianlinearweights}, \ref{cylindricallinearweights}, and \ref{sphericallinearweights} respectively. The analytical solutions for the spherical$-$meridional coordinate $(\theta)$ and irregularly$-$spaced grid are highly intricate and case$-$specific respectively. Thus, they are not mentioned in this paper as they need to be dealt numerically. \end{enumerate}

The weights and the stencil are denoted by $w_{i,l,k}^{p\pm}$ and $S_{l}^{p\pm}$ respectively, where $k$ is sequence of the weight$-$applied cell with respect to the cell considered for reconstruction $(i)$, $p$ is the order of reconstruction ($p=i_L+i_R+1$), $l$ is the stencil number, and `$\pm$' represents the positive and negative weights i.e. weights for reconstructing right ($+$) and left ($-$) interface values respectively. The derivation of middle (mid$-$value) weights ($w_{i,l,k}^{pM}$) also follow the same procedure.
 
The reconstructed values ${q}_{i,l}^{p\pm}$ represents the ${p^{th}}-$order reconstructed value at right ($+$) or left ($-$) interface of $i^{th}$ cell on stencil $l$. The formulation for the interpolated values at the interface for the WENO reconstruction are given by the linear system of Eq. (\ref{eq:28}), where $i_L$ and $i_R$ depend on the stencil $l$.

\begin{equation} \label{eq:28}
q_{i,l}^{p\pm}=\sum\limits_{s=-i_L}^{i_R}w_{i,l,s}^{p\pm}\bar{Q}_{i+s}
\end{equation}

\subsection{Optimal weights} \label{optimalweights}
The weights which optimize the sum of the lower order interpolated variables into a higher order accurate variable, are known as optimal weights \cite{Jiang-1996,Liu-1994}. For the case of fifth order WENO interpolation, the third order interpolated variables are optimally weighed in order to achieve fifth order accurate interpolated values as given in Eq. (\ref{eq:29}) for the case of $p=3$.
\begin{equation} \label{eq:29}
q_{i,0}^{(2p-1)\pm}=\sum\limits_{l=0}^{p-1}C_{i,l}^\pm q_{i,l}^{p\pm}
\end{equation}
where $C_{i,l}^\pm$ is the optimal weight for the positive/negative cases on the $i^{th}$ finite volume. $C_{i,l}^M$ for mid$-$value weights also follow the same procedure.
So, Eqs. (\ref{eq:24}) and (\ref{eq:26}) are used again to evaluate the weights for the fifth order ($2p-1=5$) interpolation ($i_L=2,i_R=2$). The fifth order interpolated variable at the interface is equated with the sum of optimally weighed third order interpolated variables, as given in Eq. (\ref{eq:29}). The optimal weights $C_{i,l}^\pm$ are evaluated by equating the coefficients of $\bar{Q}$ resulting in ($2p-1$) equations with $p$ unknowns. For the fifth order WENO$-$C reconstruction, the case is simplified to a system of linear equations as given in Eq. (\ref{eq:30}), by selecting $\bar{Q}_{i-2}$, $\bar{Q_i}$, and $\bar{Q}_{i+2}$ coefficients to reduce the computational cost.

\begin{equation} \label{eq:30}
C_{i,0}^\pm=\frac{w_{i,0,-2}^{5\pm}}{{w_{i,0,-2}^{3\pm}}};\quad C_{i,2}^\pm=\frac{w_{i,0,+2}^{5\pm}}{{w_{i,2,+2}^{3\pm}}};\quad C_{i,1}^\pm=\frac{w_{i,0,0}^{5\pm}-C_{i,0}^\pm w_{i,0,0}^{3\pm}-C_{i,2}^\pm w_{i,2,0}^{3\pm}}{w_{i,1,0}^{3\pm}}
\end{equation}

Some remarks regarding the optimal weights are given below:
\begin{enumerate}

    \item The summation of the optimal weights always yield unity value and their value is independent of the coefficients of $\bar{Q}$ equated in Eq. (\ref{eq:29}).  
    \item Since weights are independent of the conserved variables, optimal weights are also constants for a selected orthogonally$-$curvilinear mesh and can be computed in advance with a little storage cost.
    \item The analytical values in the Cartesian, cylindrical$-$radial, and spherical$-$radial coordinates for a regularly$-$spaced grid are provided in \ref{Cartesianoptimalweights}, \ref{cylindricaloptimalweights}, and \ref{sphericaloptimalweights} respectively.
    \item The only case where the optimal weights are mirror$-$symmetric is of the regularly$-$spaced grid in Cartesian coordinates. The optimal weights are the same as of the conventional fifth order WENO reconstruction \cite{Titarev-2004,Jiang-1996} in this case and also when $i \to \infty$ (limiting curvature) in the case of regularly$-$spaced grid cases in the cylindrical$-$radial and spherical$-$radial coordinates.
    \item The weights for spherical$-$radial coordinates are much more complex. For spherical coordinates, it is advised to use the fifth order weights and linear weights to evaluate the optimal weights or use direct numerical operation after mesh generation since the analytical values of optimal weights contain high order ($i^{16}$) terms. Moreover, the concept of optimal weights can be completely removed with the aid of WENO$-$AO type modification by Balsara et al. \cite{balsara2016efficient} to the present work. However, the present work remains general and provides the backbone to such construction techniques.
\end{enumerate}

\subsection{Smoothness indicators and the nonlinear weights} \label{smoothnesslimiter}
The smoothness indicators are the nonlinear tools employed to differentiate in between a smooth and a discontinuous flows \cite{Jiang-1996,Liu-1994} on a stencil. They are employed in order to discard the discontinuous stencils and maintain a high order accuracy even for the discontinuous flows. From the original idea of \cite{Jiang-1996}, the present analysis is performed. Jiang and Shu \cite{Jiang-1996} proposed a novel technique of evaluating the smoothness indicators ($IS_{i,l}$). Since, for a regularly$-$/irregularly$-$spaced grid, ($IS_{i,l}$) varies with the grid index $i$, therefore we will use ($IS_{i,l}$) later in this paper. The idea involves minimization of the $L_2-$norm of the derivatives of the reconstruction polynomial, thus, emulating the idea of minimizing the total variation of the approximation. The mathematical definition of the smoothness indicator is given in Eq. (\ref{eq:31}) \cite{Titarev-2004,Jiang-1996}.
\begin{equation} \label{eq:31}
IS_{i,l}=\sum \limits_{m=1}^{p-1}\int_{\xi_{j-\frac{1}{2}}}^{\xi_{j+\frac{1}{2}}}\bigg(\frac{d^m}{d\xi^m}Q_{i,l}(\xi)\bigg)^2\Delta{\xi_i^{2m-1}d\xi}, \quad l=0,...,p-1
\end{equation}
To evaluate the value of $IS_{i,l}$, a third order polynomial interpolation on $i^{th}$ cell is required using positive and negative reconstructed values by stencil $S_l$, as given in Eq. (\ref{eq:32}).
\begin{equation} \label{eq:32}
Q_{i,l}(\xi)=a_{i,l,0}+a_{i,l,1}(\xi_i-\xi_i^c)+a_{i,l,2}(\xi_i-\xi_i^c)^2
\end{equation}
Let $\xi_{i+1/2}-\xi_i^c=\xi_i^+$, $\xi_{i-1/2}-\xi_i^c=-\xi_i^-$, and $\xi_i^++\xi_i^-=\Delta{\xi_i}$. The polynomial will satisfy the constraints (\ref{eq:33}) for all kinds of finite volumes.
\begin{equation} \label{eq:33}
\int_{{\xi}_{i-\frac{1}{2}}}^{{\xi}_{i+\frac{1}{2}}}Q_{i,l}(\xi)d\xi=\bar{Q}_i\quad,\quad
q_{i,l}^\pm=Q_{i,l}(\xi_{i\pm\frac{1}{2}})
\end{equation}
Finally, we get the values of the $a_{i,l,0}, a_{i,l,1},$ and $a_{i,l,2}$.
\begin{equation} \label{eq:34}
\centering
\begin{split}
a_{i,l,0}=\frac{6 \bar{Q}_i \xi_i^- \xi_i^++q_{i,l}^+ \xi_i^-(\xi_i^--2 \xi_i^+)+q_{i,l}^- \xi_i^+(\xi_i^+ - 2 \xi_i^-)}{(\xi_i^++\xi_i^-)^2}\\
 a_{i,l,1}=\frac{2 q_{i,l}^-(\xi_i^- - 2 \xi_i^+)-6 \bar{Q}_i (\xi_i^- - \xi_i^+) - 2 q_{i,l}^+ (\xi_i^+ - 2\xi_i^-)}{(\xi_i^++\xi_i^-)^2}\\
 a_{i,l,2}=3\frac{(q_{i,l}^\pm-2 \bar{Q}_i+q_{i,l}^\pm)}{(\xi_i^++\xi_i^-)^2} 
\end{split}
\end{equation}
For the regularly$-$spaced grids, the values of $\xi^+$ and $\xi^-$ are constant throughout the grid, which are given below for the standard coordinates.
\begin{itemize}
\item Cartesian coordinates: \newline ($x,y,z$) direction:\quad$\xi^+=\xi^-=\frac{\Delta{\xi}}{2}$
\item Cylindrical coordinates: \newline Radial ($R$) direction:\quad$\xi^+=\Delta{R}\bigg(\frac{1}{2}-\frac{1}{12i-6}\bigg)$, $\xi^-=\Delta{R}\bigg(\frac{1}{2}+\frac{1}{12i-6}\bigg)$ \newline where $i=\Delta R/R_{i+1/2}$ \newline ($\theta,z$) direction:\quad$\xi^+=\xi^-=\frac{\Delta{\xi}}{2}$
\item Spherical coordinates: \newline Radial ($r$) direction:\quad$\xi^+=\Delta{r}\bigg(\frac{1}{2}-\frac{2i-1}{4(3i^2-3i+1)}\bigg)$, $\xi^-=\Delta{r}\bigg(\frac{1}{2}+\frac{2i-1}{4(3i^2-3i+1)}\bigg)$  \newline where $i=\Delta r/r_{i+1/2}$  \newline Meridional ($\theta$) direction:\quad$\xi^+=\theta_{i+\frac{1}{2}}-\theta_i^c$, $\xi^-=-(\theta_i^c-\theta_{i-\frac{1}{2}})$ \newline where $\theta_i^c=\frac{\theta_{i-\frac{1}{2}}cos\theta_{i-\frac{1}{2}}-sin\theta_{i-\frac{1}{2}}-\theta_{i+\frac{1}{2}}cos\theta_{i+\frac{1}{2}}+sin\theta_{i+\frac{1}{2}}}{cos\theta_{i-\frac{1}{2}}-cos\theta_{i+\frac{1}{2}}}$ \newline ($\phi$) direction:\quad$\xi^+=\xi^-=\frac{\Delta{\phi}}{2}$
\end{itemize}

These values on a regularly$-$spaced grid in Cartesian coordinates ($\xi^+=\xi^-=\frac{\Delta{\xi}}{2}$) transform relation (\ref{eq:31}) into the one given in \cite{Jiang-1996,Luo-2013}. 

Now, putting the values of $a_{i,l,0}, a_{i,l,1},$ and $a_{i,l,2}$ obtained from Eq. (\ref{eq:34}) in Eq. (\ref{eq:32}) and then finally evaluating the smoothness indicator from Eq. (\ref{eq:31}) yields the following fundamental relation (\ref{eq:35}) for evaluating the smoothness indicators in the proposed scheme.
\begin{equation} \label{eq:35}
IS_{i,l}=4(39\bar{Q}_i^2-39\bar{Q}_i(q_{i,l}^-+q_{i,l}^+)+10((q_{i,l}^-)^2+(q_{i,l}^+)^2)+19q_{i,l}^-q_{i,l}^+)
\end{equation}

Some remarks regarding the smoothness indicators are as follows:
\begin{itemize}
\item Eq. (\ref{eq:35}) is a general relation for every standard grid and depends only on the third order reconstructed variables at the interface ($q_i^\pm$).
\item $q_i^\pm$ are the third order reconstructed variables obtained from Eq. (\ref{eq:28}) after using suitable grid dependent linear weights.
\item For a regularly$-$spaced grid in Cartesian coordinates, the formulation for fifth order WENO$-$C is the same as of WENO$-$JS \cite{Titarev-2004,Jiang-1996,Luo-2013} after the linear weights are substituted.
\end{itemize}

The nonlinear weight ($\omega_{i,l}^\pm$) for the WENO$-$C interpolation is defined as follows \cite{Titarev-2004,Jiang-1996}.
\begin{equation} \label{eq:36}
\omega_{i,l}^\pm=\frac{\alpha_{i,l}^\pm}{\sum_{l=0}^{p-1}\alpha_{i,l}^\pm} \quad \quad l=0,1,2
\end{equation}
where
\begin{equation} \label{eq:37}
\alpha_{i,l}^\pm=\frac{C_{i,l}^\pm}{(\epsilon+IS_{i,l})^2} \quad \quad l=0,1,2
\end{equation}
where $\epsilon$ is a small positive number used to avoid denominator becoming zero \cite{shu2009high}. Its value is a small percentage of the typical size of the reconstructed variable $\bar{Q}_i$ in such a way that Eq. (\ref{eq:37}) stays scale invariant \cite{shu2009high}. Typically, its value is chosen to be $10^{-6}$ \cite{Jiang-1996,Luo-2013,shu2009high}. The choice of non-linear weight is not unique. There is another set of non-linear weight formulation proposed by \cite{henrick2005mapped,borges2008improved} using the same smoothness indicator definitions, which can enhance the accuracy at smooth points especially at smooth extrema \cite{shu2009high,henrick2005mapped,borges2008improved}. The final interpolated interface values are evaluated from Eq. (\ref{eq:38}).
\begin{equation} \label{eq:38}
q_i^{(2p-1)\pm}={\sum_{l=0}^{p-1}\omega_{i,l}^{p\pm}q_{i,l}^{p\pm}}
\end{equation}

\subsection{Extension to multi-dimensions} \label{extensiontomultid}
The interface values calculated after the initial application are the point values only when the domain is 1D. For 2D and 3D domains, the reconstructed variables are line and area average values respectively \cite{Mignone-2014,zhang2011order,buchmuller2014improved}. If these values are used to evaluate flux, the scheme drops down to the second order of accuracy \cite{Mignone-2014,zhang2011order,buchmuller2014improved}. Buchmuller and Helzel \cite{buchmuller2014improved} proposed a very simple and effective way of achieving the original order of accuracy, just by using one point at each boundary. In this section, we are simply extending their work from Cartesian grids to general grids in orthogonally$-$curvilinear coordinates.

\begin{figure}
\centering
\begin{subfigure}{0.5\linewidth}
  \centering
  \includegraphics[width=\linewidth]{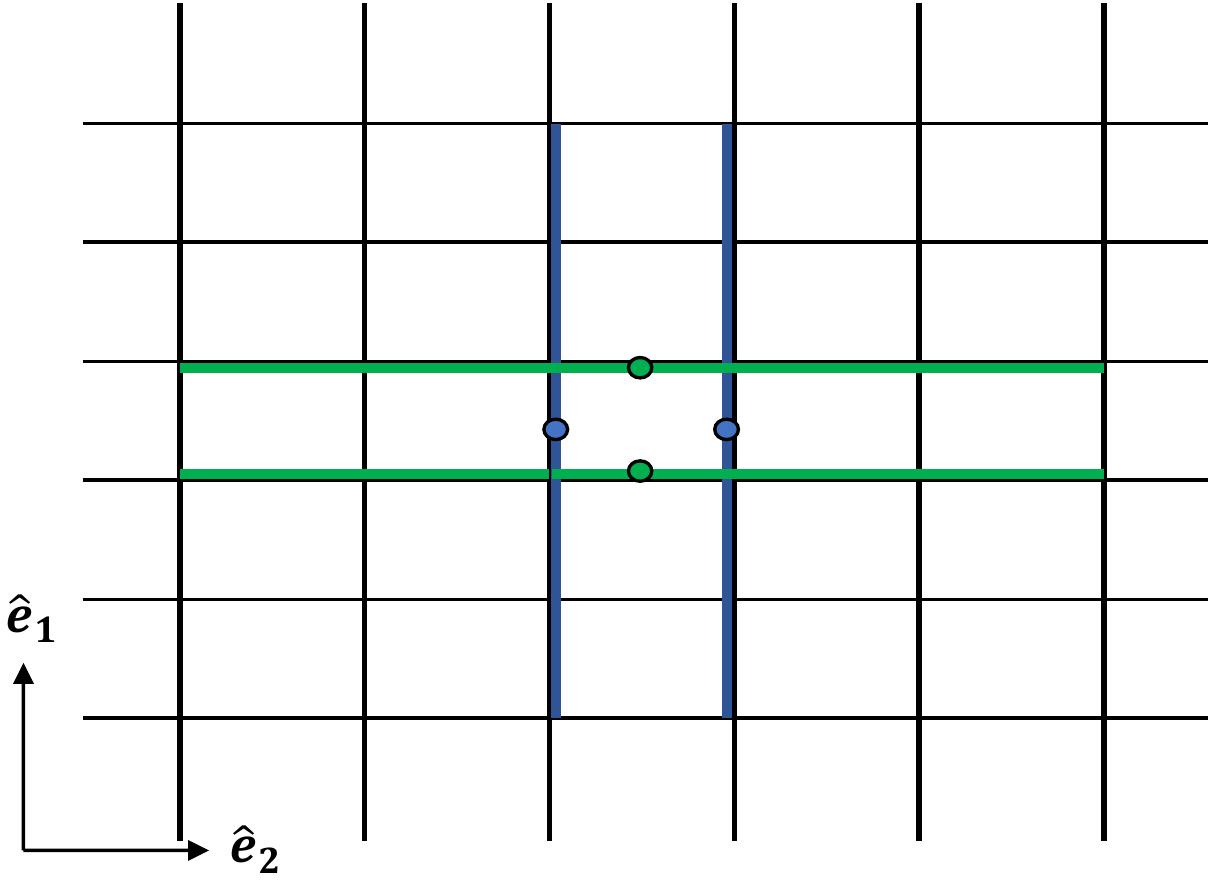}
\end{subfigure}%
\begin{subfigure}{0.5\linewidth}
  \centering
  \includegraphics[width=\linewidth]{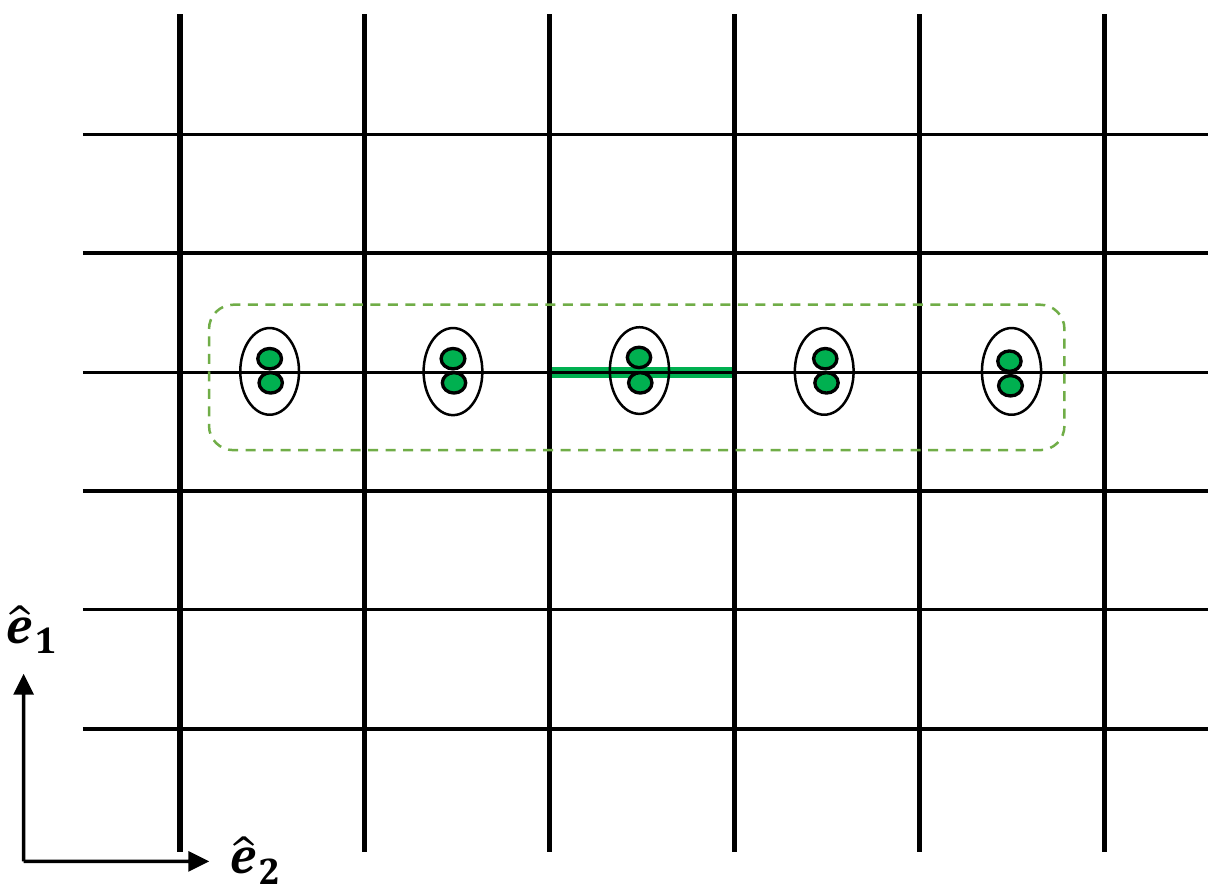}
\end{subfigure}
\caption{High order interface flux evaluation procedure. Left: Mid$-$point value reconstruction at each interface inside a cell using adjacent interface average values. 
Right: Line averaged flux evaluation by solving the Riemann problem at each mid$-$point and averaging using five adjacent points}
\label{fig:1}
\end{figure}

For the sake of simplicity, a 2D grid in orthogonally-curvilinear coordinates having unit vectors {\bf{$\bf{{\hat{e}_1}}$}} and {\bf{$\bf{{\hat{e}_2}}$}} in the corresponding orthogonal directions is considered, as shown in Fig.\ref{fig:1}. After reconstructing the left and the right interface averaged values in the first WENO sweep, the second sweep is performed to yield the point values. For the 3D case, line averaged values are yielded at this point and thus, require another reconstruction of line averaged values in the direction orthogonal previous reconstructions to obtain the point values. The Jacobian values for the conversion from volume averaged value to point values are summarized in Table \ref{tab:1}. Since this is the same principle as what we have already described in Sections \ref{linearweights} and \ref{optimalweights}, the theory and derivation are not discussed again. However, this time, the line average values are converted to the point values at the mid$-$point of the interface with the aid of adjacent interfaces' line averaged values. Also, since the quantities have been reconstructed using WENO scheme in the first face$-$normal sweep (blue$-$colored left face in {\bf{$\bf{{\hat{e}_2}}$}} direction), as shown in Fig.\ref{fig:1} (left), the second sweep of interface in the tangential direction {\bf{$\bf{{\hat{e}_1}}$}} doesn't require WENO procedure because it already contains the required smoothness information. Thus, fifth order accurate weights required for the mid$-$point value evaluation can be directly calculated by considering $\xi$ in {\bf{$\bf{{\hat{e}_1}}$}} direction with the same fifth order centered stencil, $\xi_i^c=0$, and substituting $\xi_{i}$ in the place of $\xi_{i\pm\frac{1}{2}}$ in Eq. (\ref{eq:24}). The values of the weights are the fifth order weights in the corresponding direction as evaluated earlier in Section \ref{optimalweights}. 
Then, the fluxes can be evaluated from the left and the right hand side conserved variables at the interface by solving the Riemann problem \cite{toro2013riemann}. In the future, the method will be extended to gas$-$kinetic scheme (GKS) \cite{xu2001gas}.

 The evaluated fluxes at the mid$-$points of the interfaces are averaged using polynomial interpolation, as shown in Fig. \ref{fig:1}. One-dimensional Jacobians for flux integration are coordinate specific. Since the final integrated value is a surface averaged value, it is inherently related only to the corresponding two dimensions of that surface. For example, while integrating in spherical ($r-\theta$) plane, the one$-$dimensional Jacobians are $\xi$ (not $\xi^2$) and unity (not $sin \xi$) in $r$ and $\theta$ directions respectively. This is because the averaging procedure is independent of the third dimension $\phi$ which adds $rd\phi$ term to the integration. So, the altered one$-$dimensional Jacobians for 2D planar averaging are summarized in Table \ref{tab:2}.

\begin{table}[h!]
\centering
    \caption{One$-$dimensional Jacobian $\big(\frac{\partial{\mathcal{V}}}{\partial\xi}\big)$ values for interface flux reconstruction for the regularly$-$spaced 3D grids }
    \begin{tabular}{ |c | c | c | c |}
    \hline
    \mbox{Grid type}& {Face coordinates ($i-j$)} & {$\frac{\partial{\mathcal{V}_i}}{\partial\xi_i}$}& {$\frac{\partial{\mathcal{V}_j}}{\partial\xi_j}$}\\
    \hline
    Cartesian & ($x-y$),($y-z$),($x-z$) & 1 & 1\\
    \hline
    \multirow{2}{*}{Cylindrical} & ($r-\theta$) & $\xi$ & $1$\\
    \cline{2-4}
      & ($r-z$),($\theta-z$) & $1$ & $1$\\
    \hline
    \multirow{2}{*}{Spherical}  & ($r-\theta$),($r-\phi$) & $\xi$ & $1$\\
    \cline{2-4}
      & ($\theta-\phi$) & $sin\xi$ & $1$\\   
        \hline
    \end{tabular}
\label{tab:2}
\end{table}

Consider a $p^{th}$ order accurate polynomial of any variable, say flux $Q$ in this case, joining $p$ consecutive points, say mid$-$points of the interface as represented in Fig. \ref{fig:1} (right). It can be expressed in the same form as provided in Eq. (\ref{eq:13}), which takes the matrix form given in Eq. (\ref{eq:39}).

\begin{equation} \label{eq:39}
Q_{i}(\xi)=
\begin{pmatrix}
    1       & (\xi-\xi_i^c) & \dots  & (\xi-\xi_i^c)^{p-1} \\
\end{pmatrix}
\begin{pmatrix}

    a_{i,0} \\
    a_{i,1} \\
    \vdots \\
    a_{i,p-1} 
\end{pmatrix}
\end{equation}

But this time, instead of calculating the point values from the line averaged values, vice$-$versa operation is performed. Eq. (\ref{eq:13}) is valid for the values from $i-i_L$ (leftmost value) to $i+i_R$ (rightmost value), where $i_L+i_R+1=p$. A system of $p$ equations is obtained after substituting the values at each considered point, the matrix form of which is given in Eq. (\ref{eq:40}). 

\begin{equation} \label{eq:40}
\begin{pmatrix}

    Q_{i,-i_L} \\
    Q_{i,-i_L+1} \\
    \vdots \\
    Q_{i,i_R} 
\end{pmatrix}
=
\begin{pmatrix}
    1       & (\xi_{i-i_L}-\xi_i^c) & \dots  & (\xi_{i-i_L}-\xi_i^c)^{p-1} \\
    1       & (\xi_{i-i_L+1}-\xi_i^c) & \dots  & (\xi_{i-i_L+1}-\xi_i^c)^{p-1} \\    
    \vdots                & \dots & \ddots  & \vdots \\
    1       & (\xi_{i+i_R}-\xi_i^c) & \dots  & (\xi_{i+i_R}-\xi_i^c)^{p-1} \\
\end{pmatrix}
\begin{pmatrix}

    a_{i,0} \\
    a_{i,1} \\
    \vdots \\
    a_{i,p-1} 
\end{pmatrix}
\end{equation}
where $Q$ is any-arbitrary variable which needs to be averaged in $\big[\xi_{i-\frac{1}{2}},\xi_{i+\frac{1}{2}}\big]$. It can be written in a much simpler matrix form given in Eq. (\ref{eq:41}).

\begin{equation}\label{eq:41}
[\bf{Q}]=[\bf{XI}][\bf{A}]
\end{equation}

where $[{\bf{Q}}]=[Q_{i,-i_L},Q_{i,-i_L+1},...,Q_{i,i_R}]^T$, $[{\bf{XI}}]=\begin{pmatrix}
    1       & (\xi_{i-i_L}-\xi_i^c) & \dots  & (\xi_{i-i_L}-\xi_i^c)^{p-1} \\
    1       & (\xi_{i-i_L+1}-\xi_i^c) & \dots  & (\xi_{i-i_L+1}-\xi_i^c)^{p-1} \\    
    \vdots                & \dots & \ddots  & \vdots \\
    1       & (\xi_{i+i_R}-\xi_i^c) & \dots  & (\xi_{i+i_R}-\xi_i^c)^{p-1} \\
\end{pmatrix}$, and $[{\bf{A}}]=[a_{i,0},a_{i,1},...,a_{i,p-1}]^T$

Using the same procedure as described in Sections \ref{linearweights} and \ref{optimalweights} and performing the average of the polynomial as given in Eq. (\ref{eq:39}) similar to Eq. (\ref{eq:11}) over the domain $[\xi_{i-1/2},\xi_{i+1/2}]$, Eq. (\ref{eq:42}) is obtained.

\begin{equation} \label{eq:42}
{\bar{Q}_i}=[\bf{\widetilde{XI}}][\bf{A}]
\end{equation}
where $[{\bf{\widetilde{XI}}}]=\bigg[\frac{1}{\Delta{\mathcal{V}}_{i}}{\int_{{\xi}_{i-\frac{1}{2}}}^{{\xi}_{i+\frac{1}{2}}}({\xi-\xi_i^c})^{0}\frac{\partial{\mathcal{V}}}{\partial\xi}d\xi}, \frac{1}{\Delta{\mathcal{V}}_{i}}{\int_{{\xi}_{i-\frac{1}{2}}}^{{\xi}_{i+\frac{1}{2}}}({\xi-\xi_i^c})^{1}\frac{\partial{\mathcal{V}}}{\partial\xi}d\xi},...,\frac{1}{\Delta{\mathcal{V}}_{i}}{\int_{{\xi}_{i-\frac{1}{2}}}^{{\xi}_{i+\frac{1}{2}}}({\xi-\xi_i^c})^{p-1}\frac{\partial{\mathcal{V}}}{\partial\xi}d\xi}\bigg]$

From Eqs. (\ref{eq:41}) and (\ref{eq:42}), a general form of equation for integration from a lower dimension to a higher dimension can be derived, as given by Eq. (\ref{eq:43}).

\begin{equation} \label{eq:43}
{\bar{Q}_i}=\{[\bf{\widetilde{XI}}][\bf{XI}]^{-1}\}[\bf{Q}]
\end{equation}

The term $\{[\bf{\widetilde{XI}}][\bf{XI}]^{-1}\}$ includes the weights essential for converting the mid$-$point interface flux values to the line averaged interface flux values, as shown in Fig. \ref{fig:1} (right). The next integration sweep in the transverse direction yields the area$-$averaged flux values at the interface.
The weights for integrations in the corresponding directions are provided in \ref{cart_int_wt}, \ref{cyl_int_wt}, and \ref{sph_int_wt} for the standard cases. Integration is preferred to be performed in the exact vice$-$versa fashion as of reconstruction from the surface averages. 

\subsection{Source term integration} \label{sourcetermavg}
The source terms need to be dealt with extreme accuracy since any contamination in it might deteriorate the high order accuracy. The source term integration is performed based on the works by Mignone \cite{Mignone-2014}. For 1D test cases, it is preferred to reconstruct the mid$-$point of each cell using WENO procedure, weights of which are provided in \ref{cart_source_wt}, \ref{cyl_source_wt}, and \ref{sph_source_wt}. Reconstructing at Gauss$-$Lobatto 4 points (fifth order) instead of mid$-$point and performing quadrature also yields the same results (not shown in the paper), therefore, mid$-$point reconstruction with 3 point Simpson quadrature is advised.

The present work is a significant extension to \cite{Mignone-2014} since point values are considered for the source term evaluation, unlike the constant radius averages \cite{Mignone-2014}, which can only achieve second order of accuracy in multi$-$dimensional problems \cite{zhang2011order,buchmuller2014improved}. The theory for deriving the weights for the source term integration is exactly the same as of flux integration given in Section \ref{extensiontomultid}. However, reconstruction of the source$-$term integration is performed in every dimension, so the original one$-$dimensional Jacobians given in Table \ref{tab:1} can be used for the integration. If non$-$radial integration is performed in the first place, `${1/R}$' factor in all of the tangential terms at $R=0$ will yield an infinite value, so only numerators are integrated with the original weights. Moreover, since the source terms contain `$1/R$' factor, the radial integration weights need to be regularized \cite{Mignone-2014}, by reconsidering the integration of Eq. (\ref{eq:41}) with a regularized factor of the source term in Eq. (\ref{eq:14}) i.e. $ {\int_{{\xi}_{i-\frac{1}{2}}}^{{\xi}_{i+\frac{1}{2}}}\frac{\hat{Q}_i(\xi)}{\xi}\frac{\partial{\mathcal{V}}}{\partial\xi}d\xi} = {{\Delta\mathcal{V}_{i}}}\bar{Q}_{i}$, where $Q$ represents the original source term (e.g. if $Q_i=(p_i/R_i)$, then $\hat{Q}_i=p_i$) in this context.

\begin{figure}
\centering
\begin{subfigure}{0.5\linewidth}
  \centering
  \includegraphics[width=\linewidth]{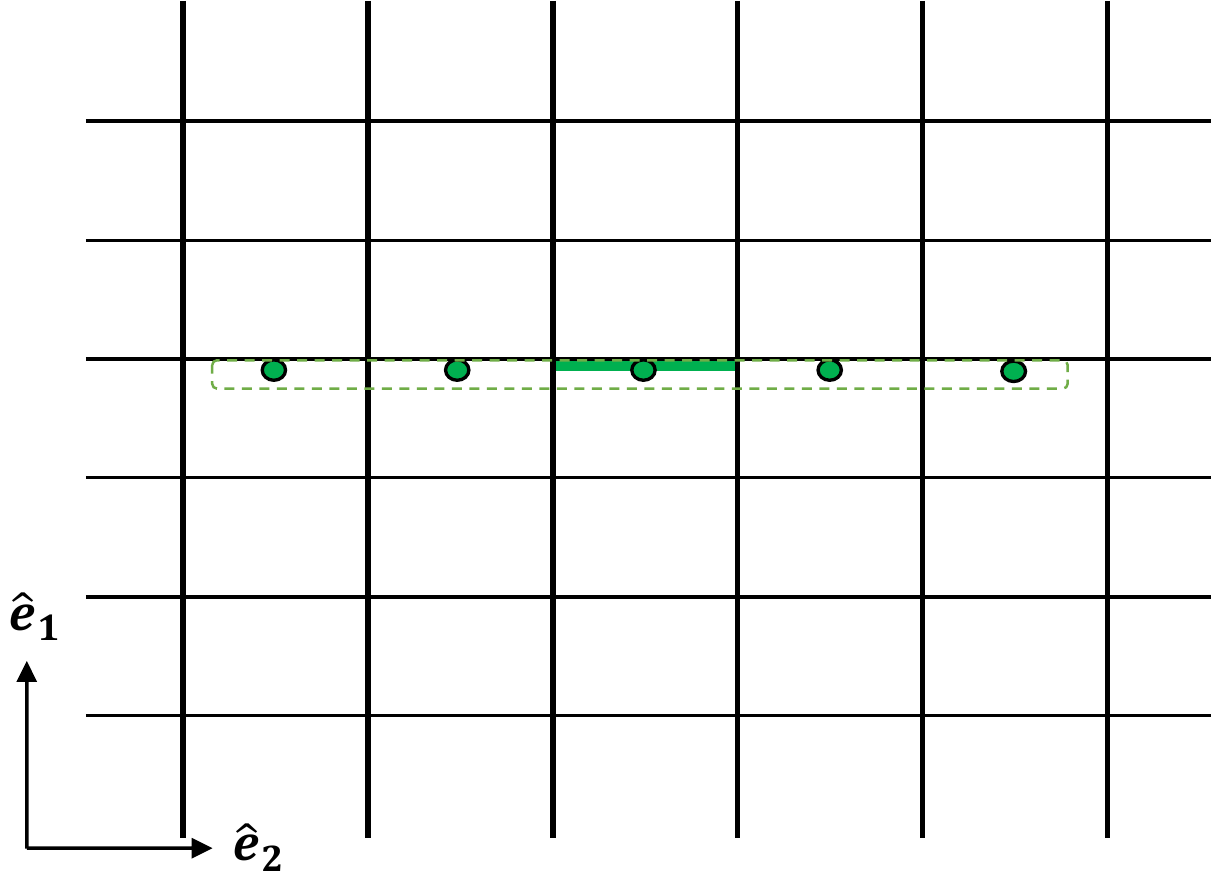}
\end{subfigure}%
\begin{subfigure}{0.5\linewidth}
  \centering
  \includegraphics[width=\linewidth]{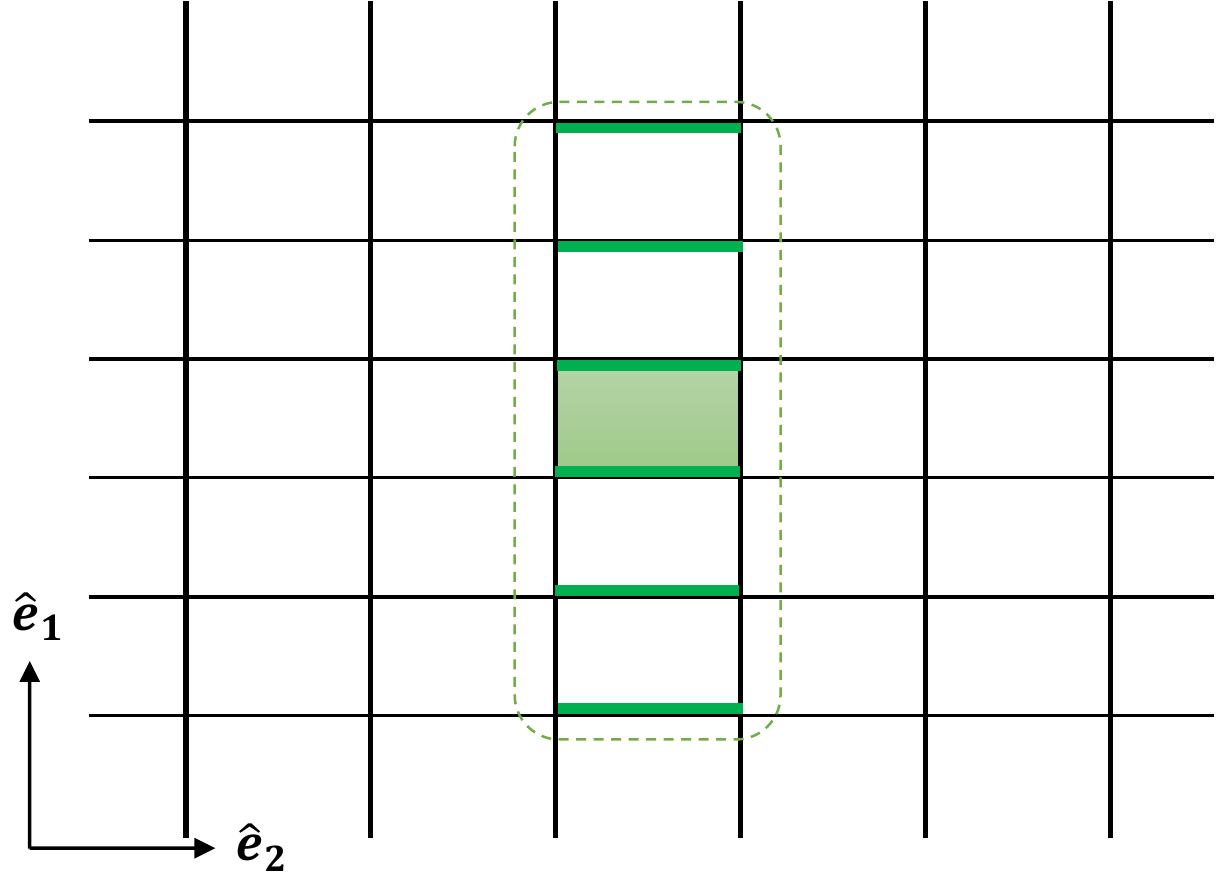}
\end{subfigure}
\caption{Fifth order source term integration procedure. Left: Fifth order using middle values. Right: Sixth order integration using face values}
\label{fig:2}
\end{figure}

First integration tangential to the surface is performed in one direction involving five points, to calculate the line average value of the source term. In the next step, five line averaged values are integrated in the transverse direction to the first sweep, tangential to the interface as shown in Fig. \ref{fig:2} (left). Finally, a face normal interpolation is performed by utilizing the face averaged source terms of six faces i.e. $(i-5/2)^+,(i-3/2)^+,(i-1/2)^+,(i+1/2)^-,(i+3/2)^-,(i+5/2)^-$ faces, as illustrated in Fig. \ref{fig:2} (right). The weights for the source term integration are provided for the standard cases in \ref{cart_source_wt}, \ref{cyl_source_wt}, and \ref{sph_source_wt}.\newline
In addition to the approach discussed above, interior points can also be used to evaluate the source terms. For 1D tests, it is feasible to utilize the mid$-$point values and perform Simpson quadrature to achieve fifth order accuracy using the weights given in the appendix. However, evaluation at the interior points becomes very expensive in multi$-$dimensions.

\subsection{WENO$-$C final algorithm} \label{finalalgo}
The final algorithm for WENO$-$C reconstruction is as follows:
\begin{itemize}
    \item After mesh$-$generation, calculate the values of linear and optimal weights, fifth order middle (mid$-$value) interpolation weights, weights for interface flux and source term integration in every dimension. For standard uniform grids, weights are provided in the appendix.
    \item Convert the volume averaged conservative variables into the interface averaged values by one$-$dimensional WENO sweeps in {{$\bf{{\hat{e}_1}}$}},{{$\bf{{\hat{e}_2}}$}}, and {{$\bf{{\hat{e}_3}}$}} directions using the evaluated weights and smoothness indicator given in Eq. (\ref{eq:39}). Refer to Sections \ref{linearweights}, \ref{optimalweights}, and \ref{smoothnesslimiter}.
    \item Perform reconstruction of the interface averaged variables to mid$-$line averages values in the plane of the interface. Perform another reconstruction of the mid$-$line values in the orthogonal direction to the previous reconstruction in the plane of the interface, to achieve the point value at the mid$-$point of the interface. Refer to Section \ref{extensiontomultid}.
    \item Calculate flux at the mid$-$point of each interface by solving the Riemann problem \cite{toro2013riemann}. 
    \item Perform volume and surface averaging of the source and flux terms respectively using dimensional$-$by$-$dimension approach by the weights provided in the appendix. \textit{Key tip}: If all of the source terms contain `$1/R$' factor, it is advised not to involve radius ($1/{R}$) term in the tangential averaging, if performed before the radial averaging. While radial averaging, regularized relations are preferred, if the considered points contain $R=0$ terms. Refer to Sections \ref{extensiontomultid} and \ref{sourcetermavg}.
\end{itemize}

\section{Numerical tests} \label{tests}
In this section, several tests on scalar and nonlinear system of equations are performed to analyze the performance of the WENO$-$C reconstruction scheme. The test cases include scalar advection (1D) on regularly$-$/irregularly$-$spaced grids, smooth (1D) and discontinuous inviscid flows (1D/2D) governed by a system of nonlinear equations (Euler equations) on regularly$-$spaced grids in cylindrical and spherical coordinates. For the sake of comparison solely on the grounds of the high order reconstruction, time marching in all WENO reconstructed 1D test cases is achieved by explicit third order TVD Runge$-$Kutta scheme \cite{gottlieb1998total,Mignone-2014}. For 2D test cases, explicit fifth order Runge$-$Kutta scheme cite{buchmuller2014improved}, is employed to reduce the computation time. Since high order spatial reconstruction with a lower order time marching requires a lower effective value of CFL number (or time step) to check the dominance of temporal errors over spatial errors, the empirical formula to evaluate the time step is given in Eq. (\ref{eq:44}).

\begin{equation} \label{eq:44}
\Delta t=C_a \Bigg[\max\limits_{\textbf{i}}^{} \Bigg(\frac{1}{D}\Bigg)\sum\limits_{d}^{}\frac{\lambda_{d,\textbf{i}}}{(\Delta l_{d,\textbf{i}})^{(ss/tt)}}\Bigg]^{-1}
\end{equation}

where $C_a$ is the CFL number, $D$ is the number of spatial dimensions $d$, while $\Delta l_d$ and $\lambda_d$ are the grid length and maximum signal speed inside zone $\textbf{i}$ in the direction {\bf{$\bf{{\hat{e}_d}}$}}. $ss$ and $tt$ are the spatial and temporal orders of convergence respectively.

For all tests performed in this paper, the initial condition on the conserved variables is averaged over the corresponding finite volumes $\Delta \mathcal{V}_{\textbf{i}}$ using seven$-$point Gaussian quadrature in a dimension$-$by$-$dimension fashion. Numerical benchmark test cases for the scalar conservation laws are reported in Section \ref{scalaradvactiontests}, while the verification tests for nonlinear systems are presented in Section \ref{eulertest}.
Errors $\epsilon_1$ are computed using the $L_1$ discrete norm defined in Eq. (\ref{eq:45}). In case of a linear system, $Q$ is a generic flow quantity while in case of a nonlinear system of equations, error in density $\rho$ is considered. 

\begin{equation} \label{eq:45}
\epsilon_1(Q)=\frac{\sum\limits_{\textbf{i}}^{} |\bar{Q}_{\textbf{i}}-\bar{Q}_{\textbf{i}}^{ref} | \Delta{\mathcal{V}_{\textbf{i}} }}{\sum\limits_{\textbf{i}}^{}\Delta{\mathcal{V}_{\textbf{i}} }}
\end{equation}
where summation is performed on all finite volumes $ \Delta \mathcal{V}_{\textbf{i}} $ with $\bar{Q}_{\textbf{i}}^{ref}$ to be the volume average of the reference (or exact) solution. Finally, the experimental order of convergence ($EOC$) is computed from Eq. (\ref{eq:46}).
\begin{equation} \label{eq:46}
EOC=\frac{log \Bigg(\frac{\epsilon_1^c(Q)}{\epsilon_1^f(Q)}\Bigg)}{log \Bigg(\frac{\prod \limits_{d=1}^{D}N^f_d}{\prod \limits_{d=1}^{D}N^c_d}\Bigg)}
\end{equation}
where the superscript $c$ and $f$ refer to the coarse and fine mesh respectively and $N$ is the number of finite volumes in {\bf{$\hat{e}_d$}} direction.

\subsection{Scalar advection tests} \label{scalaradvactiontests}
As a first benchmark, 1D scalar advection equations Eq. (\ref{eq:48}) in cylindrical$-$radial and spherical$-$radial coordinates, and Eq. (\ref{eq:52}) in spherical$-$meridional coordinates are solved. Two different tests (tests A and B) are performed on a regularly$-$spaced grid, while test A is also performed on an irregularly$-$spaced grid. Test A subsumes a monotonic profile while test B is a more stringent test involving a non$-$monotonic profile.
For the irregularly$-$spaced grid, the grid spacing increases linearly with the radial distance. The summation of all zone lengths is fixed, i.e., length of the computational domain and the number of cells $N$ is given. A parameter $Ratio$ is introduced in Eq. (\ref{eq:47}) which is an indicator of the level of non$-$uniformity in the computational domain.
\begin{equation} \label{eq:47}
Ratio=\frac{\text{Grid spacing of any cell in an N$-$cell uniform grid}}{\text{Grid spacing of the first cell (or the smallest cell) in an N$-$cell nonuniform grid}}
\end{equation}

\subsubsection{Advection equation in cylindrical$-$radial and spherical$-$radial coordinates}  \label{scalaradvactiontestscylsph}
The governing 1D scalar advection equation in cylindrical$-$radial and spherical$-$radial coordinates is formulated in Eq. (\ref{eq:48}).

\begin{equation} \label{eq:48}
\frac{\partial{Q}}{\partial{t}}+\frac{1}{\xi^m}\frac{\partial{}}{\partial{\xi}}(\xi^m Q v) =0
\end{equation}

where the $\xi^m$ is the one$-$dimensional Jacobian and therefore, $m=1$ and $2$ respectively correspond to cylindrical$-$radial and spherical$-$radial coordinates. Velocity $v$ varies linearly with the radial coordinate $\xi$ i.e. $v=\alpha \xi$ and $\alpha=1$. Eq. (\ref{eq:48}) admits an exact solution given in Eq. (\ref{eq:49}).

\begin{equation} \label{eq:49}
Q^{ref}(\xi,t)=e^{-(m+1) \alpha t}Q(\xi e^{-\alpha t},0)
\end{equation}

where $Q(\xi e^{-\alpha t},0)$ is the initial condition. For the present case, a Gaussian profile, given in Eq. (\ref{eq:50}), is employed. 

\begin{equation} \label{eq:50}
Q(\xi,0)=e^{-a^2(\xi-b)^2}
\end{equation}
where $a$ and $b$ are constants. For the two test cases, $\{a=10, b=0\}$ is employed for test A which yields a monotonically decreasing profile and $\{a=16, b=1/2\}$ is employed for test B corresponds to a more stringent non$-$monotonic profile having a maxima at $\xi=1/2$. The computational domain extends from $\xi=0$ to $\xi=2$ consisting of $N$ zones, where boundary conditions include symmetry at the origin ($\xi=0$) and zero$-$gradient at $\xi=2$. Computations are performed until $t=1$ with CFL number of $0.9$ and the interface flux is computed using Eq. (\ref{eq:51}).  

\begin{equation} \label{eq:51}
\tilde{F}_{i+\frac{1}{2}}=\frac{1}{2}\Bigg[v_{i+\frac{1}{2}}(Q_{i+1}^-+Q_i^+)-|v_{i+\frac{1}{2}}|(Q_{i+1}^-+Q_i^+)\Bigg]
\end{equation}

\begin{figure}[ht] 
  \begin{minipage}[b]{0.5\linewidth}
    \includegraphics[width=\linewidth]{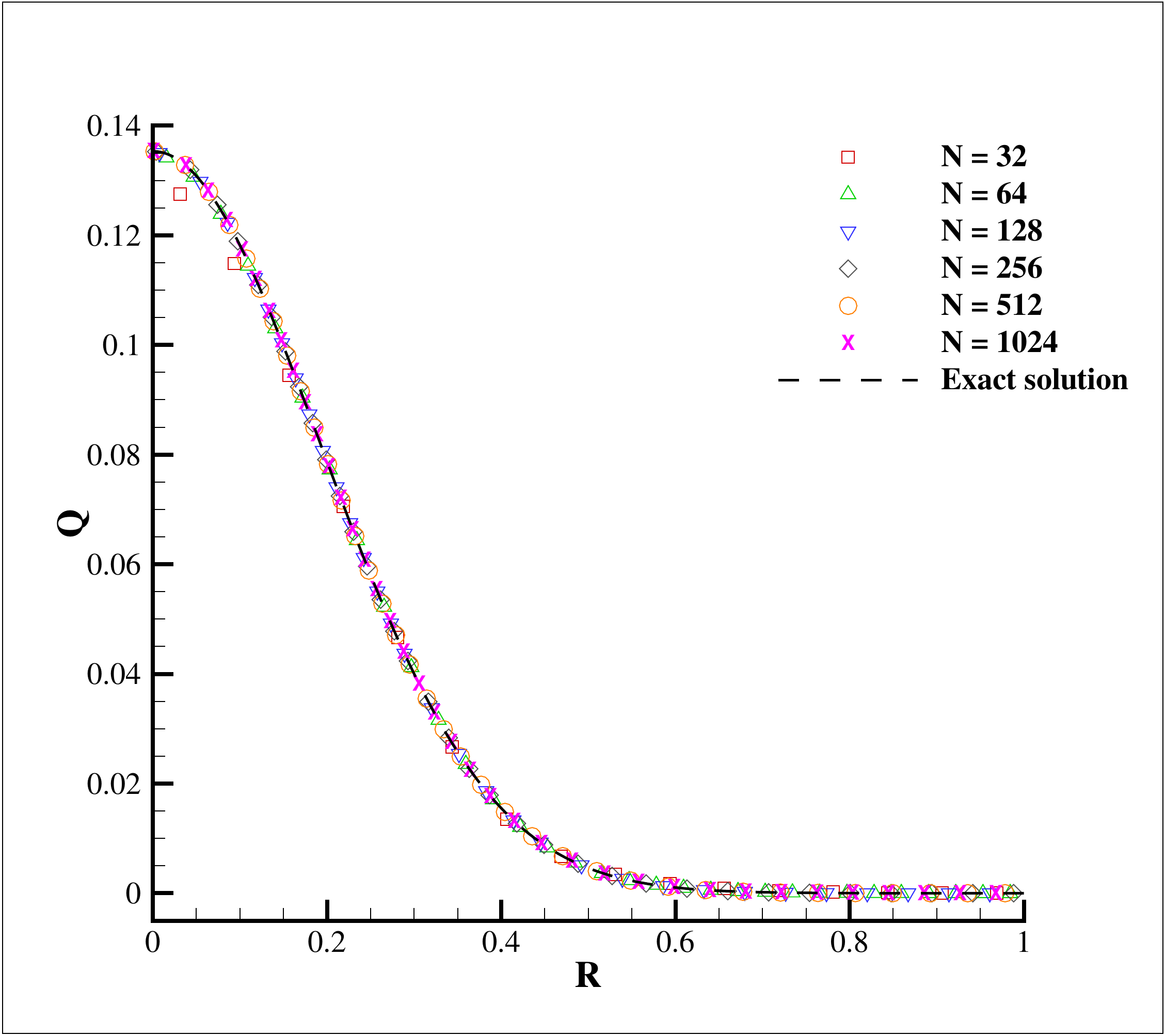} 
  \end{minipage} 
  \begin{minipage}[b]{0.5\linewidth}
    \includegraphics[width=\linewidth]{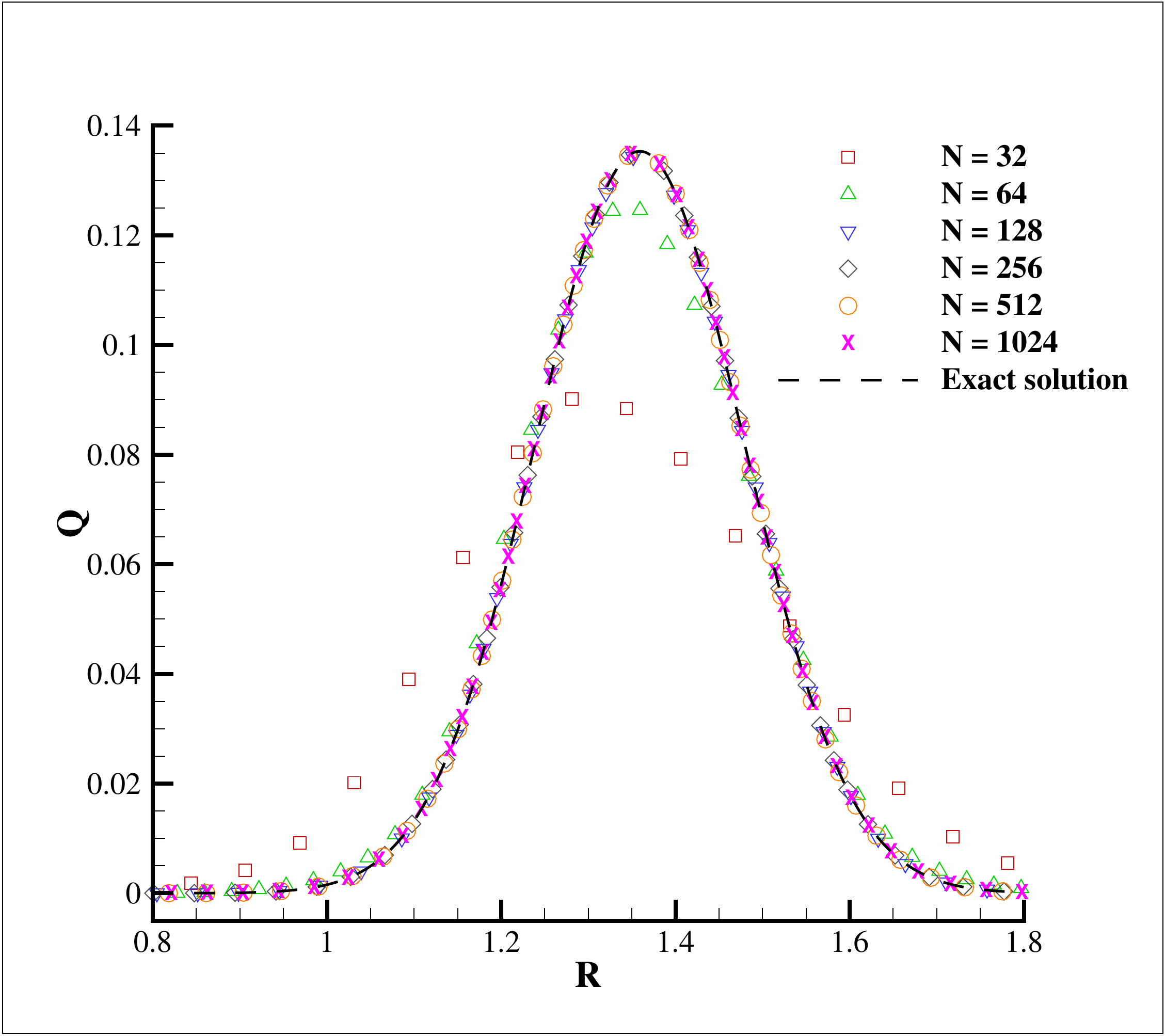} 
  \end{minipage} 
  \begin{minipage}[b]{0.5\linewidth}
    \includegraphics[width=\linewidth]{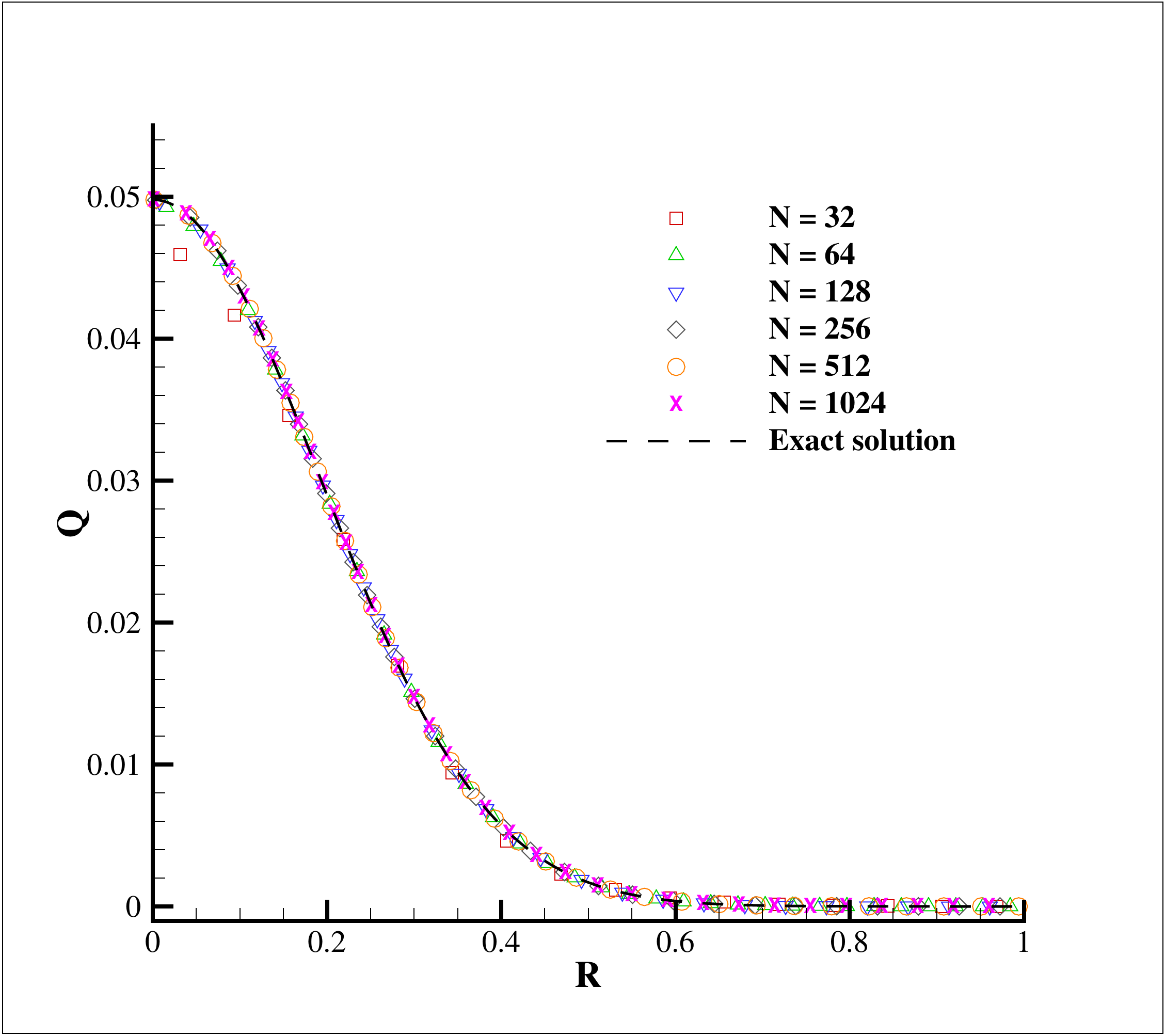} 
  \end{minipage}
  \begin{minipage}[b]{0.5\linewidth}
    \includegraphics[width=\linewidth]{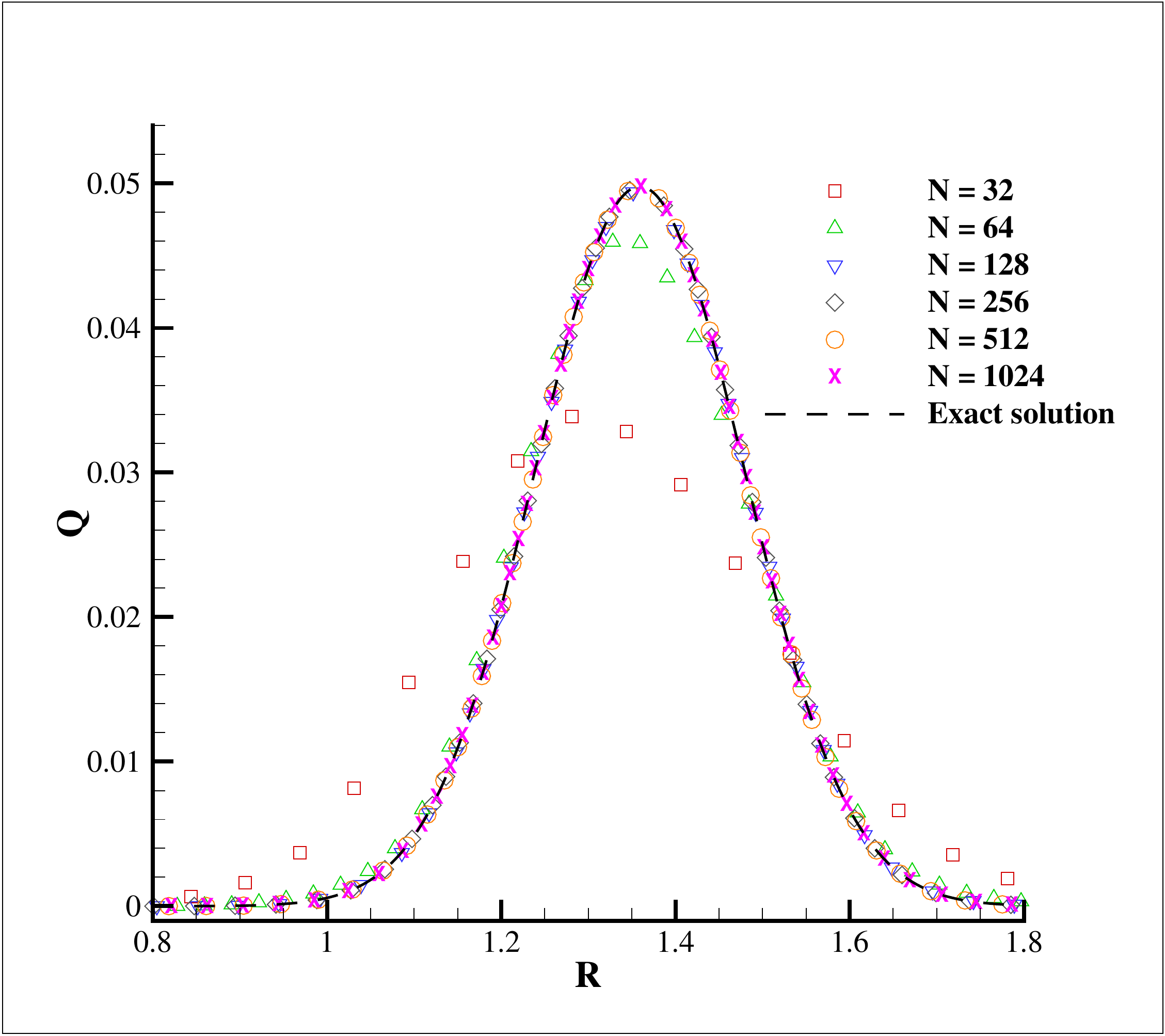} 
  \end{minipage} 
  \caption{Spatial profiles at $t=1$ for the radial advection problem in cylindrical$-$radial (top) and spherical$-$radial (bottom) coordinates. Left and right figures correspond to test A \{$a=10,b=0$\} and test B \{$a=16,b=1/2$\} respectively.}
  \label{fig:3} 
\end{figure} 

Fig. \ref{fig:3} shows the spatial variation of $Q$ with the radial distance ($\xi=R$) for the two test cases (tests A and B) on a uniform grid in cylindrical$-$radial (top) and spherical$-$radial (bottom) coordinates. For a monotonically decreasing profile (test A), even $N \ge 64$ gives accurate results for both the test cases. However, for test B, $N=64$ yields slightly lower peaks than the exact solution. 
When compared with Fig. 2 of Mignone \cite{Mignone-2014}, a slightly higher peak is observed for test A, since it is a less severe test case. The differences are much more prominent while performing test B. It can be observed that the peaks of $N=64$ for test B in Fig. \ref{fig:3} are significantly higher than earlier published results \cite{Mignone-2014}.

\begin{table}[]
\centering
\caption{$L_1$ norm errors and experimental order of convergence ($EOC$) for radial advection test in cylindrical$-$radial and spherical$-$radial coordinates at $t=1$ for test A \{$a=10,b=0$\} and test B \{$a=16,b=1/2$\}.}

\begin{tabular}{c|cc|cc|cc|cc}
\hline
& \multicolumn{4}{c}{Cylindrical} & \multicolumn{4}{|c}{Spherical}\\\cline{2-5} \cline{6-9} 
& \multicolumn{2}{c|}{Test A} &\multicolumn{2}{c|}{Test B} & \multicolumn{2}{c|}{Test A} & \multicolumn{2}{c}{Test B} \\\cline{2-3} \cline{4-5} \cline{6-7} \cline{8-9} 
$N$          & $\epsilon_1(Q)$    & $O_{L_1}$ & $\epsilon_1(Q)$    & $O_{L_1}$ & $\epsilon_1(Q)$    & $O_{L_1}$ & $\epsilon_1(Q)$    & $O_{L_1}$ \\
\hline
\hline
32         & 9.22E-05 &    $-$   & 1.07E-02 &      $-$      & 1.19E-05 &    $-$   & 3.94E-03 &   $-$    \\
64         & 1.14E-05 & 3.016 & 2.10E-03 & 2.356      & 1.28E-06 & 3.208 & 7.94E-04 & 2.312 \\
128        & 4.91E-07 & 4.537 & 1.95E-04 & 3.425      & 5.28E-08 & 4.602 & 7.44E-05 & 3.415 \\
256        & 1.94E-08 & 4.663 & 9.39E-06 & 4.378      & 2.16E-09 & 4.610 & 3.58E-06 & 4.378 \\
512        & 6.20E-10 & 4.965 & 3.14E-07 & 4.900      & 6.34E-11 & 5.093 & 1.19E-07 & 4.906 \\
1024       & 5.81E-11 & 3.415 & 1.02E-08 & 4.941      & 4.53E-12 & 3.806 & 3.88E-09 & 4.942
\end{tabular}  \label{tab:3}
\end{table}

From the experimental order of convergence ($ EOC $) Table \ref{tab:3}, it is clear that WENO$-$C approaches to the desired fifth order of convergence. The same tests performed in Cartesian coordinates using conventional WENO and present WENO$-$C (both are equivalent) showed same errors and order of convergence (not shown here), and similar behavior as of the cylindrical and spherical grid cases. When compared with Table 1 in \cite{Mignone-2014}, present results indicate a superior performance in terms of accuracy and order of convergence. Modified piecewise parabolic method (PPM$_5$) approaches the fifth order of convergence for test A. However, its order drops down to $\sim 2.4$ for test B \cite{Mignone-2014}.

\begin{figure}
\centering
\begin{subfigure}{.5\textwidth}
  \centering
  \includegraphics[width=\linewidth]{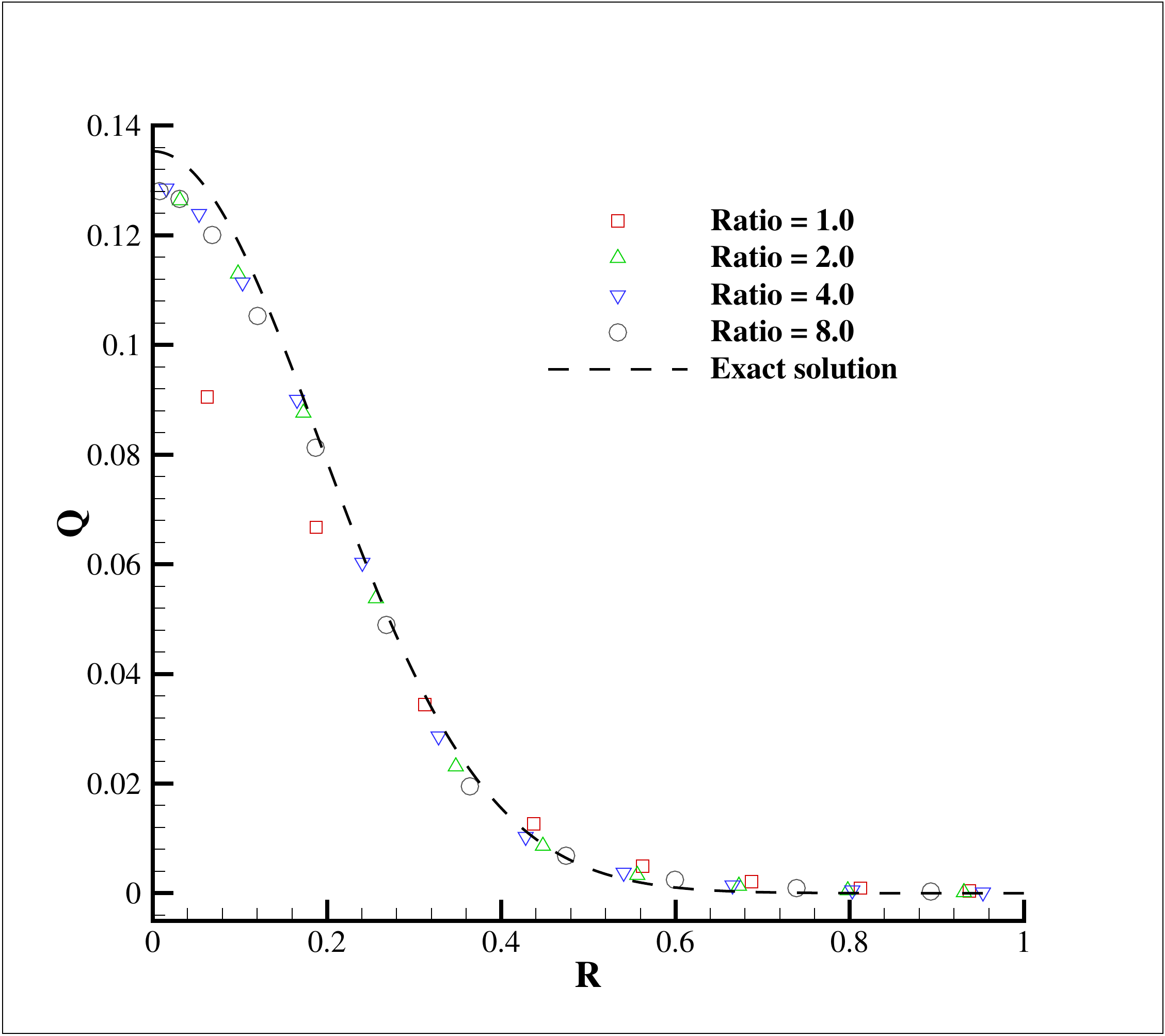}
  \label{fig:sub1.1}
\end{subfigure}%
\begin{subfigure}{.5\textwidth}
  \centering
  \includegraphics[width=\linewidth]{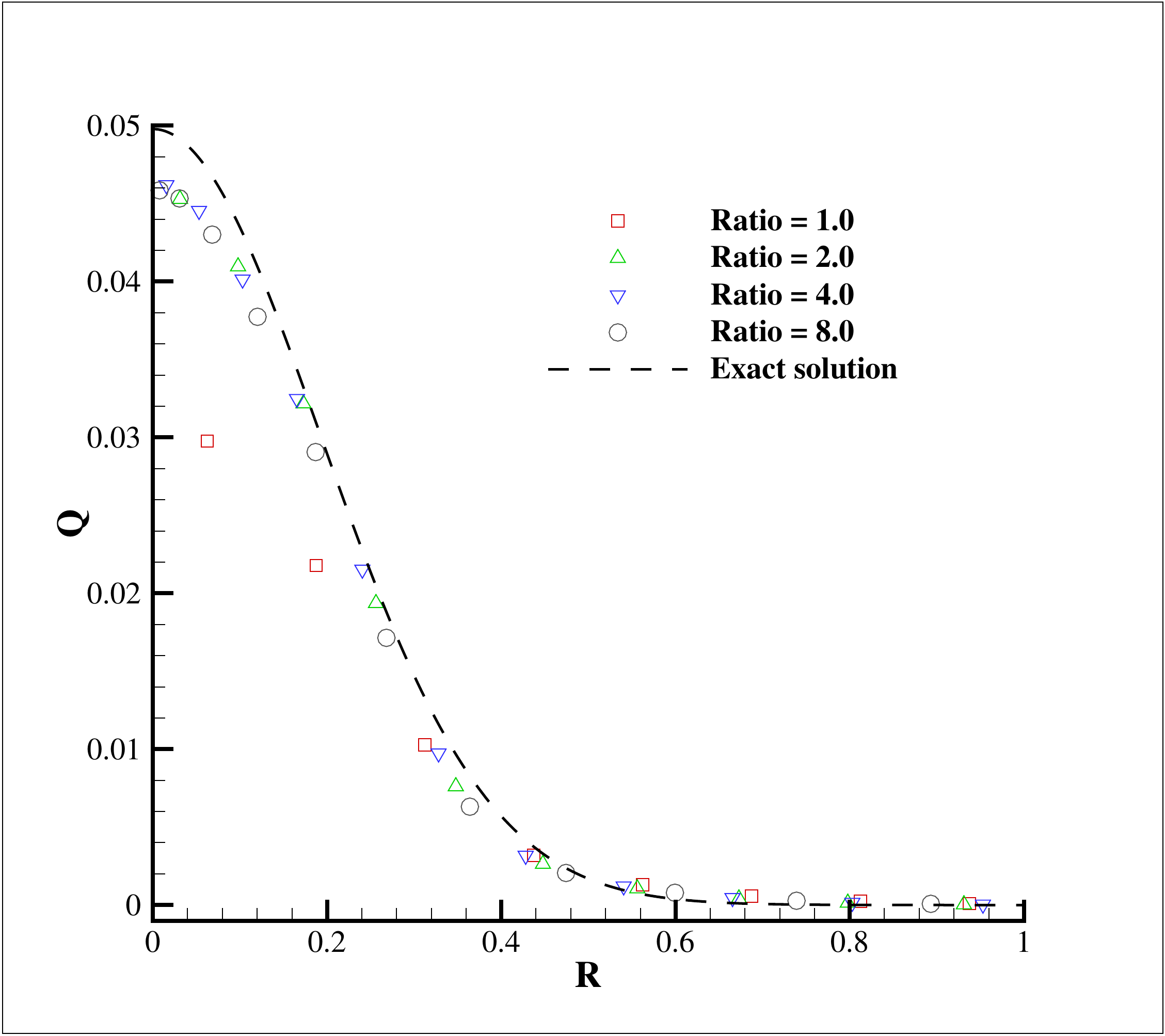}
  \label{fig:sub1.2}
\end{subfigure}
\caption{Spatial profiles at $t=1$ for the radial advection problem (test A: \{$a=10,b=0$\}) using $N=16$ with different values of $Ratio$ (degree of non$-$uniformity) in cylindrical$-$radial (left) and spherical$-$radial (right) coordinates}
\label{fig:4}
\end{figure}

Fig. \ref{fig:4} illustrates the spatial variation of the conserved variable $Q$ on a non$-$uniform grid ($N=16$) during test A. It can be clearly interpreted from the plot that the numerical results approach towards the exact solution with an increase in $Ratio$ (defined in Eq. (\ref{eq:47})), i.e., biasing towards the origin. It can be well analyzed from Table \ref{tab:4} that a considerable reduction in errors is observed along with a rapid increase of $EOC$ to desired fifth order when the grid spacing is biased towards the origin.

\begin{table}[]
\centering
\caption{$L_1$ norm errors and experimental order of convergence ($EOC$) for the radial advection problem (test A: \{$a=10,b=0$\}) with different values of $Ratio$ (degree of non$-$uniformity) in cylindrical$-$radial and spherical$-$radial coordinates}
\begin{tabular}{c|cc|cc|cc|cc}
\hline
& \multicolumn{2}{c|}{$Ratio=1$} &\multicolumn{2}{c|}{$Ratio=2$} & \multicolumn{2}{c|}{$Ratio=4$} & \multicolumn{2}{|c}{$Ratio=8$} \\\cline{2-3} \cline{4-5} \cline{6-7} \cline{8-9} 
$N$          & $\epsilon_1(Q)$    & $O_{L_1}$ & $\epsilon_1(Q)$    & $O_{L_1}$ & $\epsilon_1(Q)$    & $O_{L_1}$ & $\epsilon_1(Q)$    & $O_{L_1}$ \\
\hline
\hline
& \multicolumn{8}{c}{Cylindrical}\\
\hline
16   & 5.54E-04 &    $-$   & 1.85E-04    &    $-$   & 1.70E-04    &  $-$     & 1.80E-04  &   $-$    \\
32   & 9.22E-05 & 2.587 & 3.44E-05    & 2.429 & 2.78E-05    & 2.607 & 3.03E-05 & 2.573 \\
64   & 1.14E-05 & 3.016 & 1.81E-06    & 4.247 & 1.26E-06    & 4.468 & 1.39E-06 & 4.440 \\
128  & 4.91E-07 & 4.537 & 7.89E-08    & 4.519 & 5.47E-08    & 4.523 & 5.96E-08 & 4.548 \\
\hline
& \multicolumn{8}{c}{Spherical}\\
\hline
16   & 5.32E-05 &    $-$   & 2.40E-05 &   $-$    & 2.19E-05 &   $-$    & 2.47E-05 &     $-$  \\
32   & 1.19E-05 & 2.167 & 4.48E-06   & 2.420 & 3.81E-06 & 2.523 & 4.20E-06 & 2.557 \\
64   & 1.28E-06 & 3.208 & 2.33E-07 & 4.267 & 1.72E-07  & 4.475 & 1.92E-07  & 4.449 \\
128  & 5.28E-08 & 4.602 & 9.64E-09 & 4.594 & 6.90E-09 & 4.635 & 7.57E-09 & 4.669
\end{tabular} \label{tab:4}
\end{table}

\subsubsection{Advection equation in spherical$-$meridional coordinates}  \label{scalaradvactiontestssphmer}
The governing 1D scalar advection equation in spherical$-$meridional coordinates is given in Eq. (\ref{eq:52}).

\begin{equation} \label{eq:52}
\frac{\partial{Q}}{\partial{t}}+\frac{1}{sin \theta}\frac{\partial{}}{\partial{\theta}}(sin \theta Q v) =0
\end{equation}

where the velocity $v$ varies linearly with the $\theta$ coordinate  i.e. $v=\alpha \theta$ and $\alpha=1$. Eq. (\ref{eq:52}) admits an exact solution given in Eq. (\ref{eq:53}).

\begin{equation} \label{eq:53}
Q^{ref}(\xi,t)=e^{-\alpha t}\frac{sin\big(e^{-\alpha t \theta}\big)}{sin \theta} Q\big(e^{-\alpha t}\theta,0\big)
\end{equation}

A 1D computational grid spanning the interval $\theta \in [0,\pi/2]$ is divided into $N$ zones. Initial condition ($t=0$) for the problem is given in Eq. (\ref{eq:54}).

\begin{equation} \label{eq:54}
  Q(\theta,0) =
    \begin{cases}
       \text{$\Bigg[\frac{1+cos(a(\theta- b))}{2}\Bigg]^2$} & \text{$|\theta - b|<\frac{\pi}{a}$}\\
       0 & \text{otherwise}\\
    \end{cases}
\end{equation}

where $a$ and $b$ are constants. Two different tests are performed namely, test A with $\{a=10, b=0\}$ yielding a monotonically decreasing profile and a more stringent test B $\{a=16, b=\pi/a\}$ resulting in a non$-$monotonic profile having a maxima at $\theta=\pi/a$. The computational domain extends from $\theta=0$ to $\theta=\pi/2$, where the boundary conditions include symmetry at the origin ($\theta=0$) and zero$-$derivative at $\theta=\pi/2$. Computations are performed till $t=1$ with CFL number of $0.9$ and the interface flux is computed using Eq. (\ref{eq:51}).

\begin{figure}
\centering
\begin{subfigure}{.5\textwidth}
  \centering
  \includegraphics[width=\linewidth]{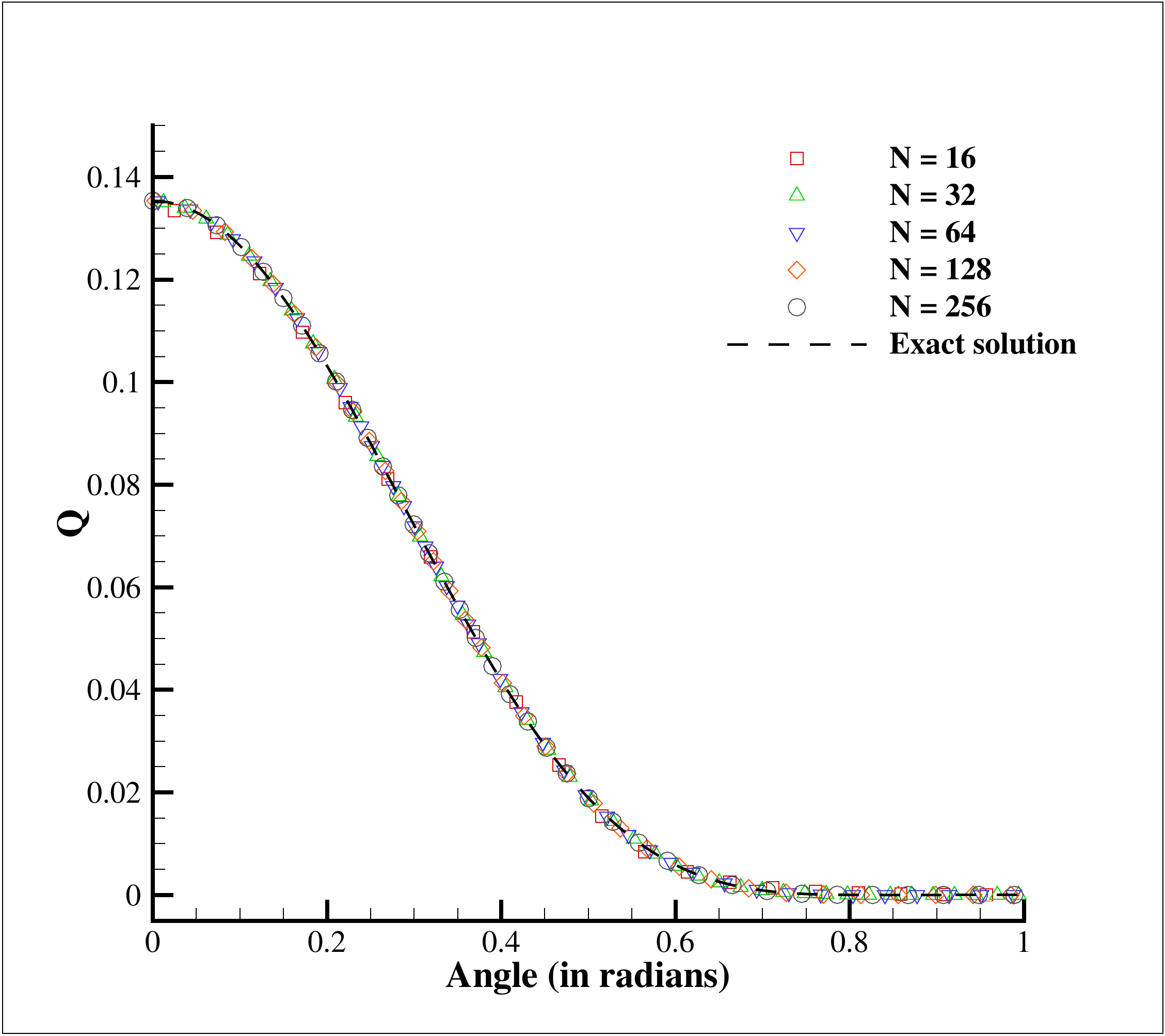}
\end{subfigure}%
\begin{subfigure}{.5\textwidth}
  \centering
  \includegraphics[width=\linewidth]{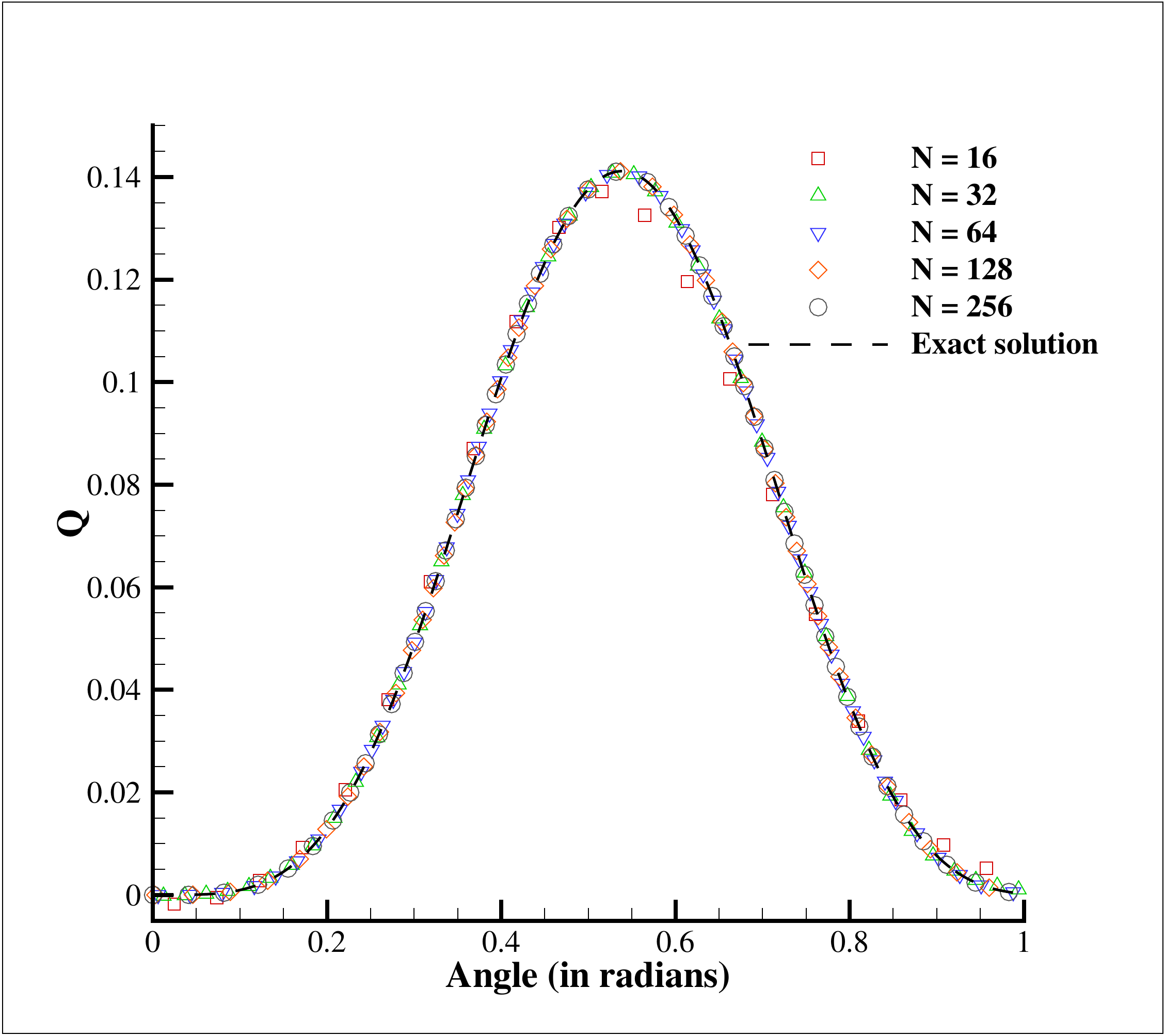}
\end{subfigure}
\caption{Spatial profiles at $t=1$ for the scalar advection problem in spherical$-$meridional coordinates with different mesh points. Left and right subfigures refer to test A $\{a=10, b=0\}$ and B \{$a=15,b=\pi/a$\} respectively.}
\label{fig:5}
\end{figure}

\begin{figure}
\centering
  \includegraphics[width=0.5\linewidth]{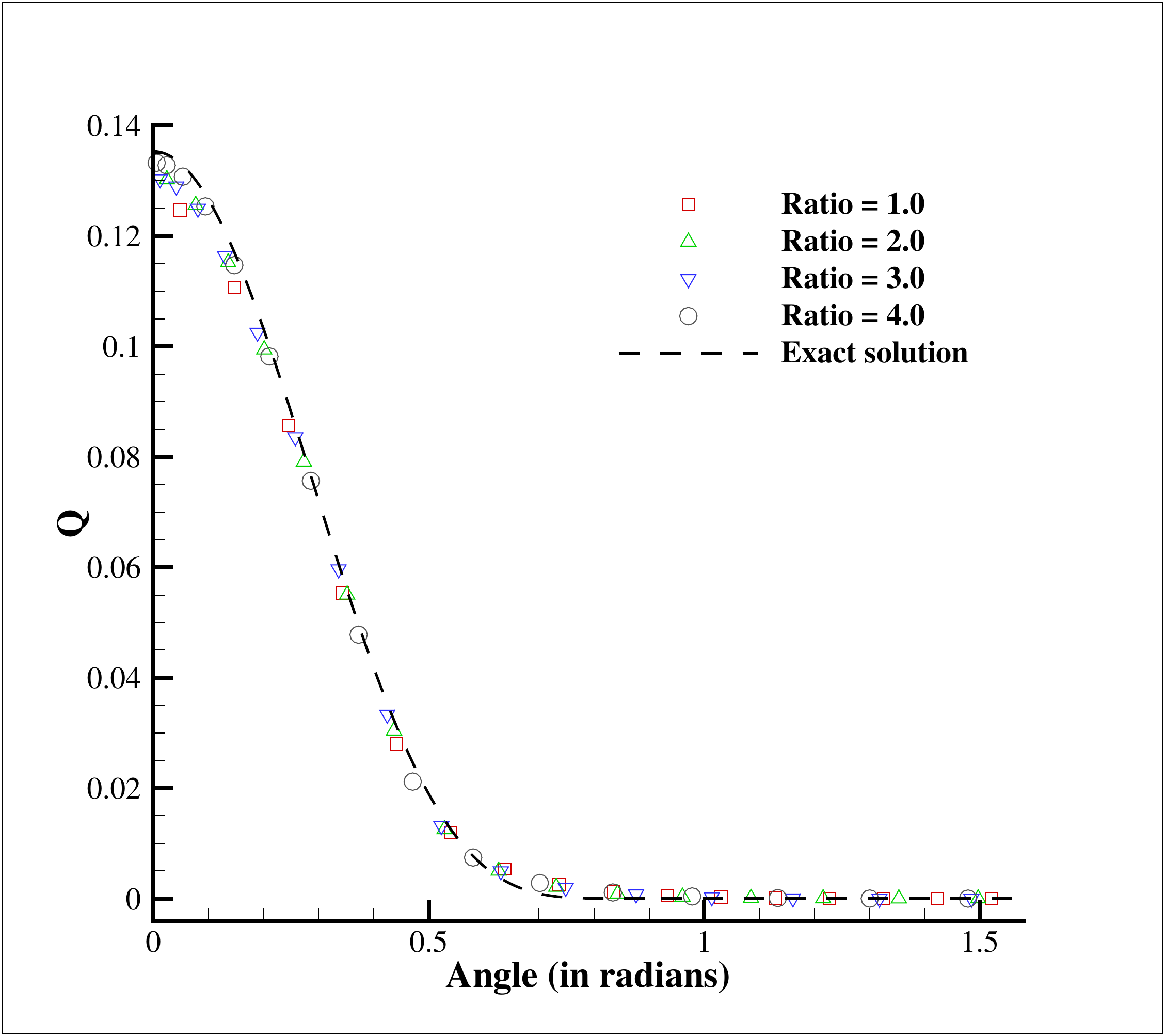}
\caption{Spatial profiles at $t=1$ for the scalar advection problem (test A: \{$a=10,b=0$\}) using $N=16$ with different values of $Ratio$ (degree of non$-$uniformity) in spherical$-$meridional coordinates.}
\label{fig:6}
\end{figure}

Fig. \ref{fig:5} shows the variation of conserved variable $ Q$ with angle $\theta$ for both the tests. For test A, even $N=16$ give accurate results, while for test B, $N \ge 32$ provide a good approximation of the exact solution. Table \ref{tab:5} illustrates the achievement of the desired fifth order of convergence for both the test cases. When the results obtained by the present scheme are compared with the previously proposed schemes (Table 2 of \cite{Mignone-2014}), it can be realized that WENO$-$C shows superior performance. For the non$-$uniform mesh case, a fifth order of convergence is still preserved with a rapid achievement, as summarized in Table \ref{tab:6}. Moreover, Fig. \ref{fig:6} shows that mesh biasing leads to a significant reduction in the errors when compared with a uniform mesh of the same number of cells.

\begin{table}[h!]
\centering
    \caption{$L_1$ norm errors and experimental order of convergence ($EOC$) for scalar advection test in spherical$-$meridional coordinates coordinates at $t=1$ for test A \{$a=10,b=0$\} and test B \{$a=16,b=\pi/a$\} respectively.}
\begin{tabular}{c|cc|cc}
\hline
&\multicolumn{2}{c|}{Test A} &\multicolumn{2}{c}{Test B} \\\cline{2-3} \cline{4-5} 
$N$          & $\epsilon_1(Q)$    & $O_{L_1}$ & $\epsilon_1(Q)$    & $O_{L_1}$\\
\hline
\hline
32  & 1.71E-04 &   $-$    & 1.57E-03 &   $-$    \\
64  & 1.99E-05 & 3.103 & 2.11E-04 & 2.894 \\
128 & 7.10E-07 & 4.808 & 1.62E-05 & 3.699 \\
256 & 2.25E-08 & 4.978 & 4.81E-07 & 5.078
\end{tabular} \label{tab:5}
\end{table}

\begin{table}[]
\centering
\caption{$L_1$ norm errors and experimental order of convergence ($EOC$) for the scalar advection problem (test A: \{$a=10,b=0$\}) in spherical$-$meridional coordinates with different values of $Ratio$ (degree of non$-$uniformity)}
\begin{tabular}{c|cc|cc|cc|cc}
\hline
& \multicolumn{2}{c|}{$Ratio=1$} &\multicolumn{2}{c|}{$Ratio=2$} & \multicolumn{2}{c|}{$Ratio=4$} & \multicolumn{2}{|c}{$Ratio=8$} \\\cline{2-3} \cline{4-5} \cline{6-7} \cline{8-9} 
$N$          & $\epsilon_1(Q)$    & $O_{L_1}$ & $\epsilon_1(Q)$    & $O_{L_1}$ & $\epsilon_1(Q)$    & $O_{L_1}$ & $\epsilon_1(Q)$    & $O_{L_1}$ \\
\hline
\hline
16  & 7.43E-04 &   $-$    & 4.27E-04 &    $-$    & 4.61E-04 &    $-$    & 5.05E-04 &    $-$    \\
32  & 1.71E-04 & 2.120 & 9.18E-05 & 2.217 & 1.01E-04 & 2.195 & 1.16E-04 & 2.128 \\
64  & 1.99E-05 & 3.103 & 8.33E-06 & 3.463 & 9.13E-06 & 3.465 & 1.07E-05 & 3.438 \\
128 & 7.10E-07 & 4.808 & 2.45E-07 & 5.085 & 2.72E-07 & 5.069 & 3.24E-07 & 5.040
\end{tabular} \label{tab:6}
\end{table}

\subsection{Euler equations based tests} \label{eulertest}
The present reconstruction scheme is now tested for more challenging test cases involving nonlinear systems of equations, i.e., Euler equations. Although primitive variable reconstruction is preferred in the past due to the well$-$behaved results, in the case of curvilinear coordinates, the involvement of the higher order derivatives in the extraction of the primitive variables causes spurious oscillations \cite{Mignone-2014}. Therefore, we restrict our work to the reconstruction of the conserved variables instead of computationally expensive and intricate primitive variable reconstruction. Maximum characterstic speed is employed to evaluate the time step from Eq. (\ref{eq:44}). Several tests are performed in cylindrical and spherical coordinates to investigate the essentially non$-$oscillatory property of WENO$-$C for discontinuous flows and the convex combination property for smooth flows.

\subsubsection{Isothermal radial wind problem}

The isothermal 1D radial wind problem is performed to analyze the deviations of spatial reconstruction schemes near the origin in curvilinear coordinates \cite{Mignone-2014}.
 The general form of Euler equation in 1D Cartesian / cylindrical$-$radial / spherical$-$radial coordinates can be written in the form of Eq. (\ref{eq:55}).
 
 \begin{equation} \label{eq:55}
\frac{\partial{}}{\partial{t}}
\begin{pmatrix}

    \rho \\
    \rho v \\
     E 
\end{pmatrix}
+\frac{1}{\xi^m}\frac{\partial{}}{\partial{\xi}}
\begin{pmatrix}

    \rho v \xi^m \\
    (\rho v^2+p) \xi^m \\
    (E+p)v \xi^m
\end{pmatrix}
=
\begin{pmatrix}

    0\\
    mp/\xi\\
    0
\end{pmatrix}
\end{equation}

where $\rho$ is the mass density, $v$ is the radial velocity, $p$ is the pressure, $E$ is the total energy, and $m=0,1,$ and $2$ for Cartesian, cylindrical$-$radial ($\xi=R$), and spherical$-$radial ($\xi=r$) coordinates respectively. For an isothermal flow, the energy equation is discarded whereas Eq. (\ref{eq:56}) serves as the adiabatic equation of state (EOS).

\begin{equation} \label{eq:56}
E=\frac{p}{\gamma - 1}+\frac{1}{2} \rho v^2
\end{equation}
where $\gamma=5/3$ is assumed for this case. At $\xi=0$, axisymmetric boundary conditions apply, while at the outer edge, density, pressure, and scaled velocity (${v}/\bar{\xi}$) have zero gradients. The initial conditions are provided in Eq. (\ref{eq:57}) and the interface flux is evaluated with Lax-Friedrichs scheme with local speed estimate \cite{rusanov1962calculation}.

\begin{equation} \label{eq:57}
\rho(\xi,0)=1;\quad\quad v(\xi,0)=100\xi;\quad\quad p(\xi,0)=1/\gamma
\end{equation}

\begin{figure}
\centering
\begin{subfigure}{.5\textwidth}
  \centering
  \includegraphics[width=\linewidth]{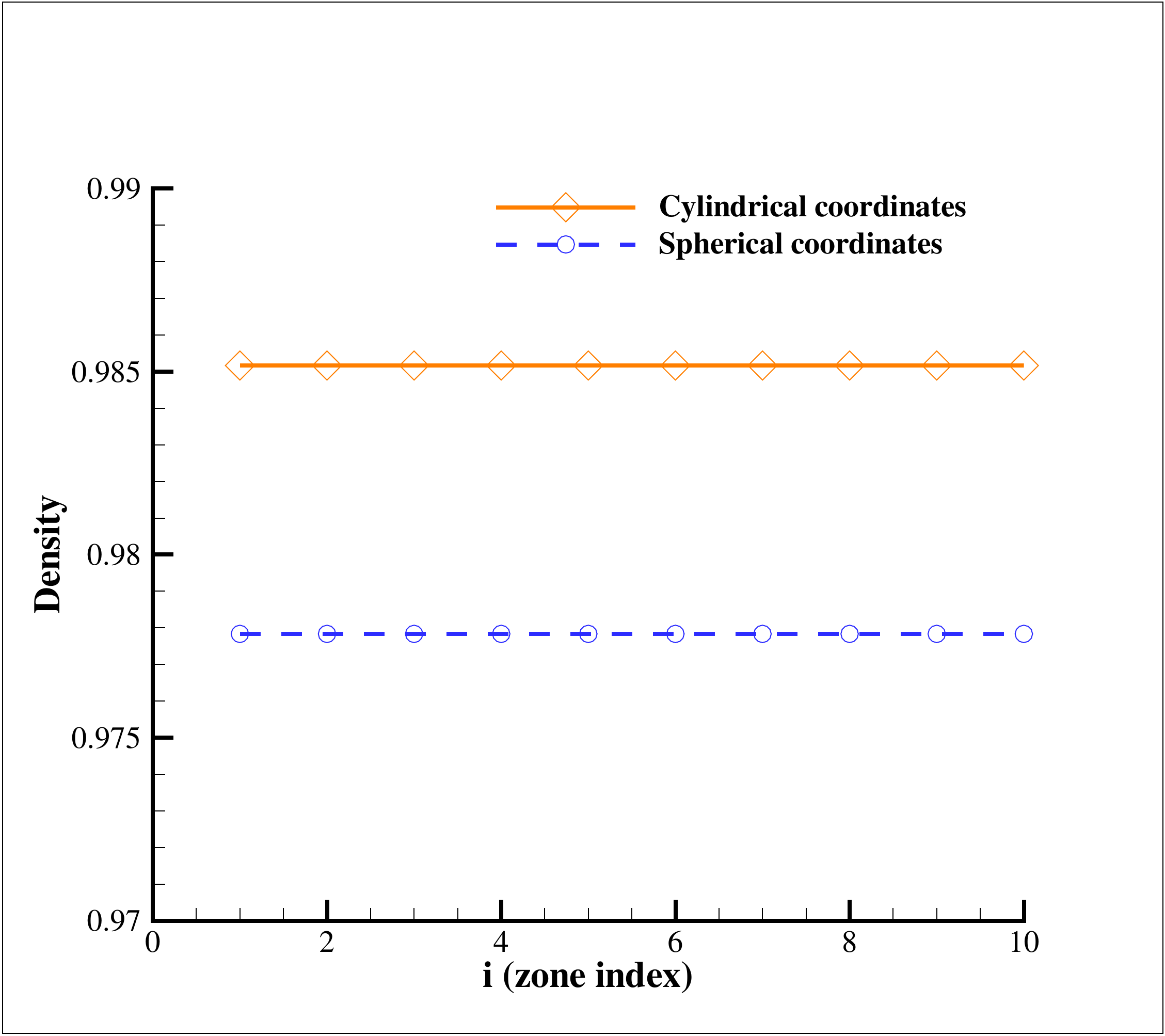}
\end{subfigure}%
\begin{subfigure}{.5\textwidth}
  \centering
  \includegraphics[width=\linewidth]{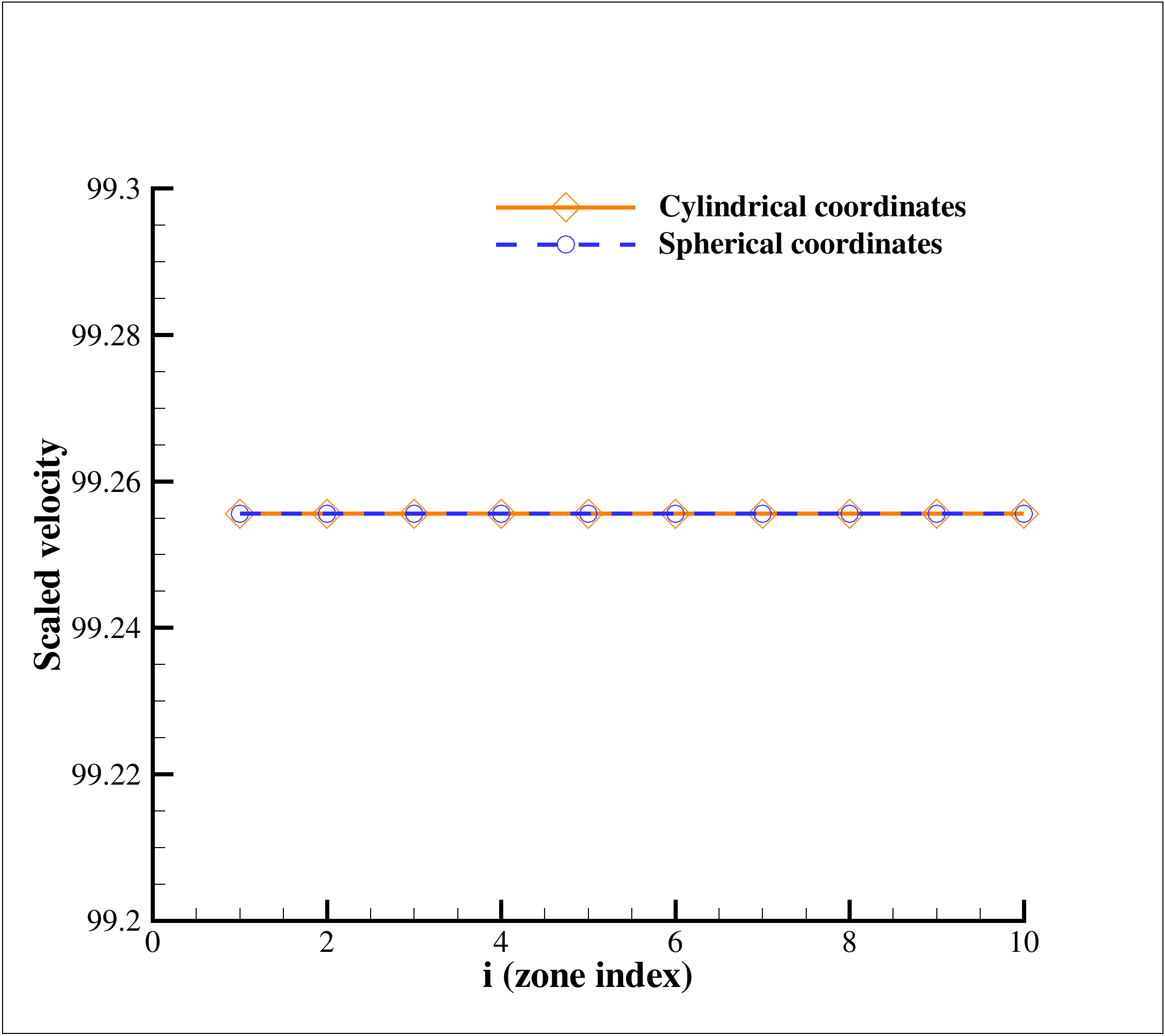}
\end{subfigure}
\caption{Spatial profiles of density $\rho$ (left) and scaled radial velocity $v/\bar{\xi}$ (right) for the isothermal radial wind problem \cite{blondin1993piecewise, Mignone-2014} with constant density after one timestep in cylindrical$-$radial (orange, diamonds) and spherical$-$radial (blue, circles) coordinates. Only the region close to the origin shown.}
\label{fig:7}
\end{figure}

The computational domain spanning $0\le \xi \le 2$ is divided into $N=100$ points. The spatial profiles of density ($\rho$; left) and scaled velocity ($v/\bar{\xi}$; right) are plotted in Fig. \ref{fig:7} after one integration step $\Delta t=7 \times 10^{-5}$ for the case of cylindrical and spherical grid. Here, $\bar{\xi}$ represents the location of the centroid as discussed in section \ref{smoothnesslimiter}. By comparing it with the previously published results \cite{Mignone-2014,blondin1993piecewise}, it can be noted that the density and the scaled velocity remain linear and no signs of deviations are observed near the origin.

\subsubsection{Acoustic wave propagation}
A smooth problem involving a nonlinear system of 1D gas dynamical equations is solved to test fifth order accuracy. The original problem, introduced by Johnsen and Colonius \cite{johnsen2006implementation}, is adapted to cylindrical and spherical coordinates \cite{wang2017high}. The governing equations and the initial conditions for this test are provided in Eqs. (\ref{eq:55}, \ref{eq:56}) and (\ref{eq:58}) respectively.
\begin{equation} \label{eq:58}
\rho(r,0)=1+\varepsilon f(r), \quad
u(r,0) = 0,\quad
p(r,0)=1/\gamma+\varepsilon f(r)
\end{equation}

with the perturbation,

\begin{equation} \label{eq:59}
f(r) =
    \begin{cases}
       \text{$\frac{sin^4(5\pi r)}{r}$} & \text{if $0.4\le r \le0.6$}\\
       0 & \text{otherwise}\\
    \end{cases}
\end{equation}
where $\gamma =1.4$. A sufficiently small $\varepsilon$  ($\varepsilon=10^{-4}$) yields a smooth solution. The interface flux is evaluated using Lax$-$Friedrichs scheme with local speed estimate \cite{rusanov1962calculation} with a CFL number of $0.3$.

\begin{figure}
\centering
\begin{subfigure}{.5\textwidth}
  \centering
  \includegraphics[width=\linewidth]{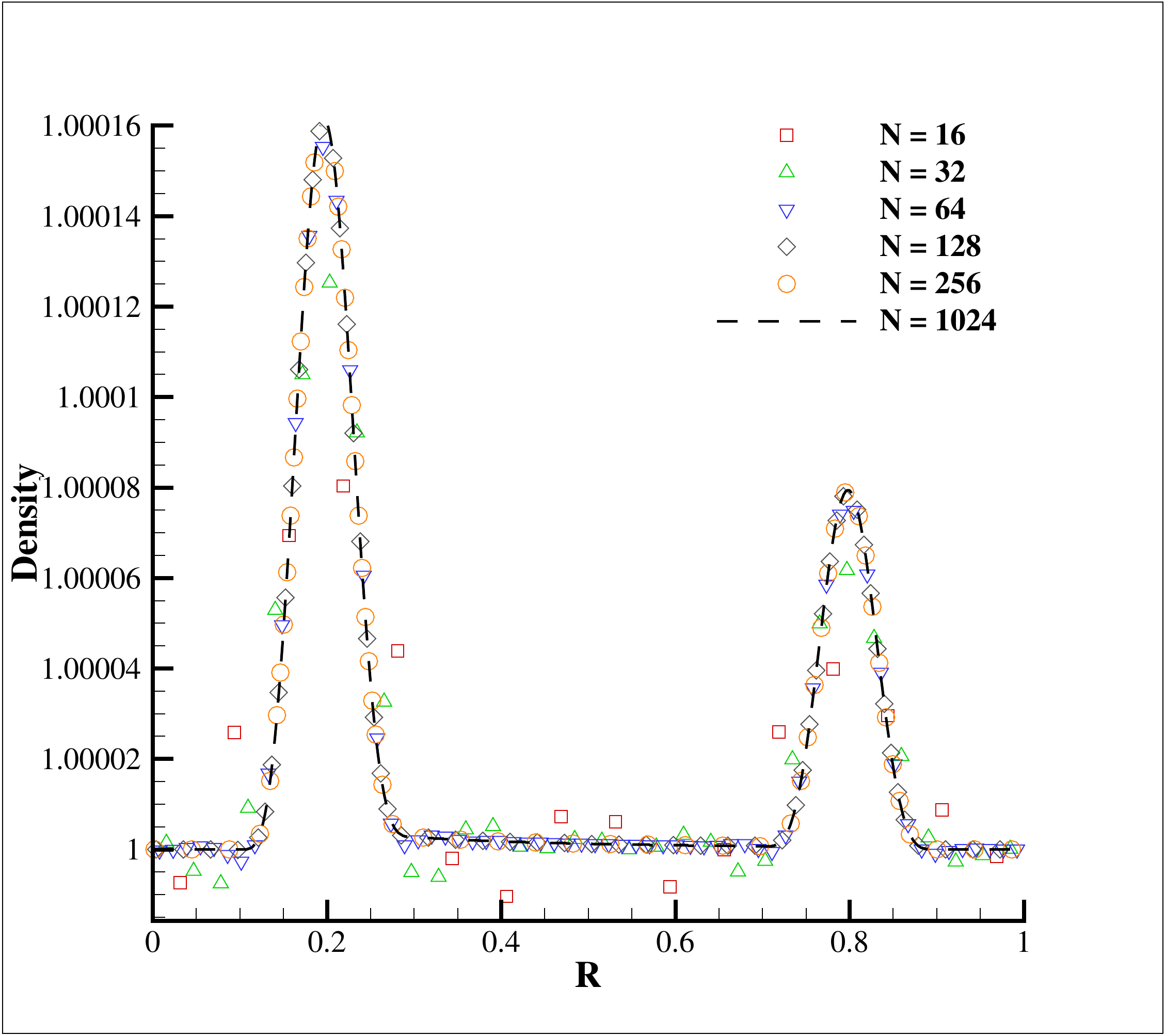}
\end{subfigure}%
\begin{subfigure}{.5\textwidth}
  \centering
  \includegraphics[width=\linewidth]{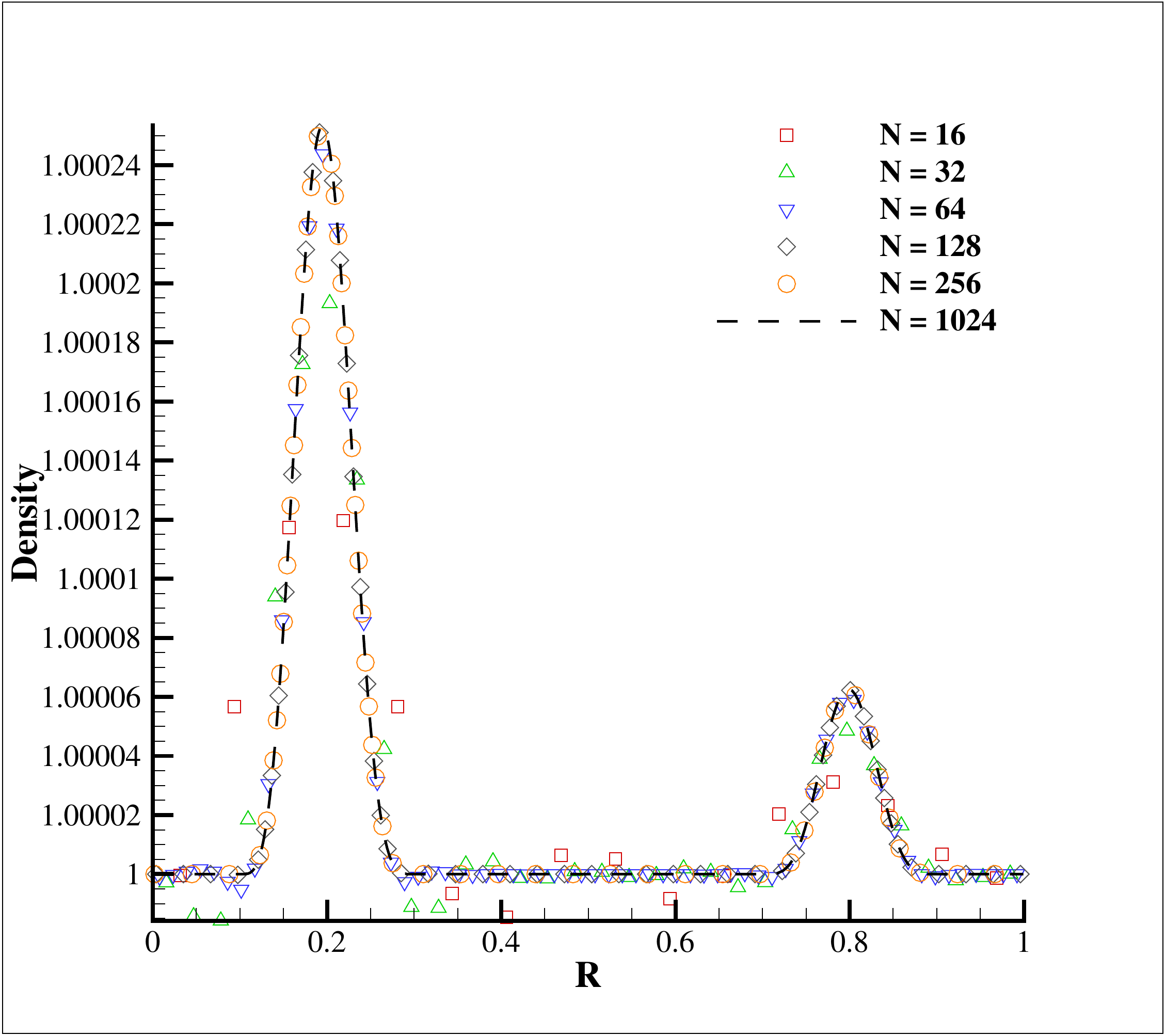}
\end{subfigure}
\caption{Spatial profiles of density ($\rho$) acoustic wave propagation problem \cite{johnsen2006implementation,wang2017high} at time $t=0.3$ in cylindrical$-$radial (left) and spherical$-$radial (right) coordinates.}
\label{fig:8}
\end{figure}

\begin{table}[h!]
\centering
    \caption{$L_1$ norm errors and experimental order of convergence ($EOC$) for acoustic wave propagation test in cylindrical$-$radial and spherical$-$radial coordinates coordinates at $t=0.3$.}
\begin{tabular}{c|cc|cc}
\hline
& \multicolumn{2}{c|}{Cylindrical} &\multicolumn{2}{c}{Spherical} \\\cline{2-3} \cline{4-5} 
$N$          & $\epsilon_1(Q)$    & $O_{L_1}$ & $\epsilon_1(Q)$    & $O_{L_1}$\\
\hline
\hline
16  & 1.01E-05 &    $-$   & 7.98E-06 &    $-$   \\
32  & 4.91E-06 & 1.036 & 3.90E-06 & 1.033 \\
64  & 6.74E-07 & 2.865 & 5.40E-07 & 2.852 \\
128 & 3.24E-08 & 4.380 & 2.59E-08 & 4.383 \\
256 & 1.27E-09 & 4.670 & 1.01E-09 & 4.675
\end{tabular}\label{tab:7}
\end{table}

The initial perturbation splits into two acoustic waves traveling in opposite directions. The final time ($t=0.3$) is set such that the waves remain in the domain and the problem is free from the boundary effects. The computational domain of unity length is uniformly divided into $N$ different zones i.e. $N=16, 32, 64, 128, 256$. Although an exact solution known up to O($\varepsilon^2$) is known, the solution on the finest mesh $N=1024$ is taken as the reference. Error in density is evaluated from Eq. (\ref{eq:45}). Fig. \ref{fig:8} illustrate the spatial variation of density at $t=0.3$ inside the domain in cylindrical$-$radial (left) and spherical$-$radial (right) coordinates. The location of the peaks is same. However, the height of the peaks differs due to different one$-$dimensional Jacobians for both the coordinates. From Table \ref{tab:7}, it clear that the scheme approaches the desired fifth order of convergence ($EOC$) for both the cases.

\subsubsection{Sedov explosion test}
Sedov explosion test is performed to investigate code's ability to deal with strong shocks and non$-$planar symmetry \cite{fryxell2000flash}. The problem involves a self$-$similar evolution of a cylindrical/spherical blastwave from a localized initial pressure perturbation (delta$-$function) in an otherwise homogeneous medium. Governing equations for this problem are the same as given in Eq. (\ref{eq:55}) earlier. For the code initialization, dimensionless energy $\epsilon$ ($\epsilon=1$) is deposited into a small region of radius $\delta r$, which is three times the cell size at the center. Inside this region, the dimensionless pressure $P^{'}_0$ is given by Eq. (\ref{eq:60}). 

\begin{equation}\label{eq:60}
P^{'}_0=\frac{3(\gamma-1)\epsilon}{(m+2)\pi \delta r^{(m+1)}}
\end{equation}
where $\gamma=1.4$ and $m=1,2$ for cylindrical, spherical geometries respectively. Reflecting boundary condition is employed at the center ($r=0$), whereas boundary condition at $r=1$ is not required for this problem. The initial velocity and density inside the domain are 0 and 1 respectively and the initial pressure everywhere except the kernel is $10^{-5}$. Due to reflecting boundary condition at the center, the high pressure region (kernel) consists of 6 cells, i.e., 3 ghost cells and 3 interior cells. As the source term is very stiff, the CFL number set to be $0.1$. The final time is $t=0.05$. In a self$-$similar blastwave that develops, the analytical results are available in the literature \cite{fryxell2000flash,kamm2007efficient}. 

\begin{figure}[] 
  \begin{minipage}[b]{0.5\linewidth}
    \includegraphics[width=\linewidth]{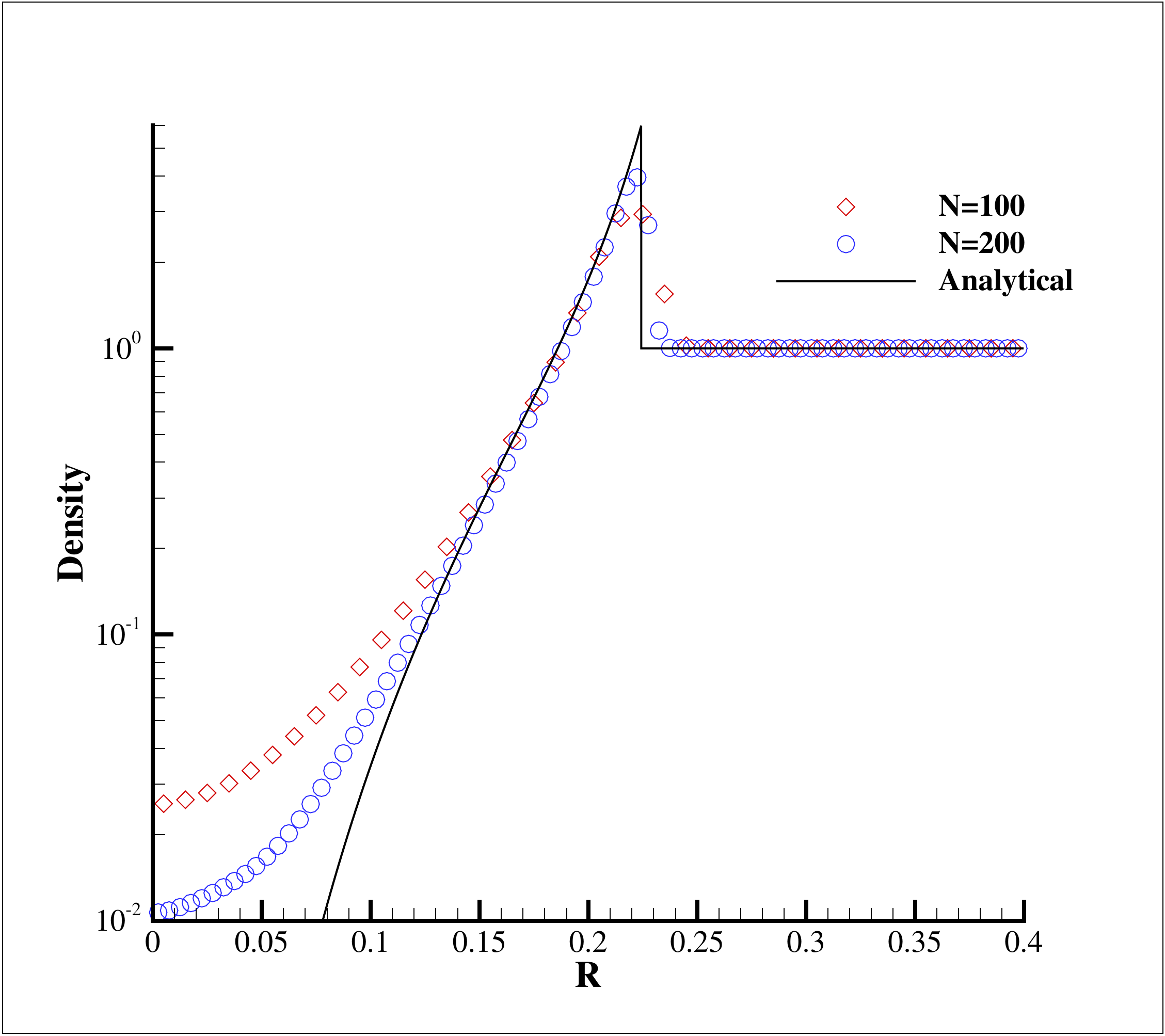} 
  \end{minipage} 
  \begin{minipage}[b]{0.5\linewidth}
    \includegraphics[width=\linewidth]{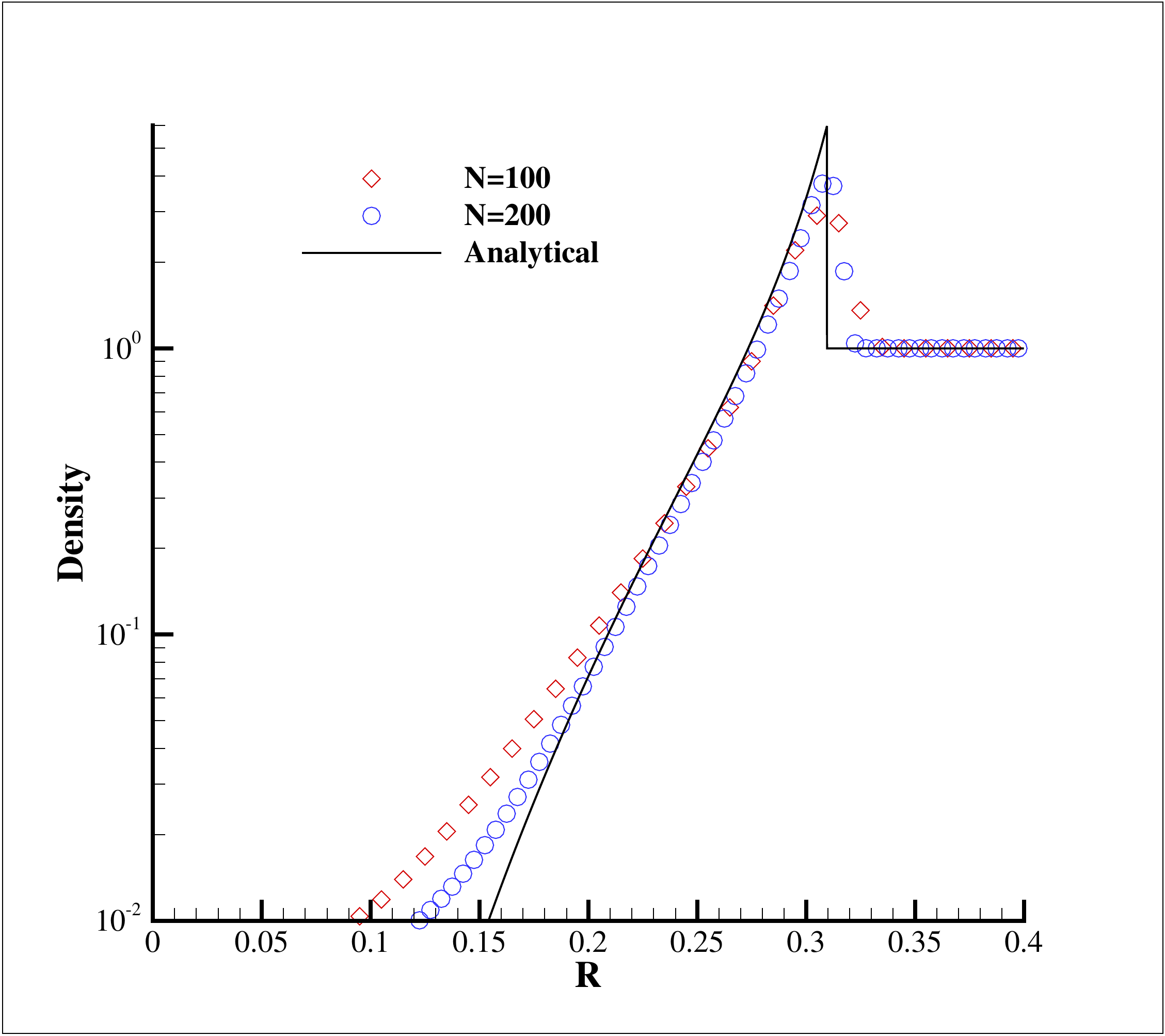} 
  \end{minipage} 
    \begin{minipage}[b]{0.5\linewidth}
    \includegraphics[width=\linewidth]{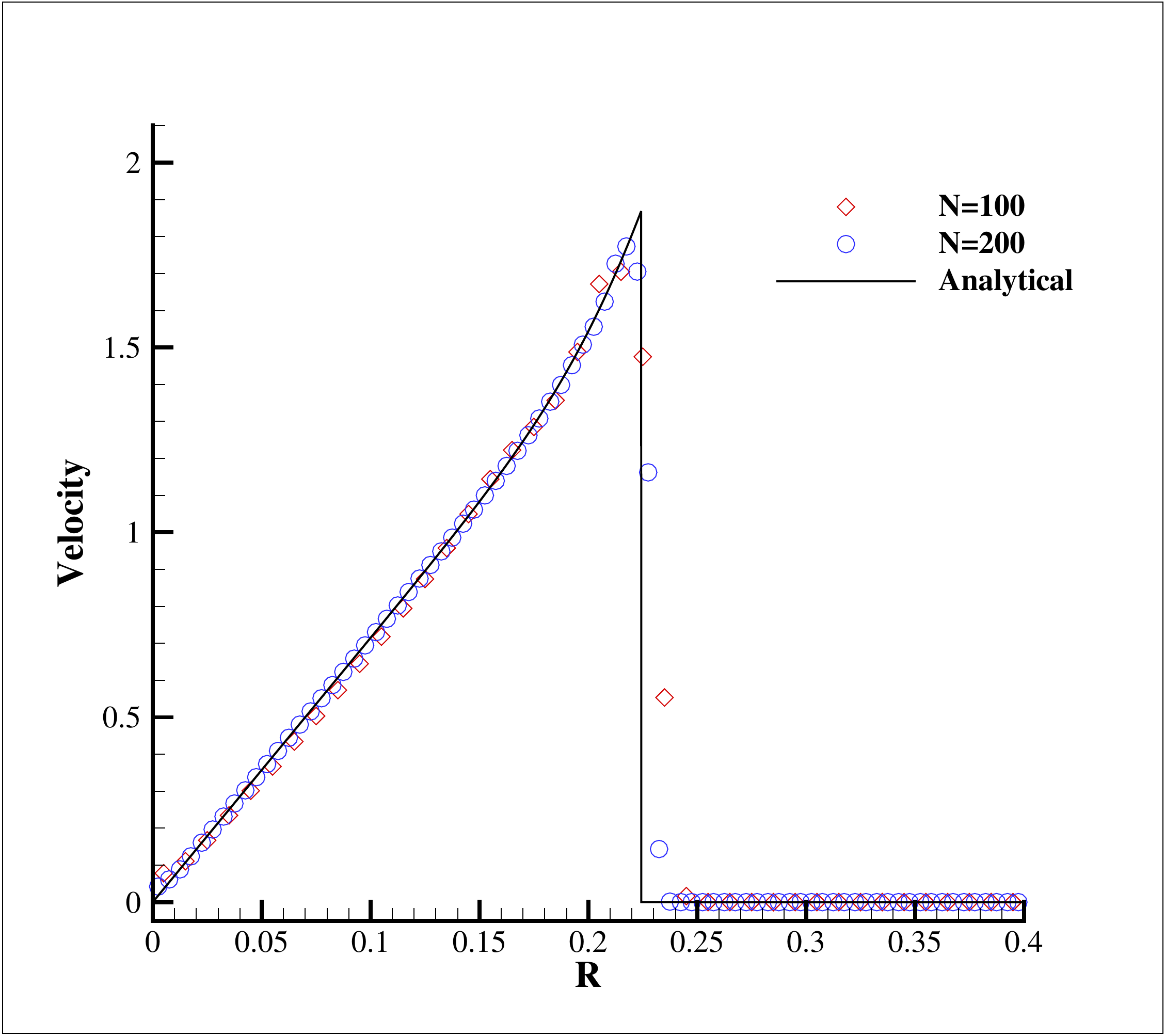} 
  \end{minipage} 
  \begin{minipage}[b]{0.5\linewidth}
    \includegraphics[width=\linewidth]{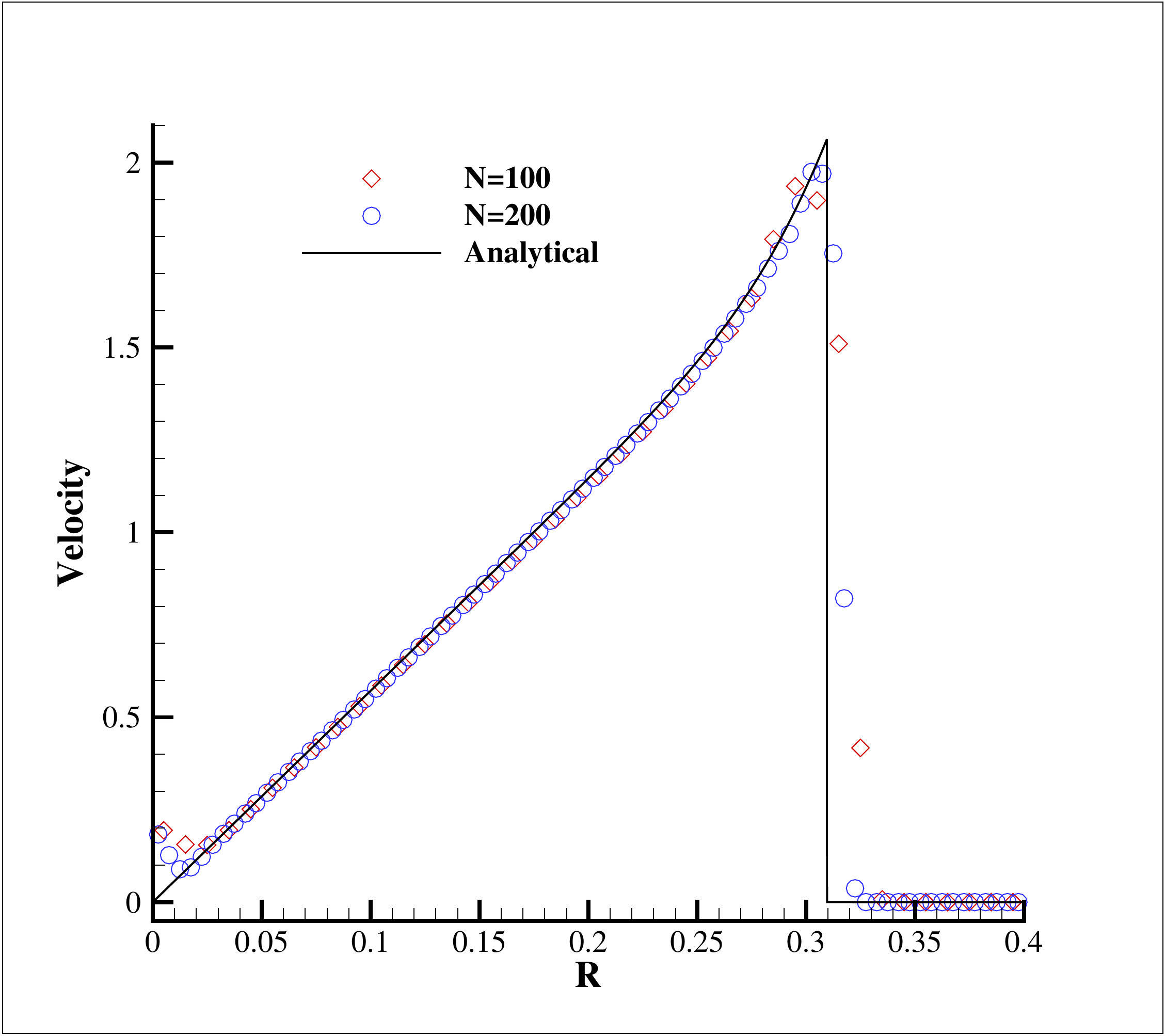} 
  \end{minipage} 
  \begin{minipage}[b]{0.5\linewidth}
    \includegraphics[width=\linewidth]{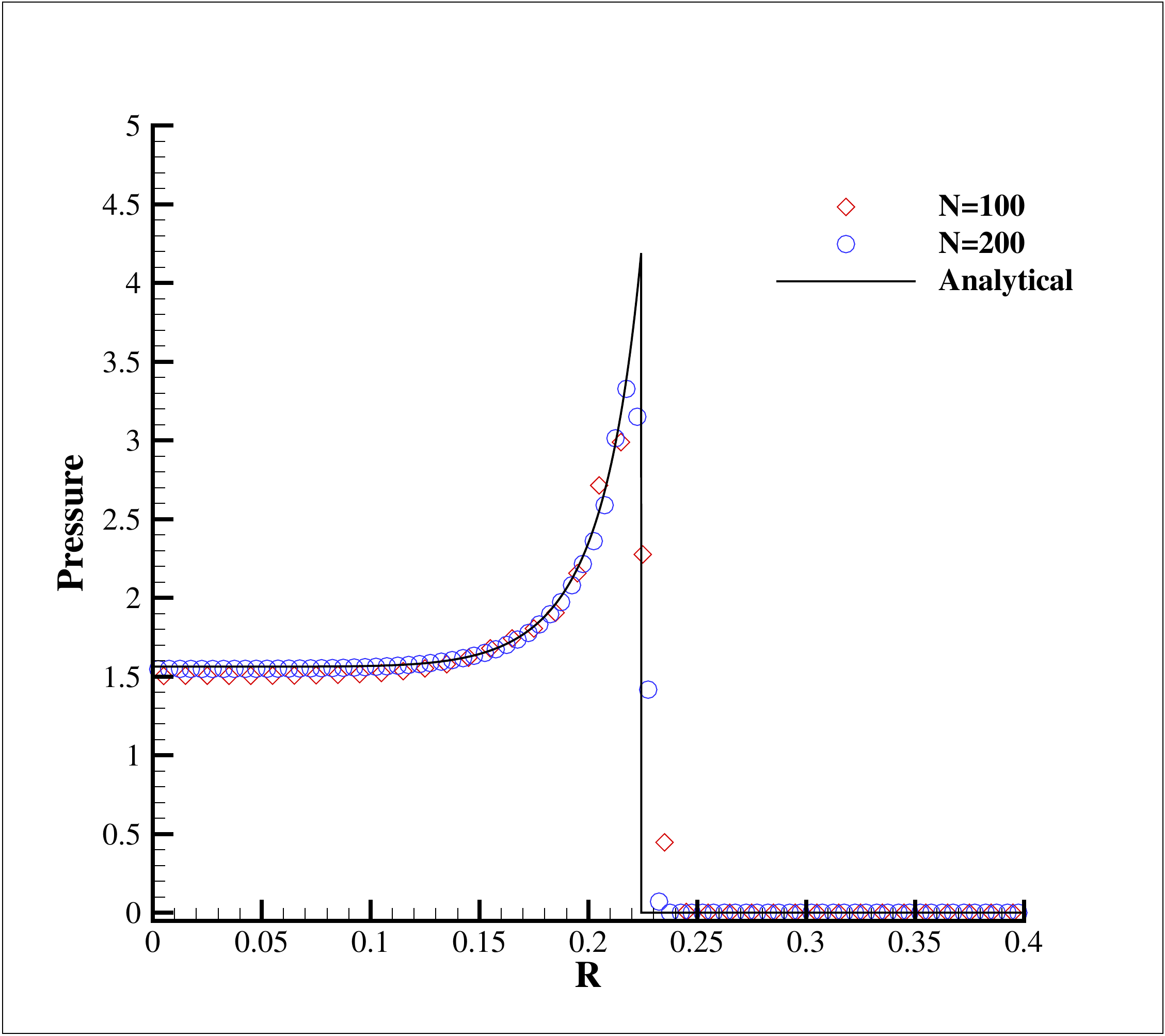} 
  \end{minipage}
  \begin{minipage}[b]{0.5\linewidth}
    \includegraphics[width=\linewidth]{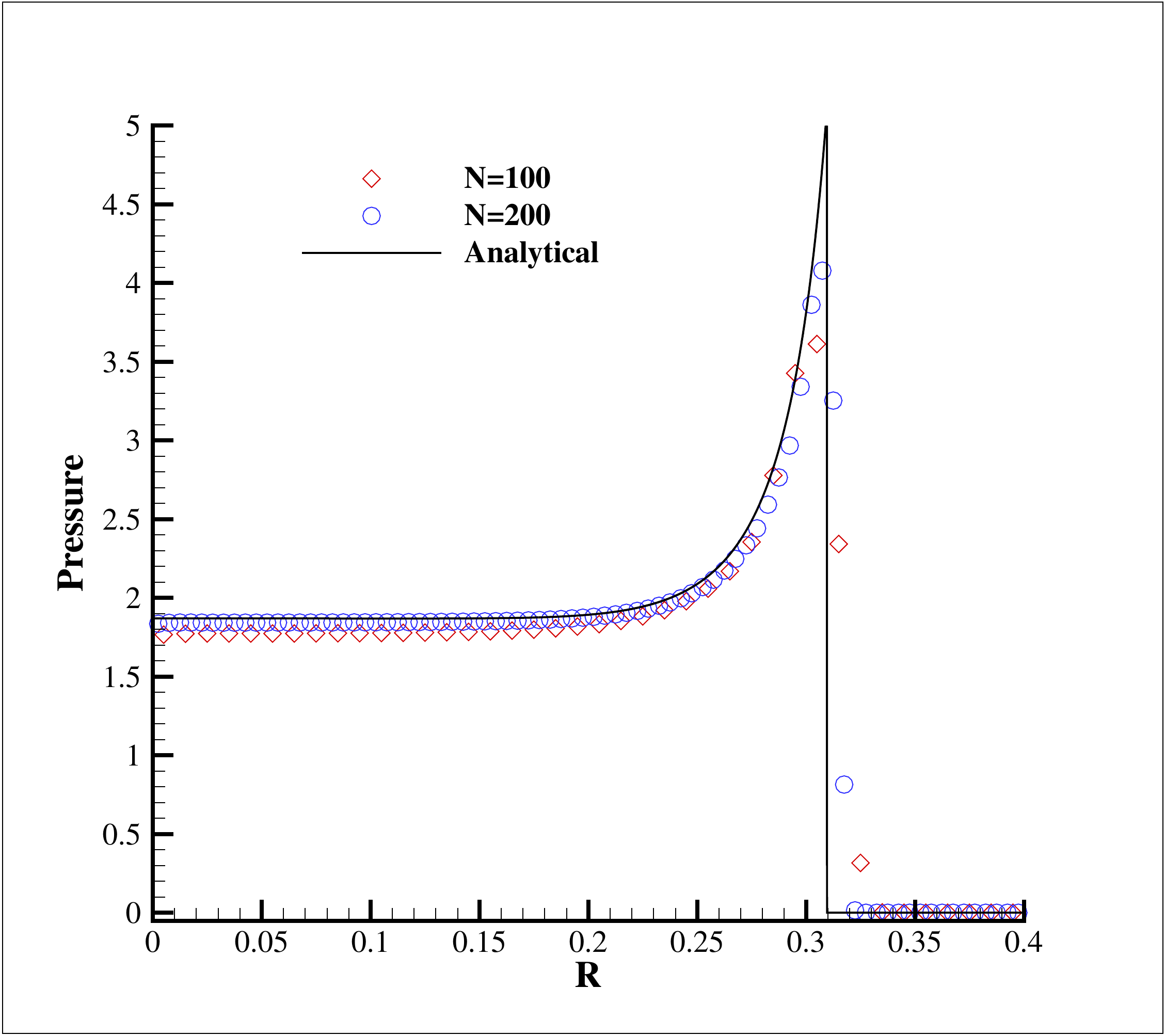} 
  \end{minipage} 
  \caption{Variation of density (first row), velocity (second row), and pressure (third row) with the radius for cylindrical$-$radial (left column) and spherical$-$radial (right column) coordinates for the Sedov explosion test \cite{fryxell2000flash,wang2017high}. Domain is restricted to $R=0.4$ for the sake of clarity.}
  \label{fig:9} 
\end{figure} 

Fig. \ref{fig:9} shows the variations in density, velocity, and pressure with radius on a uniform grid ($N=100,200$) in 1D cylindrical$-$radial and spherical$-$radial coordinates along with their analytical values \cite{kamm2007efficient}. The peak values of pressure, velocity, and density show similar behavior as given in \cite{wang2017high}, but the locations of the shocks are different due to different $\epsilon$ and final time values.

\subsubsection{Sod test}
Sod test \cite{sod1978survey} is considered in 1D cylindrical$-$radial, spherical$-$radial, and 2D cylindrical ($r-\theta$) coordinates. For 1D radial cases, governing equation is given in Eq. (\ref{eq:55}), while governing equation for cylindrical ($r-\theta$) coordinates is given in Eq. (\ref{eq:61}).

\begin{equation} \label{eq:61}
\frac{\partial{}}{\partial{t}}
\begin{pmatrix}
    \rho \\
    \rho v_r \\
    \rho v_{\theta} \\
    \rho e 
\end{pmatrix}
+
\frac{1}{r}
\frac{\partial{}}{\partial{r}}
\begin{pmatrix}
    \rho v_{r}r\\
    (\rho v_{r}^2+p)r \\
    \rho v_r v_{\theta} r \\
    (\rho e+p)v_r r 
\end{pmatrix}
+\frac{1}{r}\frac{\partial{}}{\partial{\theta}}
\begin{pmatrix}
    \rho v_{\theta}\\
    \rho v_r v_{\theta} \\
    \rho v_{\theta}^2 + p \\
    (\rho e+p)v_{\theta} 
\end{pmatrix}
=
\begin{pmatrix}
    0 \\
    (p+\rho v_{\theta}^2)/r \\
    -\rho v_r v_{\theta}/r \\
    0
\end{pmatrix}
\end{equation}
where terms $(\rho v_{\theta}^2)/r$ and $(\rho v_r v_{\theta})/r$ are related to the centrifugal and Coriolis forces respectively. In this problem, the interface flux is evaluated with HLL Riemann solver \cite{harten1997upstream}. The initial condition consists of two regions (left and right states) inside the domain separated by a diaphragm at $r=0.5$ as provided in Eq. (\ref{eq:62}).

 \begin{equation} \label{eq:62}
\begin{pmatrix}
    \rho\\
    v_{r} \\
    v_{\theta} \\
    p  
\end{pmatrix}_L
=
\begin{pmatrix}
    1 \\
    0 \\
    0 \\    
    1
\end{pmatrix};
\quad \quad
\begin{pmatrix}
    \rho\\
    v_{r} \\
    v_{\theta} \\
    p 
\end{pmatrix}_R
=
\begin{pmatrix}
    0.125 \\
    0 \\
    0 \\
    0.1
\end{pmatrix}
\end{equation}

The computational domain ($0\le r \le 1$) for 1D tests is uniformly divided in $N$ zones ($N=100, 500$), while for the 2D test, the computational domain ($0\le r \le 1$, $0\le \theta \le \pi/2$) is uniformly divided into $100 \times 100$ zones in the corresponding directions. The boundary conditions for 1D cases are not required, however, for 2D case, symmetry of conserved variables at $r=0$ (except radial velocity which is antisymmetric) is considered along with outflow boundary condition applied to all other boundaries ($r=1$, $\theta=0$, and $ \theta=\pi/2$). The computation is performed till $t=0.2$ with a CFL number of $0.3$. For first order and second order (MUSCL \cite{van1979towards}) spatial reconstruction, Euler time marching and Maccormack (predictor$-$corrector) schemes \cite{maccormack1982numerical} are respectively employed.

\begin{figure}[] 
  \begin{minipage}[b]{0.5\linewidth}
    \includegraphics[width=\linewidth]{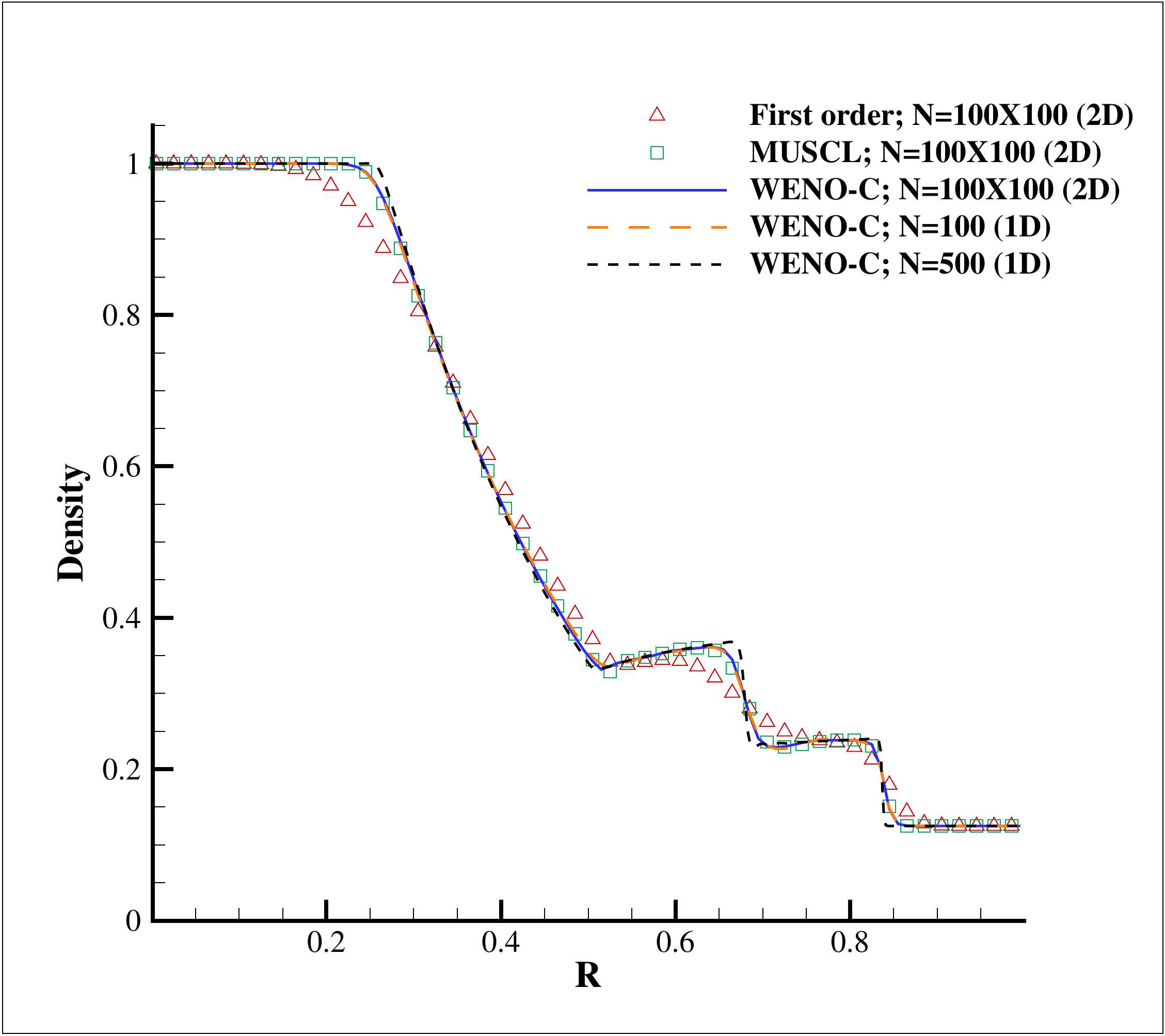} 
  \end{minipage} 
  \begin{minipage}[b]{0.5\linewidth}
    \includegraphics[width=\linewidth]{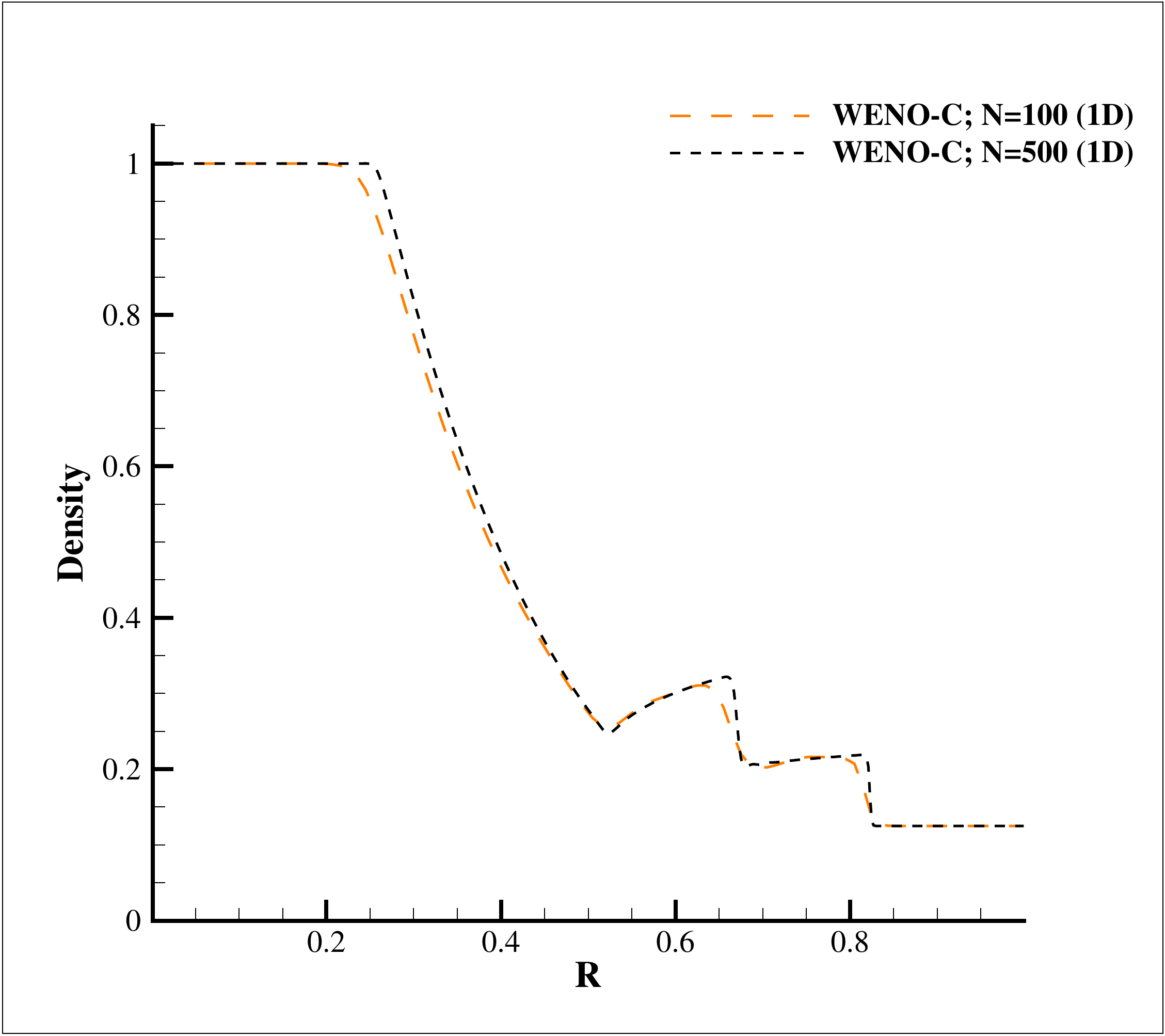} 
  \end{minipage} 
    \begin{minipage}[b]{0.5\linewidth}
    \includegraphics[width=\linewidth]{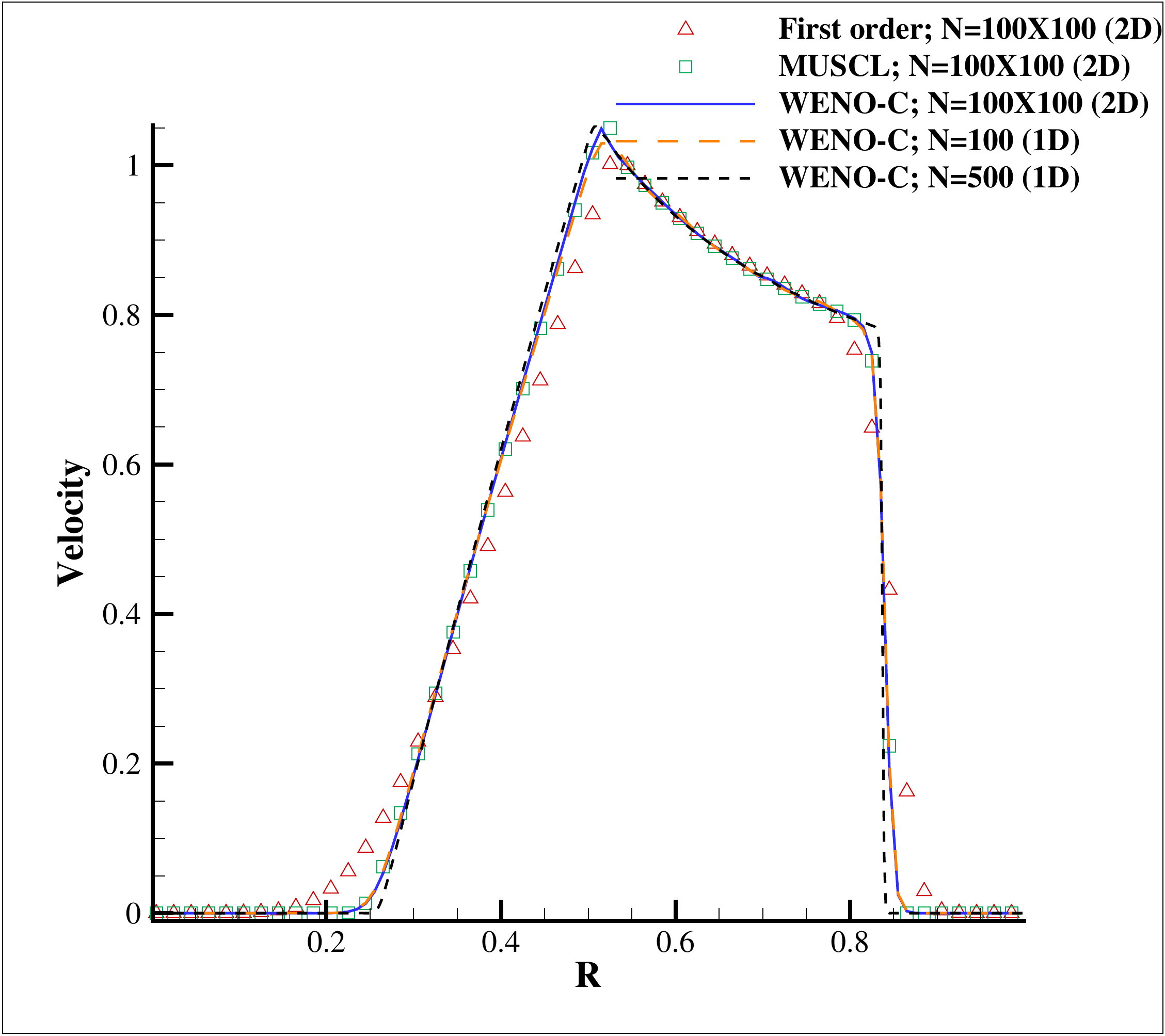} 
  \end{minipage} 
  \begin{minipage}[b]{0.5\linewidth}
    \includegraphics[width=\linewidth]{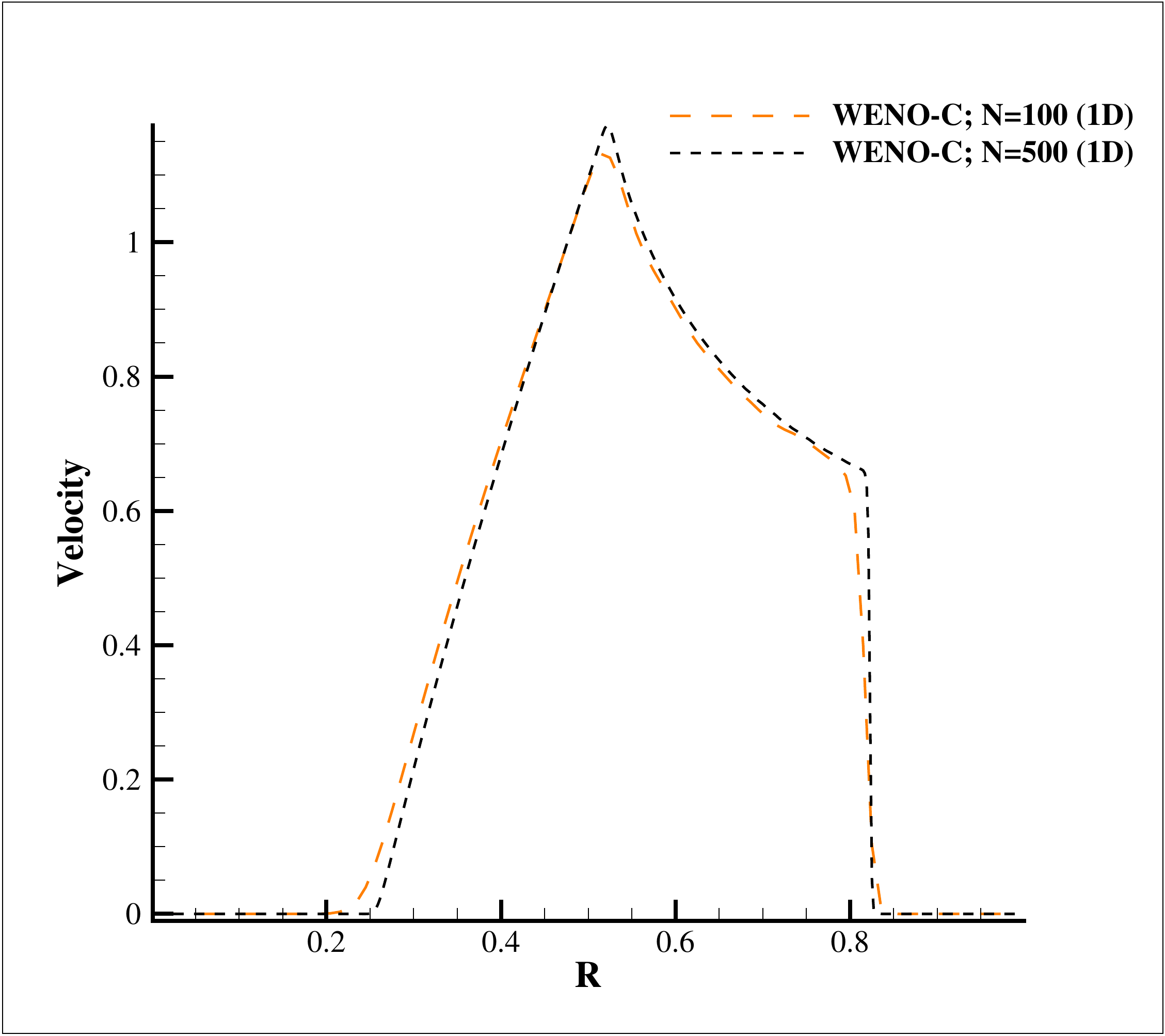} 
  \end{minipage} 
  \begin{minipage}[b]{0.5\linewidth}
    \includegraphics[width=\linewidth]{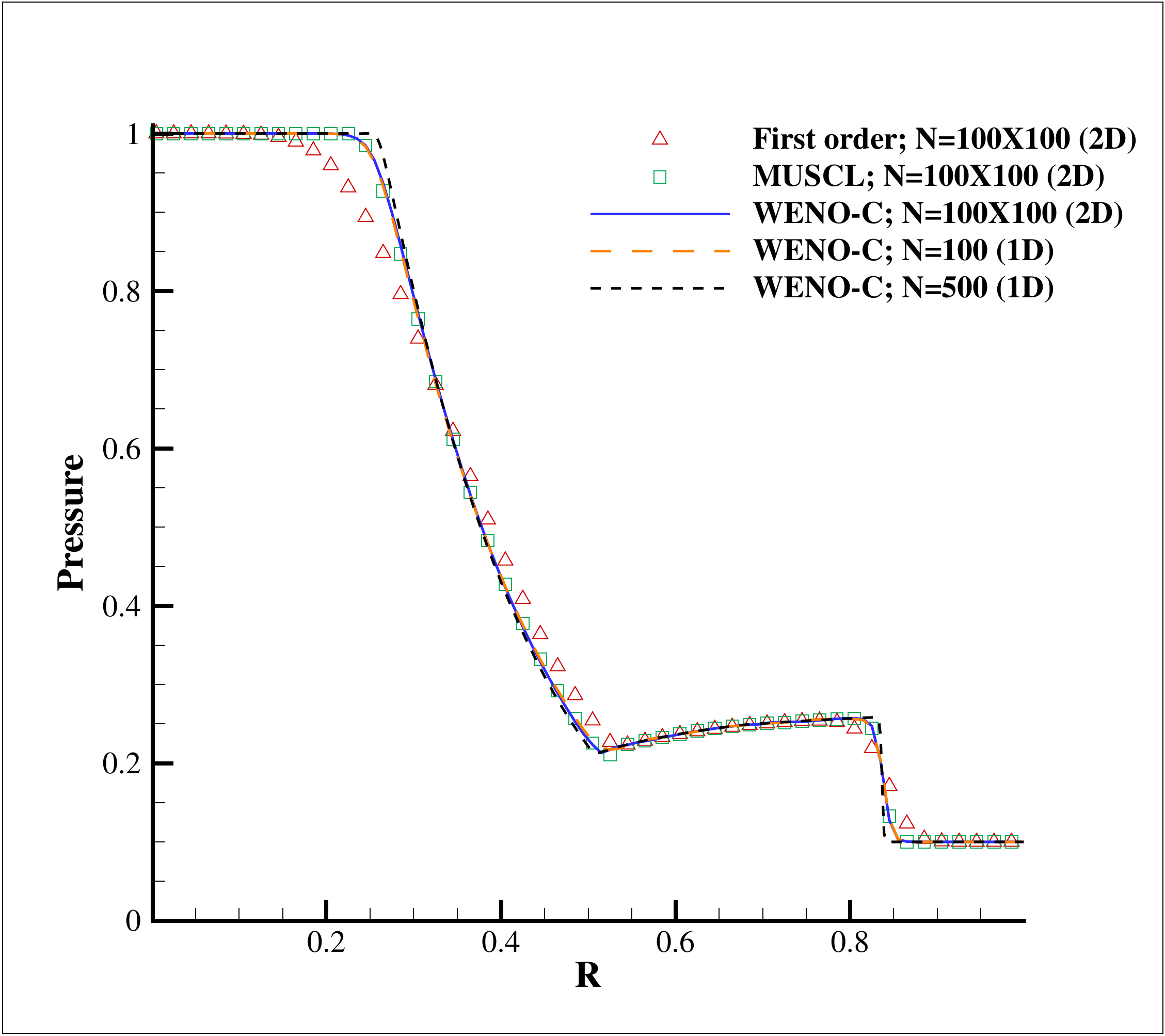} 
  \end{minipage}
  \hfill
  \begin{minipage}[b]{0.5\linewidth}
    \includegraphics[width=\linewidth]{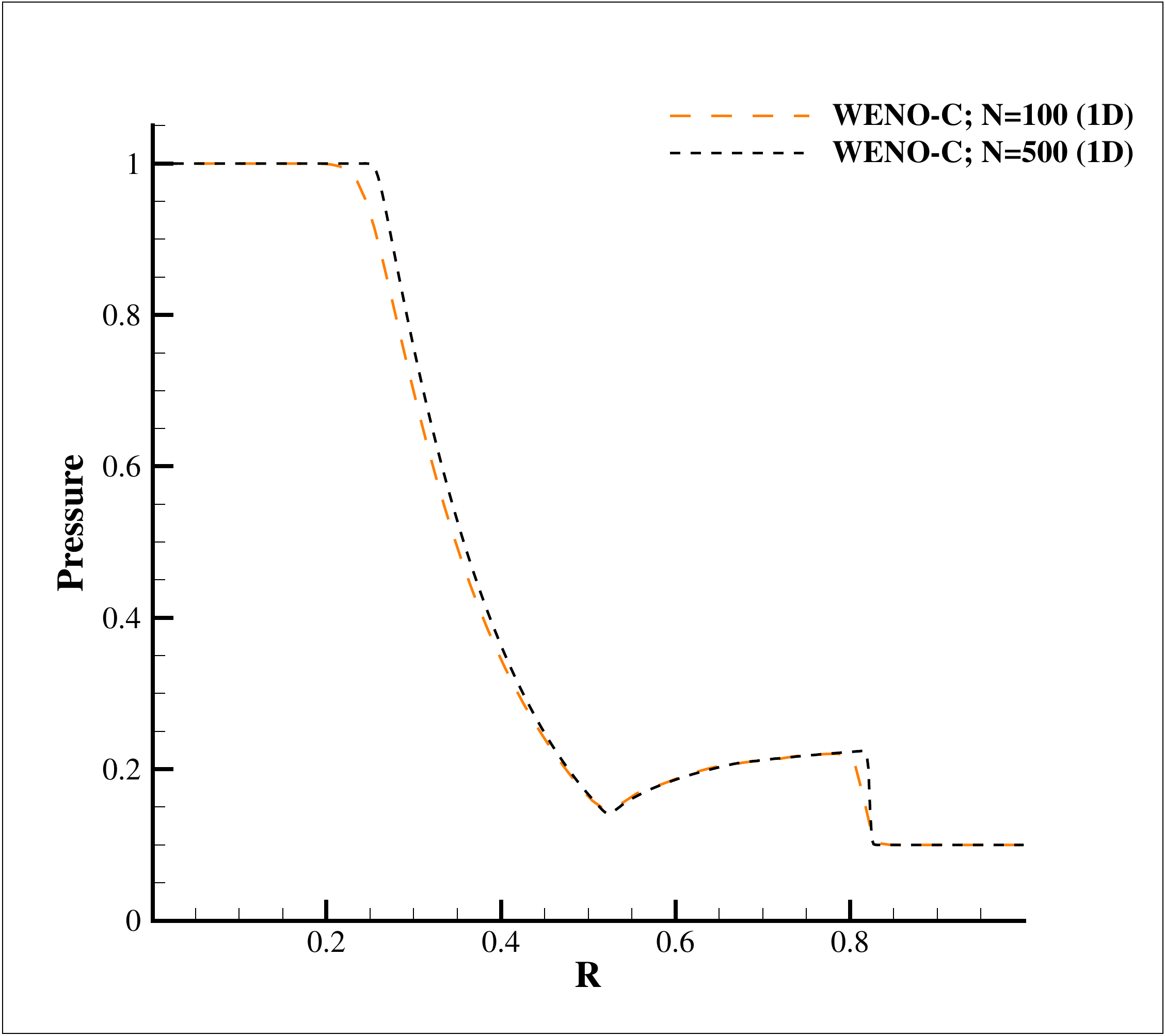} 
  \end{minipage} 
  \caption{Variation of density (first row), velocity (middle row), and pressure (third row) with the radius at $t=0.2$ for cylindrical (left column) and spherical$-$radial (right column) coordinates for the modified Sod test \cite{sod1978survey,wang2017high}.}
  \label{fig:10} 
\end{figure} 

\begin{figure}
\centering
  \includegraphics[width=0.5\linewidth]{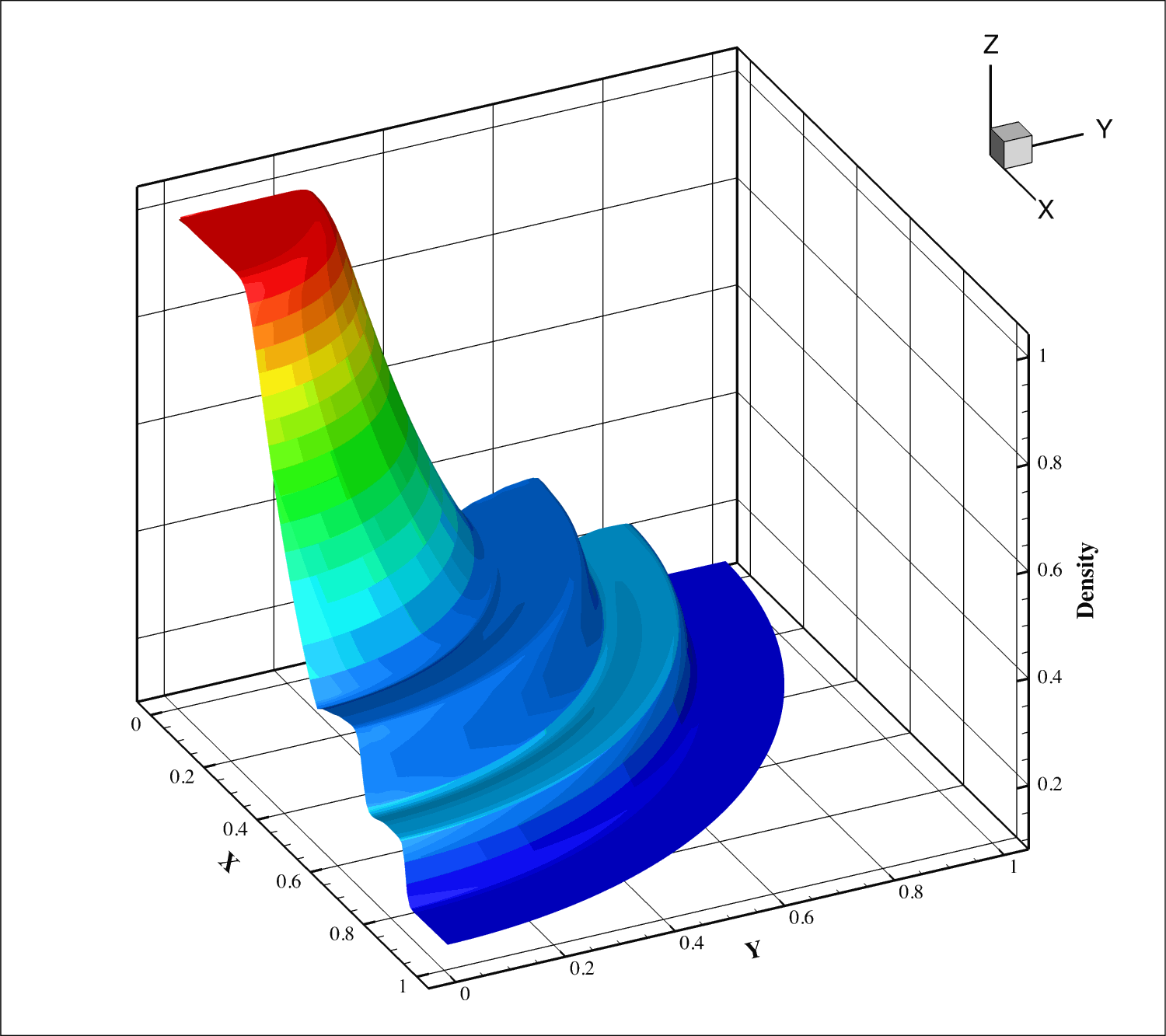}
\caption{Variation of density with the radius at $t=0.2$ for cylindrical ($r-\theta$) coordinates in the Cartesian plane for the modified Sod test \cite{sod1978survey,wang2017high}.}
\label{fig:11}
\end{figure}

Fig. \ref{fig:10} shows the spatial profiles of density, velocity, and pressure for Sod test case in 1D/2D cylindrical coordinates (left) and 1D spherical$-$radial (right) coordinates. WENO$-$C performs better than first order and second order (MUSCL \cite{van1979towards}) reconstruction techniques. The 2D test results exactly overlap with the 1D test results in cylindrical coordinates. Fig. \ref{fig:11} shows the spatial variation of the density in the 2D Cartesian plane at time $t=0.2$. When compared with the results obtained from fifth order finite difference WENO \cite{wang2017high}, it is clear that WENO$-$C yields similar but less oscillatory results.

\subsubsection{Modified 2D Riemann problem in cylindrical (R$-$z) coordinates}
\begin{figure}
\centering
  \includegraphics[width=0.5\linewidth]{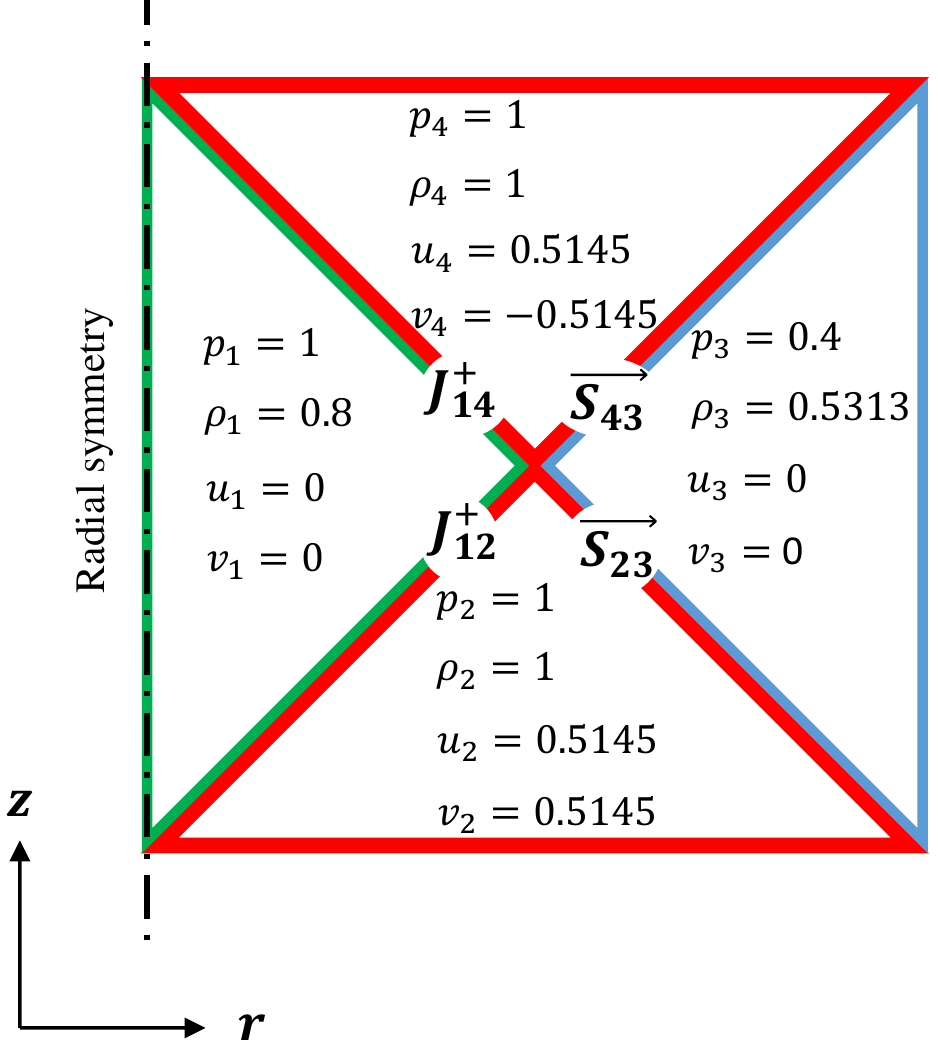}
\caption{A schematic of modified 2D Riemann problem in cylindrical ($r-z$) coordinates.}
\label{fig:12}
\end{figure}

The final test for the present scheme involves a modified 2D Riemann problem in cylindrical ($R-z$) coordinates, as illustrated in Fig. \ref{fig:12}. The problem corresponds to configuration 12 of \cite{lax1998solution} involving two contact discontinuity and two shocks as the initial condition, resulting in the formation of a self$-$similar structure propagating towards the low density$-$low pressure region (region 3). To make the problem symmetric about the origin, the original problem \cite{lax1998solution} is rotated by an angle of 45 degrees in the clockwise direction. The governing equations are provided in Eq. (\ref{eq:63}).

\begin{equation} \label{eq:63}
\frac{\partial{}}{\partial{t}}
\begin{pmatrix}
    \rho \\
    \rho v_R \\
    \rho v_z \\
    \rho e 
\end{pmatrix}
+ 
\frac{1}{R}
\frac{\partial{}}{\partial{R}}
\begin{pmatrix}
    \rho v_{R} R\\
    (\rho v_{R}^2+p)R \\
    \rho v_R v_{z}R \\
    (\rho e+p)v_R R 
\end{pmatrix}
+\frac{\partial{}}{\partial{z}}
\begin{pmatrix}
    \rho v_{z}\\
    \rho v_R v_{z} \\
    \rho v_{z}^2 + p \\
    (\rho e+p)v_{z} 
\end{pmatrix}
=
\begin{pmatrix}
    0 \\
    p/R \\
    0 \\
    0
\end{pmatrix}
\end{equation}

\begin{figure}[]
\centering
  \begin{minipage}[b]{\linewidth}
  \centering
    \includegraphics[width=0.5\linewidth]{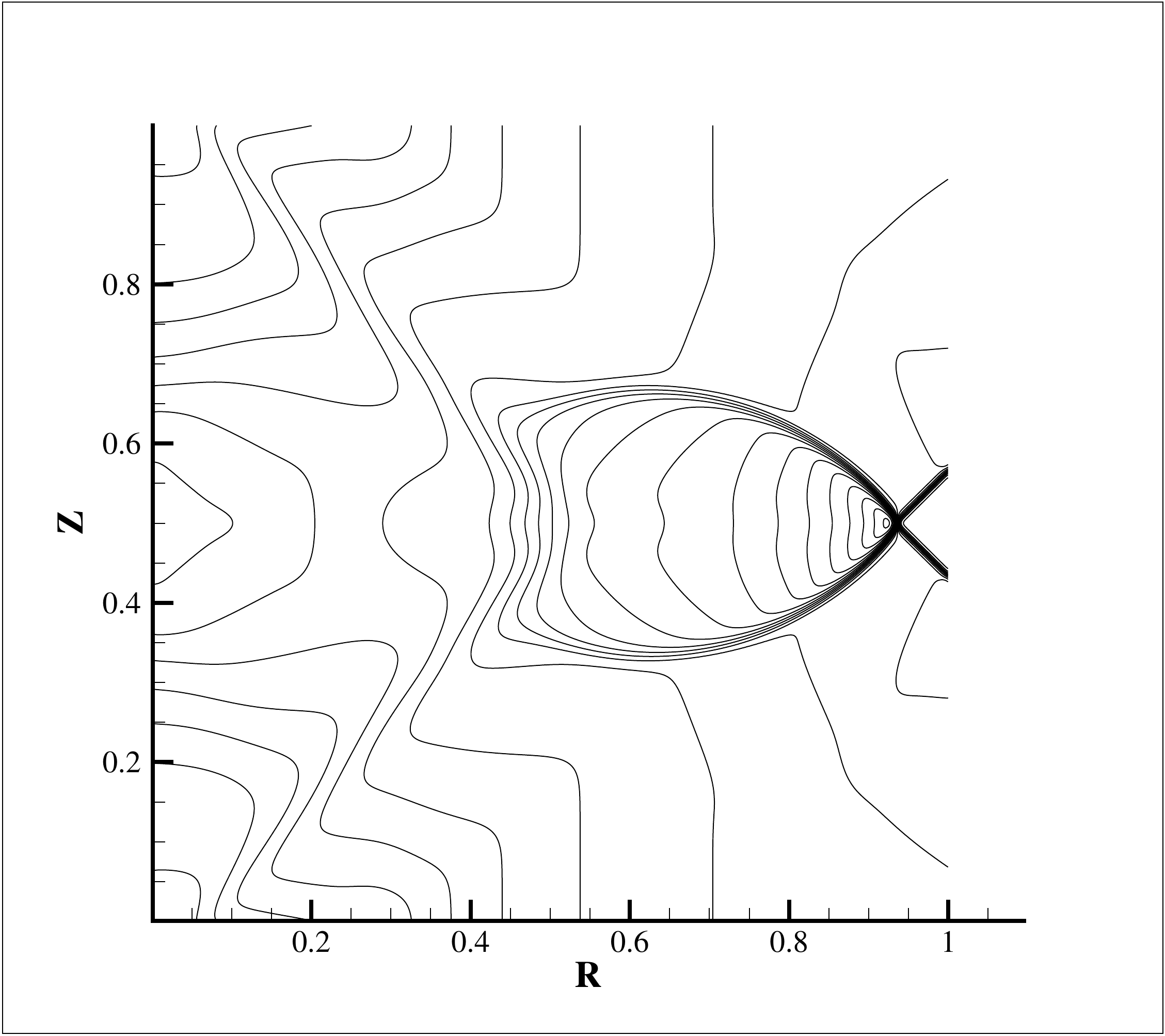} 
  \end{minipage} 
  \begin{minipage}[b]{\linewidth}
  \centering
    \includegraphics[width=0.5\linewidth]{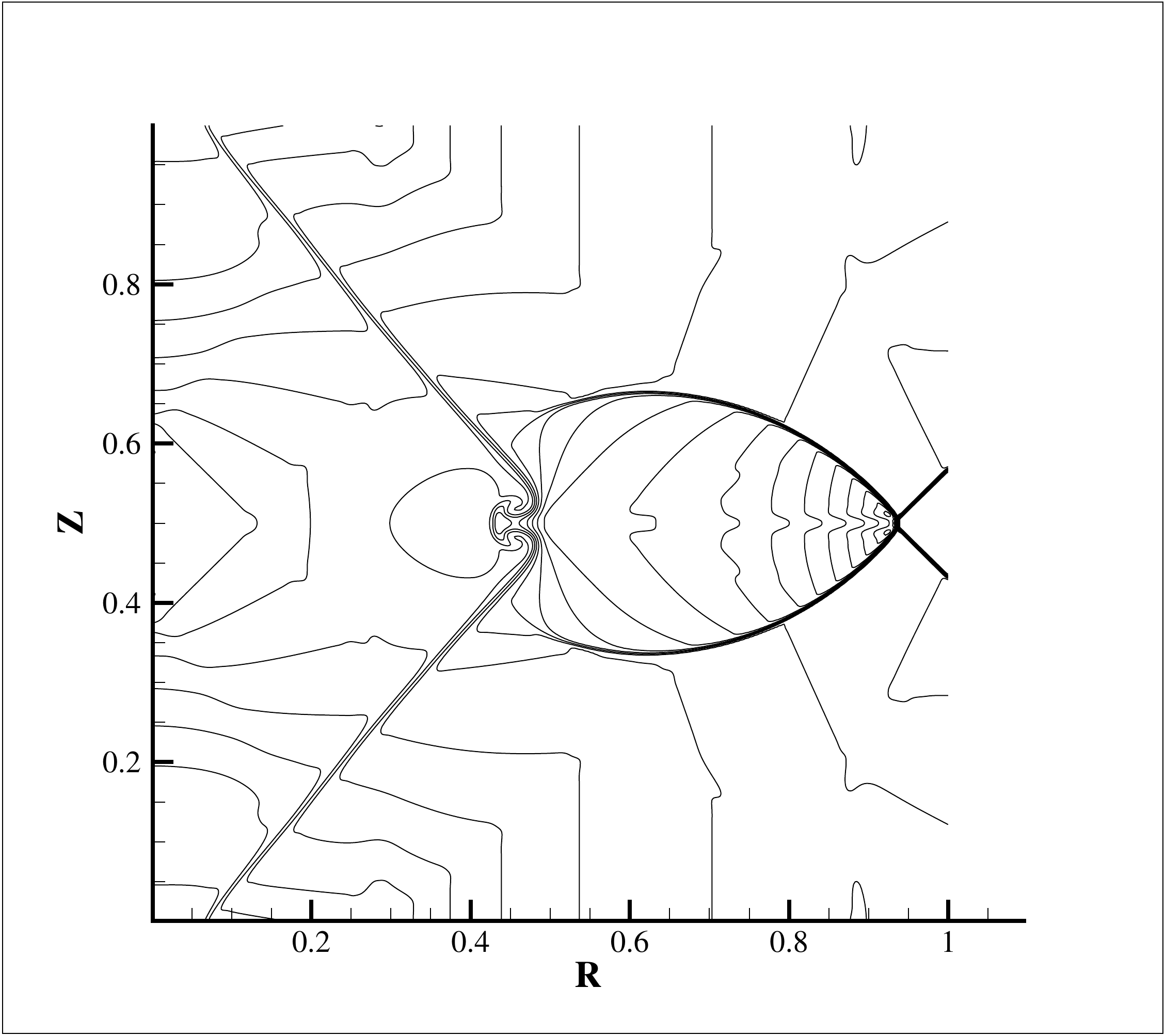} 
  \end{minipage} 
  \begin{minipage}[b]{\linewidth}
  \centering
    \includegraphics[width=0.5\linewidth]{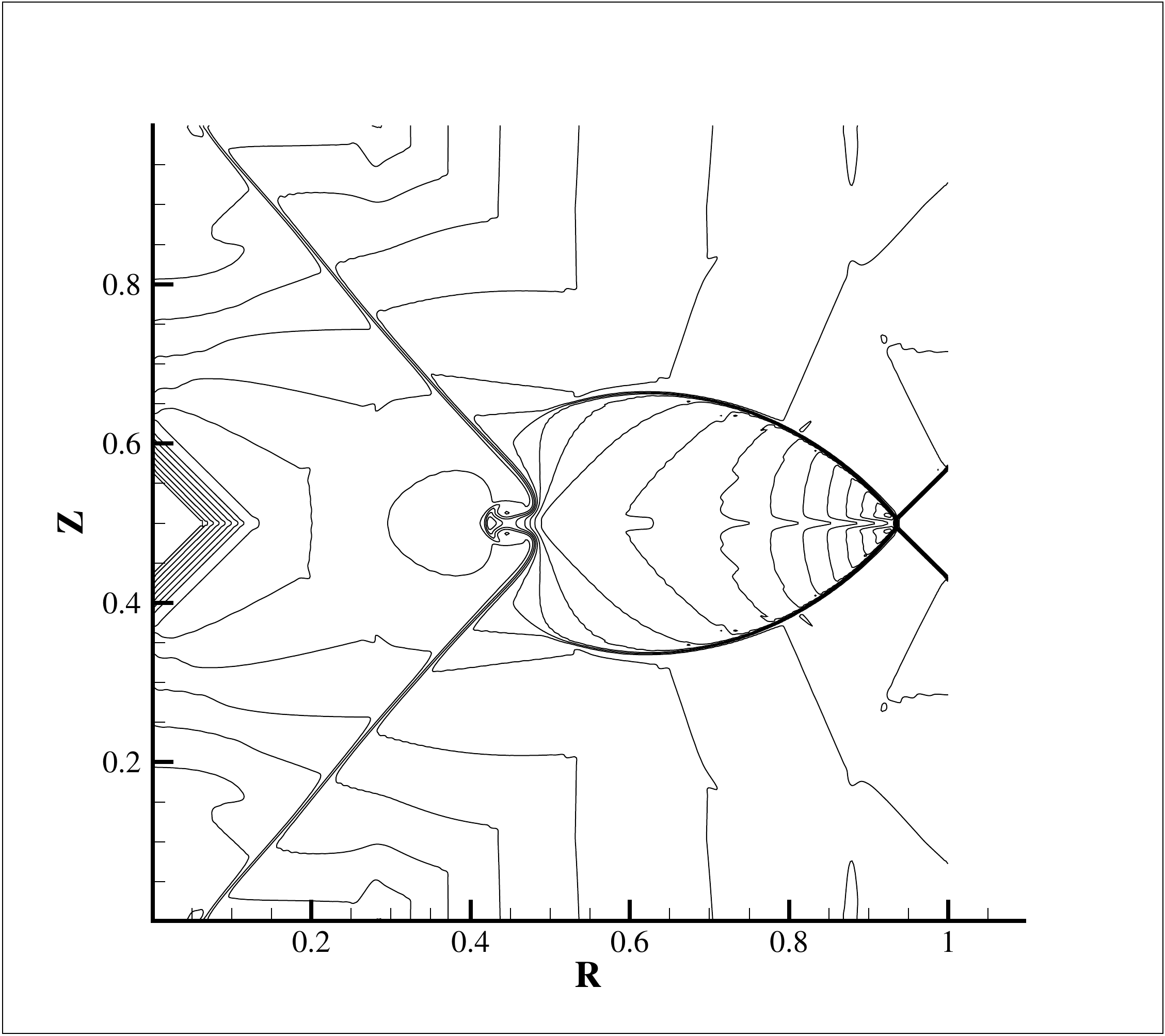} 
  \end{minipage}
  \hfill
  \caption{Density contours with different reconstruction techniques (first order (top), second order MUSCL \cite{van1979towards} (middle), and WENO$-$C (bottom)) at $t=0.2$ for the modified Riemann problem in cylindrical ($r-z$) coordinates}
  \label{fig:13} 
\end{figure} 

The computations are performed until $t=0.2$ with a CFL number of $0.5$ on a domain ($r,z$)$=$[0,1]$\times$[0,1] divided into 500$\times$500 zones. The boundary conditions include symmetry at the center (except for the antisymmetric radial velocity) and outflow elsewhere. For the first order and second order (MUSCL \cite{van1979towards}) spatial reconstructions, Euler time marching and Maccormack (predictor$-$corrector) schemes \cite{maccormack1982numerical}, are respectively employed. Rich small$-$scale structures in the contact$-$contact region (region 1) can be observed from Fig. \ref{fig:13} for WENO$-$C reconstruction, when compared with first and second order MUSCL reconstruction. Structures are highly smeared for the case of first order reconstruction.

\section{Conclusions} \label{conclusions}
The fifth order finite volume WENO$-$C reconstruction scheme provides a more general framework in orthogonally$-$curvilinear coordinates to achieve high order spatial accuracy with minimal computational cost. Analytical values of linear weights, optimal weights, weights for mid$-$point interpolation, and flux/source term integration are derived for the standard grids. The proposed reconstruction scheme can be applied to both regularly$-$spaced and irregularly$-$spaced grids. A grid independent smoothness indicator is derived from the basic definition. For uniform grids, the analytical values in Cartesian, cylindrical$-$radial, and spherical$-$radial coordinates for $R\to \infty$ conform to WENO$-$JS. A simple and computationally efficient extension to multi$-$dimensions is employed. 1D Scalar advection tests are performed in curvilinear coordinates on regularly$-$spaced and irregularly$-$spaced grids followed by several smooth and discontinuous flow test cases in 1D spherical coordinates and 1D/2D cylindrical coordinates, which testify for the fifth order accuracy and ENO property of the scheme. For a multi$-$dimensional test case, only the interface values are considered to integrate the source term, while for 1D test cases, mid$-$point values are also used. As a final note, it is emphasized that the present scheme can be extended to arbitrary order of accuracy and different techniques of reconstruction in multi$-$dimensions. 

\section{Acknowledgement}
The current research is supported by Hong Kong Research Grant Council (16207715, 16206617) and National Science Foundation of China (11772281, 91530319).

\appendix 
\section{WENO$-$C reconstruction weights}\label{Appendix:A}
\subsection{Cartesian coordinates}
Weights for a uniform grid in Cartesian coordinates are provided for the sake of completeness of the present scheme and ease in understanding of the reader. Also, cylindrical ($z$,$\theta$) and spherical ($\phi$) coordinates discussed in the later sections require same weights as of Cartesian coordinates.

\subsubsection{Linear weights}  \label{Cartesianlinearweights}
 In case of Cartesian coordinates ($x,y,z$), the linear weights are obtained by putting $m=0$ in Eq. (\ref{eq:26}) and then inverting the $\beta-$matrix in Eq. (\ref{eq:24}).
 
 \begin{itemize}
 
     \item Positive (right) weights: \newline $S_0^{3+}(i-2, i-1,i)::\quad(w_{i,0,-2}^{3+},w_{i,0,-1}^{3+},w_{i,0,0}^{3+}) =\bigg(\frac{1}{3},-\frac{7}{6},\frac{11}{6}\bigg)$
     \newline
     $S_1^{3+}(i-1,i,i+1)::\quad(w_{i,1,-1}^{3+},w_{i,1,0}^{3+},w_{i,1,+1}^{3+}) = \bigg(-\frac{1}{6},\frac{5}{6},\frac{1}{3}\bigg)$
     \newline
     $S_2^{3+}(i,i+1,i+2)::\quad(w_{i,2,0}^{3+},w_{i,2,+1}^{3+},w_{i,2,+2}^{3+}) = \bigg(\frac{1}{3},\frac{5}{6},-\frac{1}{6}\bigg)$
     
     \item Middle (mid$-$value) weights: \newline $S_0^{3M}(i-2,i-1,i)::\quad(w_{i,0,-2}^{3M},w_{i,0,-1}^{3M},w_{i,0,0}^{3M}) =\noindent\bigg(-\frac{1}{24},\frac{1}{12},\frac{23}{24}\bigg)$
     \newline
     $S_1^{3M}(i-1,i,i+1)::\quad(w_{i,1,-1}^{3M},w_{i,1,0}^{3M},w_{i,1,+1}^{3M}) = \noindent\bigg(-\frac{1}{24},\frac{13}{12},-\frac{1}{24}\bigg)$
     \newline
     $S_2^{3M}(i,i+1,i+2)::\quad(w_{i,2,0}^{3M},w_{i,2,+1}^{3M},w_{i,2,+2}^{3M}) = \noindent\bigg(\frac{23}{24},\frac{1}{12},-\frac{1}{24}\bigg)$
     
     \item Negative (left) weights:\newline
     $S_0^{3-}(i-2,i-1,i)::\quad(w_{i,0,-2}^{3-},w_{i,0,-1}^{3-},w_{i,0,0}^{3-}) = \noindent\bigg(-\frac{1}{6},\frac{5}{6},\frac{1}{3}\bigg)$ \newline
     $S_1^{3-}(i-1,i,i+1)::\quad(w_{i,1,-1}^{3-},w_{i,1,0}^{3-},w_{i,1,+1}^{3-}) =\noindent\bigg(\frac{1}{3},\frac{5}{6},-\frac{1}{6}\bigg)$ \newline
     $S_2^{3-}(i,i+1,i+2)::\quad(w_{i,2,0}^{3-},w_{i,2,+1}^{3-},w_{i,2,+2}^{3-}) = \noindent\bigg(\frac{11}{6},-\frac{7}{6},\frac{1}{3}\bigg)$
 \end{itemize}

\subsubsection{Fifth order interpolation weights}
\label{cart_fifth_order}
\begin{itemize}
    \item Positive (right) weights:\newline
    $S_0^{5+}::\quad(w_{i,0,-2}^{5+},w_{i,0,-1}^{5+},w_{i,0,0}^{5+},w_{i,0,+1}^{5+},w_{i,0,+2}^{5+})=\noindent\bigg(\frac{1}{30},-\frac{13}{60},\frac{47}{60},\frac{9}{20},-\frac{1}{20}\bigg)$
    
    \item Middle (mid$-$value) weights:\newline
    $S_0^{5M}::\quad(w_{i,0,-2}^{5M},w_{i,0,-1}^{5M},w_{i,0,0}^{5M},w_{i,0,+1}^{5M},w_{i,0,+2}^{5M})=\noindent\bigg(\frac{3}{640},-\frac{29}{480},\frac{1067}{960},-\frac{29}{480},\frac{3}{640}\bigg)$     
    
    \item Negative (left) weights:\newline
    $S_0^{5-}::\quad(w_{i,0,-2}^{5-},w_{i,0,-1}^{5-},w_{i,0,0}^{5-},w_{i,0,+1}^{5-},w_{i,0,+2}^{5-})=\noindent\bigg(-\frac{1}{20},\frac{9}{20},\frac{47}{60},-\frac{13}{60},\frac{1}{30}\bigg)$
\end{itemize}    
    
\subsubsection{Optimal weights} \label{Cartesianoptimalweights}
The linear weights in Cartesian coordinates in ($x,y,z$) coordinates are constants, thus, the optimal weights are also constants. Moreover, positive and negative weights are mirror$-$symmetric for this case.
\begin{itemize}
\item Positive (right) weights:: {$(C_{i,0}^+,C_{i,1}^+,C_{i,2}^+)=\noindent\bigg(\frac{1}{10},\frac{3}{5},\frac{3}{10}\bigg)$}

\item Middle (mid$-$value) weights:: {$(C_{i,0}^M,C_{i,1}^M,C_{i,2}^M)=\noindent\bigg(-\frac{9}{80},\frac{49}{40},-\frac{9}{80}\bigg)$}

\item Negative (left) weights:: {$(C_{i,0}^-,C_{i,1}^-,C_{i,2}^-)=\noindent\bigg(\frac{3}{10},\frac{3}{5},\frac{1}{10}\bigg)$}
\end{itemize}

\subsubsection{Weights for interface value integration} \label{cart_int_wt}
Weights for the interface value integration to yield line$-$/face$-$averaged flux with different integration points are provided as follows:
\begin{itemize}
\item Fifth order quadrature (all middle values)::$\quad(w_{i,-2}^M,w_{i,-1}^M,w_{i,0}^M,w_{i,+1}^M,w_{i,+2}^M)= \\ \noindent\bigg(-\frac{17}{5760},\frac{77}{1440},\frac{863}{960},\frac{77}{1440},-\frac{17}{5760}\bigg)$

\item Sixth order quadrature (all interface values)::$\quad(w_{i,-5/2}^+,w_{i,-3/2}^+,w_{i,-1/2}^+,w_{i,+1/2}^-,w_{i,+3/2}^-,w_{i,+5/2}^-)=\\ \noindent\bigg(\frac{11}{1440},-\frac{31}{480},\frac{401}{720},\frac{401}{720},-\frac{31}{480},\frac{11}{1440}\bigg)$
\end{itemize}

\subsubsection{Weights for source term integration} \label{cart_source_wt}
Since one$-$dimensional Jacobian is unity for Cartesian coordinates, weights for flux and source term integrations are the same. For 1D case, 3 point based Simpson quadrature can also be used to attain fifth order accuracy. Few quadratures are given below:
\begin{itemize}
\item 3 point Simpson quadrature (2 interface, 1 middle values)::$\quad(w_{i,-1/2}^+,w_{i,0}^M,w_{i,+1/2}^-)= \noindent\bigg(\frac{1}{6},\frac{2}{3},\frac{1}{6}\bigg)$ 

\item Fifth order quadrature (all middle values):: Refer to \ref{cart_int_wt}
\item Sixth order quadrature (all interface values):: Refer to \ref{cart_int_wt}
\end{itemize} 

\subsection{Cylindrical coordinates}
The weights for WENO$-$C reconstruction and integration in cylindrical ($\theta,z$) coordinates are the same as of Cartesian coordinates because the one$-$dimensional Jacobians are unity. However, the weights in the radial direction are different as the one$-$dimensional Jacobian is $\xi$. Their values are given in this section.

\subsubsection{Linear weights} \label{cylindricallinearweights}
The linear weights for the radial coordinate $R$ are independent of the grid spacing and depend only on the index number $i$ ($i={R_{i+\frac{1}{2}}}/{\Delta{R}}$), as given below. In the vanishing curvature ($R \to \infty $ and therefore $i \to \infty$), the linear weights of the conventional WENO reconstruction in Cartesian coordinates can be recovered.

 \begin{itemize}
 
     \item Positive (right) weights: \newline $S_0^{3+}(i-2,i-1,i)::\quad(w_{i,0,-2}^{3+},w_{i,0,-1}^{3+},w_{i,0,0}^{3+}) =\noindent\bigg(\frac{(-5+2 i) (4-9 i+4 i^2)}{12 (-3+2 i) (1-3 i+i^2)},\frac{-23+45 i-14 i^2}{12 (1-3
i+i^2)},\frac{(-1+2 i) (85-90 i+22 i^2)}{12 (-3+2 i) (1-3 i+i^2)}\bigg)$
     \newline
     $S_1^{3+}(i-1,i,i+1)::\quad(w_{i,1,-1}^{3+},w_{i,1,0}^{3+},w_{i,1,+1}^{3+}) = \noindent\bigg(-\frac{(-3+2 i) (-1+2 i^2)}{12 (-1+2 i) (-1-i+i^2)},\frac{11+9 i-10 i^2}{12 (1+i-i^2)},-\frac{-4+i+14
i^2-8 i^3}{12 (1-i-3 i^2+2 i^3)}\bigg)$
     \newline
     $S_2^{3+}(i,i+1,i+2)::\quad(w_{i,2,0}^{3+},w_{i,2,+1}^{3+},w_{i,2,+2}^{3+}) = \noindent\bigg(\frac{(-1+2 i) (4+9 i+4 i^2)}{12 (1+2 i) (-1+i+i^2)},\frac{-11+9 i+10 i^2}{12 (-1+i+i^2)},-\frac{(3+2
i) (-1+2 i^2)}{12 (1+2 i) (-1+i+i^2)}\bigg)
$
     
     \item Middle (mid$-$value) weights: \newline $S_0^{3M}(i-2,i-1,i)::\quad(w_{i,0,-2}^{3M},w_{i,0,-1}^{3M},w_{i,0,0}^{3M}) =\noindent\bigg(\frac{5+3 i-7 i^2+2 i^3}{72-264 i+216 i^2-48 i^3},\frac{-4-i+i^2}{12 (1-3 i+i^2)},\frac{(-1+2
i) (91-95 i+23 i^2)}{24 (-3+2 i) (1-3 i+i^2)}\bigg)
$
     \newline
     $S_1^{3M}(i-1,i,i+1)::\quad(w_{i,1,-1}^{3M},w_{i,1,0}^{3M},w_{i,1,+1}^{3M}) = \noindent\bigg(\frac{3-2 i}{-24+48 i},\frac{13}{12},\frac{1+2 i}{24-48 i}\bigg)$
     \newline
     $S_2^{3M}(i,i+1,i+2)::\quad(w_{i,2,0}^{3M},w_{i,2,+1}^{3M},w_{i,2,+2}^{3M}) = \noindent\bigg(\frac{(-1+2 i) (19+49 i+23 i^2)}{24 (1+2 i) (-1+i+i^2)},\frac{-4-i+i^2}{12 (-1+i+i^2)},-\frac{(3+2
i) (-1-i+i^2)}{24 (1+2 i) (-1+i+i^2)}\bigg)$
     
     \item Negative (left) weights:\newline
     $S_0^{3-}(i-2,i-1,i)::\quad(w_{i,0,-2}^{3-},w_{i,0,-1}^{3-},w_{i,0,0}^{3-}) = \noindent\bigg(-\frac{(-5+2 i) (1-4 i+2 i^2)}{12 (-3+2 i) (1-3 i+i^2)},\frac{8-29 i+10 i^2}{12 (1-3
i+i^2)},\frac{(-1+2 i) (17-17 i+4 i^2)}{12 (-3+2 i) (1-3 i+i^2)}\bigg)$ \newline
     $S_1^{3-}(i-1,i,i+1)::\quad(w_{i,1,-1}^{3-},w_{i,1,0}^{3-},w_{i,1,+1}^{3-}) =\noindent\bigg(\frac{(-3+2 i) (-1+i+4 i^2)}{12 (-1+2 i) (-1-i+i^2)},\frac{10+11 i-10 i^2}{12 (1+i-i^2)},\frac{-1+2
i+6 i^2-4 i^3}{12 (1-i-3 i^2+2 i^3)}\bigg)$ \newline
     $S_2^{3-}(i,i+1,i+2)::\quad(w_{i,2,0}^{3-},w_{i,2,+1}^{3-},w_{i,2,+2}^{3-}) = \noindent\bigg(\frac{(-1+2 i) (17+46 i+22 i^2)}{12 (1+2 i) (-1+i+i^2)},\frac{8-17 i-14 i^2}{12 (-1+i+i^2)},\frac{(3+2
i) (-1+i+4 i^2)}{12 (1+2 i) (-1+i+i^2)}\bigg)
$
 \end{itemize}

\subsubsection{Fifth order interpolation weights}
\label{cyl_fifth_order}
\begin{itemize}
    \item Positive (right) weights:\newline
    $S_0^{5+}::\quad(w_{i,0,-2}^{5+},w_{i,0,-1}^{5+},w_{i,0,0}^{5+},w_{i,0,+1}^{5+},w_{i,0,+2}^{5+})=\noindent\bigg(\frac{(-5+2 i) (4-10 i^2+3 i^4)}{30 (-1+2 i) (12+16 i-13 i^2-6 i^3+3 i^4)},\\-\frac{(-3+2
i) (164+45 i-380 i^2-75 i^3+78 i^4)}{120 (-1+2 i) (12+16 i-13 i^2-6 i^3+3 i^4)},\frac{1276+1395 i-1300 i^2-525
i^3+282 i^4}{120 (12+16 i-13 i^2-6 i^3+3 i^4)},\frac{(1+2 i) (-228+465 i-60 i^2-175 i^3+54 i^4)}{40 (-1+2 i)
(12+16 i-13 i^2-6 i^3+3 i^4)},\\-\frac{(3+2 i) (-12+15 i+20 i^2-25 i^3+6 i^4)}{40 (-1+2 i) (12+16 i-13 i^2-6
i^3+3 i^4)}\bigg)$
    
    \item Middle (mid$-$value)  weights:\newline
    $S_0^{5M}::\quad(w_{i,0,-2}^{5M},w_{i,0,-1}^{5M},w_{i,0,0}^{5M},w_{i,0,+1}^{5M},w_{i,0,+2}^{5M})=\\ \noindent\bigg(\frac{3 (-5+2 i)}{640 (-1+2 i)},-\frac{29 (-3+2 i)}{480 (-1+2 i)},\frac{1067}{960},\frac{29+58
i}{480-960 i},\frac{3 (3+2 i)}{640 (-1+2 i)}\bigg)$     
    
    \item Negative (left) weights:\newline
    $S_0^{5-}::\quad(w_{i,0,-2}^{5-},w_{i,0,-1}^{5-},w_{i,0,0}^{5-},w_{i,0,+1}^{5-},w_{i,0,+2}^{5-})=\noindent\bigg(-\frac{(-5+2 i) (4-4 i-19 i^2+i^3+6 i^4)}{40 (-1+2 i) (12+16 i-13 i^2-6 i^3+3 i^4)},\\ \frac{(-3+2
i) (56-36 i-261 i^2-41 i^3+54 i^4)}{40 (-1+2 i) (12+16 i-13 i^2-6 i^3+3 i^4)},\frac{1128+1652 i-1183 i^2-603
i^3+282 i^4}{120 (12+16 i-13 i^2-6 i^3+3 i^4)},-\frac{(1+2 i) (-168+628 i-137 i^2-237 i^3+78 i^4)}{120 (-1+2
i) (12+16 i-13 i^2-6 i^3+3 i^4)},\\ \frac{(3+2 i) (-3+8 i+8 i^2-12 i^3+3 i^4)}{30 (-1+2 i) (12+16 i-13 i^2-6
i^3+3 i^4)}\bigg)$
\end{itemize}    
    
\subsubsection{Optimal weights} 
\label{cylindricaloptimalweights}
The optimal weights in cylindrical$-$radial $R$ coordinates are given below. It is observed that the weights are not mirror$-$symmetric and are independent of the grid spacing but depend only on the index number $i$ ( $i={R_{i+\frac{1}{2}}}/{\Delta{R}}$).
\begin{itemize}
\item Positive (right) weights:: $(C_{i,0}^+,C_{i,1}^+,C_{i,2}^+)=\noindent\bigg(\frac{2 (-3+2 i) (1-3 i+i^2) (4-10 i^2+3 i^4)}{5 (-1+2 i) (4-9 i+4 i^2) (12+16 i-13
i^2-6 i^3+3 i^4)},\\ \frac{3 (-1-i+i^2) (96-192 i-191 i^2+500 i^3-83 i^4-154 i^5+48 i^6)}{10 (-1+2
i^2) (4-9 i+4 i^2) (12+16 i-13 i^2-6 i^3+3 i^4)},\frac{3 (1+2 i) (-1+i+i^2) (-12+15
i+20 i^2-25 i^3+6 i^4)}{10 (-1+2 i) (-1+2 i^2) (12+16 i-13 i^2-6 i^3+3 i^4)}\bigg)$

\item Middle (mid$-$value) weights:: $(C_{i,0}^M,C_{i,1}^M,C_{i,2}^M)=\noindent\bigg(-\frac{9 (-3+11 i-9 i^2+2 i^3)}{80 (-1+2 i) (-1-i+i^2)},\frac{22+49 i-49 i^2}{40
(1+i-i^2)},-\frac{9 (1+2 i) (-1+i+i^2)}{80 (-1+2 i) (-1-i+i^2)}\bigg)$

\item Negative (left) weights:: $(C_{i,0}^-,C_{i,1}^-,C_{i,2}^-)=\noindent\bigg(\frac{3 (-3+2 i) (1-3 i+i^2) (4-4 i-19 i^2+i^3+6 i^4)}{10 (-1+2 i) (1-4 i+2 i^2) (12+16
i-13 i^2-6 i^3+3 i^4)},\\ \frac{3 (-1-i+i^2) (24-112 i-9 i^2+412 i^3-133 i^4-134 i^5+48 i^6)}{10 (1-4
i+2 i^2) (-1+i+4 i^2) (12+16 i-13 i^2-6 i^3+3 i^4)},\frac{2 (1+2 i) (-1+i+i^2) (-3+8
i+8 i^2-12 i^3+3 i^4)}{5 (-1+2 i) (-1+i+4 i^2) (12+16 i-13 i^2-6 i^3+3 i^4)}\bigg)
$
\end{itemize}    

\subsubsection{Weights for interface value integration} \label{cyl_int_wt}
For 2D cases, one$-$dimensional Jacobian is the same as of source$-$term integration, given in table \ref{tab:1}. The weights for quadrature in the radial direction are given below, where $R_i$ is the radius of cell center. 
\begin{itemize}
\item Fifth order quadrature (all middle values)::$\quad(w_{i,-2}^M,w_{i,-1}^M,w_{i,0}^M,w_{i,+1}^M,w_{i,+2}^M)= \\ \noindent\bigg(\frac{17 (2{\Delta R}-R_i)}{5760 R_i}, -\frac{77 ({\Delta R}-R_i)}{1440 R_i},\frac{863}{960},\frac{77 ({\Delta R}+R_i)}{1440 R_i},-\frac{17
(2{\Delta R}+R_i)}{5760 R_i}\bigg)$ 

\item Sixth order quadrature (all interface values)::\\$\quad(w_{i,-5/2}^+,w_{i,-3/2}^+,w_{i,-1/2}^+,w_{i,+1/2}^-,w_{i,+3/2}^-,w_{i,+5/2}^-)=\noindent\bigg(\frac{154-\frac{3{\Delta R}}{R_i}}{20160},-\frac{31}{480}+\frac{43 {\Delta R}}{20160 R_i},\frac{401}{720}-\frac{299 {\Delta R}}{3360
R_i},\frac{401}{720}+\frac{299 {\Delta R}}{3360 R_i},\\ -\frac{31}{480}-\frac{43 {\Delta R}}{20160 R_i},\frac{154+\frac{3 {\Delta R}}{R_i}}{20160}\bigg)$ 
\end{itemize} 

From table \ref{tab:2}, it is clear that for 3D cases, one$-$dimensional Jacobian is altered for surface integrals. Therefore, the weights for surface averaging are different. For ($R-z$) and ($\theta-z$) coordinates, the one$-$dimensional Jacobians are unity for both the sweeps. But for ($R-\theta$) case, the $R-$directional integration can be performed by the weights given earlier in this section and $\theta-$directional integration using the same weights as of Cartesian case, given in \ref{cart_int_wt}.   

\subsubsection{Weights for source term integration} \label{cyl_source_wt}
For source term integration, the one$-$dimensional Jacobian is the original value as summarized in table \ref{tab:1}. But in this case, regularization is performed to get rid of `$1/R$' factor. Apart from the radial integration, the weights for $\theta-$ and $z-$directional integration are the same as of Cartesian weights given in \ref{cart_source_wt}. Weights for $r-$directional integration are given below:

\begin{itemize}
\item 3 point Simpson quadrature (2 interface, 1 middle values)::
\begin{enumerate}
\item Original weights: {$\quad(w_{i,-1/2}^+,w_{i,0}^M,w_{i,+1/2}^-)= \noindent\bigg(\frac{1}{6}-\frac{{\Delta R}}{12 R_i},\frac{2}{3},\frac{{\Delta R}+2 R_i}{12 R_i}\bigg)$}

\item Regularized weights: $\quad(\hat{w}_{i,-1/2}^+,\hat{w}_{i,0}^M,\hat{w}_{i,+1/2}^-)= \noindent\bigg(\frac{1}{6 R_i},\frac{2}{3 R_i},\frac{1}{6 R_i}\bigg)$ 
\end{enumerate}

\item Fifth order quadrature (all middle values):: 
\begin{enumerate}
\item Original weights: Refer to \ref{cyl_int_wt}

\item Regularized weights: $\quad(\hat{w}_{i,-2}^M,\hat{w}_{i,-1}^M,\hat{w}_{i,0}^M,\hat{w}_{i,+1}^M,\hat{w}_{i,+2}^M)= \\ \noindent\bigg(-\frac{17}{5760 R_i},\frac{77}{1440 R_i},\frac{863}{960 R_i},\frac{77}{1440 R_i},-\frac{17}{5760 R_i}\bigg)$
\end{enumerate}

\item Sixth order quadrature (all interface values):: 
\begin{enumerate}
\item Original weights: Refer to \ref{cyl_int_wt}

\item Regularized weights: $\quad(\hat{w}_{i,-5/2}^+,\hat{w}_{i,-3/2}^+,\hat{w}_{i,-1/2}^+,\hat{w}_{i,+1/2}^-,\hat{w}_{i,+3/2}^-,\hat{w}_{i,+5/2}^-)=\\ \noindent\bigg(\frac{11}{1440 R_i},-\frac{31}{480 R_i},\frac{401}{720 R_i},\frac{401}{720 R_i},-\frac{31}{480 R_i},\frac{11}{1440 R_i}\bigg)$
\end{enumerate}
\end{itemize}

 \subsection{Spherical coordinates} 
The weights for WENO$-$C reconstruction and integration in spherical ($\phi$) coordinates are the same as of Cartesian coordinates because the one$-$dimensional Jacobian is unity. However, the weights in spherical$-$radial and spherical$-$meridional directions are different as the one$-$dimensional Jacobians are $\xi^2$ and $sin \xi$ respectively for the volumetric operations.
\subsubsection{Linear weights}
 \label{sphericallinearweights}
 The weights for the radial coordinate $r$ are independent of the grid spacing and depend only on the index number $i$ ($i={r_{i+\frac{1}{2}}}/{\Delta{r}}$) of the grid, as given below. Again, in the vanishing curvature ($R \to \infty $ and therefore $i \to \infty$), the linear weights of the conventional WENO reconstruction in Cartesian coordinates can be recovered. Also, for the case of spherical$-$meridional coordinate ($\theta$), analytical solutions are highly complex. Therefore, application of direct numerical inversion is advised.

 \begin{itemize}
 
     \item Positive (right) weights: \newline {$S_0^{3+}(i-2,i-1,i)::\quad(w_{i,0,-2}^{3+},w_{i,0,-1}^{3+},w_{i,0,0}^{3+}) =\noindent\bigg(\frac{(19-15 i+3 i^2) (12-48 i+72 i^2-45 i^3+10 i^4)}{9 (36-198 i+471 i^2-540 i^3+315 i^4-90
i^5+10 i^6)},\\ -\frac{(7-9 i+3 i^2) (219-768 i+963 i^2-450 i^3+70 i^4)}{18 (36-198 i+471 i^2-540 i^3+315
i^4-90 i^5+10 i^6)},\frac{(1-3 i+3 i^2) (1725-3552 i+2709 i^2-900 i^3+110 i^4)}{18 (36-198 i+471
i^2-540 i^3+315 i^4-90 i^5+10 i^6)}\bigg)$}
     \newline
     {$S_1^{3+}(i-1,i,i+1)::\quad(w_{i,1,-1}^{3+},w_{i,1,0}^{3+},w_{i,1,+1}^{3+}) = \noindent\bigg(-\frac{(7-9 i+3 i^2) (3-9 i^2+10 i^4)}{18 (4-6 i-9 i^2+20 i^3+15 i^4-30 i^5+10 i^6)},\\ \frac{(1-3
i+3 i^2) (69+96 i-63 i^2-90 i^3+50 i^4)}{18 (4-6 i-9 i^2+20 i^3+15 i^4-30 i^5+10 i^6)},\frac{(1+3
i+3 i^2) (12-48 i+72 i^2-45 i^3+10 i^4)}{9 (4-6 i-9 i^2+20 i^3+15 i^4-30 i^5+10 i^6)}\bigg)$}
     \newline
     {$S_2^{3+}(i,i+1,i+2)::\quad(w_{i,2,0}^{3+},w_{i,2,+1}^{3+},w_{i,2,+2}^{3+}) = \noindent\bigg(\frac{(1-3 i+3 i^2) (12+48 i+72 i^2+45 i^3+10 i^4)}{9 (4+6 i-9 i^2-20 i^3+15 i^4+30 i^5+10 i^6)},\\ \frac{(1+3
i+3 i^2) (69-96 i-63 i^2+90 i^3+50 i^4)}{18 (4+6 i-9 i^2-20 i^3+15 i^4+30 i^5+10 i^6)},-\frac{(7+9
i+3 i^2) (3-9 i^2+10 i^4)}{18 (4+6 i-9 i^2-20 i^3+15 i^4+30 i^5+10 i^6)}\bigg)
$}

     \item Middle (mid$-$value) weights: \newline {$S_0^{3M}(i-2,i-1,i)::\quad(w_{i,0,-2}^{3M},w_{i,0,-1}^{3M},w_{i,0,0}^{3M}) =\noindent\bigg(-\frac{(19-15 i+3 i^2) (-20+58 i-21 i^2-20 i^3+10 i^4)}{72 (36-198 i+471 i^2-540 i^3+315 i^4-90
i^5+10 i^6)},\\ \frac{(7-9 i+3 i^2) (-223+590 i-222 i^2-40 i^3+20 i^4)}{72 (36-198 i+471 i^2-540 i^3+315
i^4-90 i^5+10 i^6)},\frac{(1-3 i+3 i^2) (3773-7672 i+5781 i^2-1900 i^3+230 i^4)}{72 (36-198 i+471
i^2-540 i^3+315 i^4-90 i^5+10 i^6)}\bigg)
$}
     \newline
     {$S_1^{3M}(i-1,i,i+1)::\quad(w_{i,1,-1}^{3M},w_{i,1,0}^{3M},w_{i,1,+1}^{3M}) = \noindent\bigg(-\frac{(7-9 i+3 i^2) (7+4 i-21 i^2-20 i^3+10 i^4)}{72 (4-6 i-9 i^2+20 i^3+15 i^4-30 i^5+10 i^6)},\\ \frac{(1-3
i+3 i^2) (317+482 i-222 i^2-520 i^3+260 i^4)}{72 (4-6 i-9 i^2+20 i^3+15 i^4-30 i^5+10 i^6)},-\frac{(1+3
i+3 i^2) (-20+58 i-21 i^2-20 i^3+10 i^4)}{72 (4-6 i-9 i^2+20 i^3+15 i^4-30 i^5+10 i^6)}\bigg)$}
     \newline
     {$S_2^{3M}(i,i+1,i+2)::\quad(w_{i,2,0}^{3M},w_{i,2,+1}^{3M},w_{i,2,+2}^{3M}) = \noindent\bigg(\frac{(1-3 i+3 i^2) (212+890 i+1461 i^2+980 i^3+230 i^4)}{72 (4+6 i-9 i^2-20 i^3+15 i^4+30 i^5+10
i^6)},\\ \frac{(1+3 i+3 i^2) (125-106 i-222 i^2-40 i^3+20 i^4)}{72 (4+6 i-9 i^2-20 i^3+15 i^4+30 i^5+10
i^6)},-\frac{(7+9 i+3 i^2) (7+4 i-21 i^2-20 i^3+10 i^4)}{72 (4+6 i-9 i^2-20 i^3+15 i^4+30 i^5+10
i^6)}\bigg)$}
     
     \item Negative (left) weights:\newline
     {$S_0^{3-}(i-2,i-1,i)::\quad(w_{i,0,-2}^{3-},w_{i,0,-1}^{3-},w_{i,0,0}^{3-}) = \noindent\bigg(-\frac{(19-15 i+3 i^2) (4-22 i+51 i^2-40 i^3+10 i^4)}{18 (36-198 i+471 i^2-540 i^3+315 i^4-90
i^5+10 i^6)},\\ \frac{(7-9 i+3 i^2) (50-248 i+507 i^2-290 i^3+50 i^4)}{18 (36-198 i+471 i^2-540 i^3+315
i^4-90 i^5+10 i^6)},\frac{(1-3 i+3 i^2) (187-367 i+267 i^2-85 i^3+10 i^4)}{9 (36-198 i+471 i^2-540
i^3+315 i^4-90 i^5+10 i^6)}\bigg)$} \newline
     {$S_1^{3-}(i-1,i,i+1)::\quad(w_{i,1,-1}^{3-},w_{i,1,0}^{3-},w_{i,1,+1}^{3-}) =\noindent\bigg(\frac{(7-9 i+3 i^2) (1-i-3 i^2+5 i^3+10 i^4)}{9 (4-6 i-9 i^2+20 i^3+15 i^4-30 i^5+10 i^6)},\\ \frac{(1-3
i+3 i^2) (62+100 i-33 i^2-110 i^3+50 i^4)}{18 (4-6 i-9 i^2+20 i^3+15 i^4-30 i^5+10 i^6)}, -\frac{(1+3
i+3 i^2) (4-22 i+51 i^2-40 i^3+10 i^4)}{18 (4-6 i-9 i^2+20 i^3+15 i^4-30 i^5+10 i^6)}\bigg)$} \newline
     {$S_2^{3-}(i,i+1,i+2)::\quad(w_{i,2,0}^{3-},w_{i,2,+1}^{3-},w_{i,2,+2}^{3-}) = \noindent\bigg(\frac{(1-3 i+3 i^2) (92+394 i+669 i^2+460 i^3+110 i^4)}{18 (4+6 i-9 i^2-20 i^3+15 i^4+30 i^5+10
i^6)},\\ -\frac{(1+3 i+3 i^2) (34-88 i+33 i^2+170 i^3+70 i^4)}{18 (4+6 i-9 i^2-20 i^3+15 i^4+30 i^5+10
i^6)},\frac{(7+9 i+3 i^2) (1-i-3 i^2+5 i^3+10 i^4)}{9 (4+6 i-9 i^2-20 i^3+15 i^4+30 i^5+10 i^6)}\bigg)
$}
 \end{itemize}

\subsubsection{Fifth order interpolation weights}
\label{sph_fifth_order}
\begin{itemize}
    \item Positive (right) weights:\newline
    {$S_0^{5+}::\quad(w_{i,0,-2}^{5+},w_{i,0,-1}^{5+},w_{i,0,0}^{5+},w_{i,0,+1}^{5+},w_{i,0,+2}^{5+})=\\ \noindent\bigg(\frac{(19-15 i+3 i^2) (16-60 i^2+94 i^4-45 i^6+7 i^8)}{90 (48-48 i-164 i^2+200 i^3+390 i^4-399
i^5-161 i^6+210 i^7-35 i^9+7 i^{10})},\\ -\frac{(7-9 i+3 i^2) (508+240 i-1740 i^2-795 i^3+2417 i^4+930 i^5-780
i^6-175 i^7+91 i^8)}{180 (48-48 i-164 i^2+200 i^3+390 i^4-399 i^5-161 i^6+210 i^7-35 i^9+7 i^{10})},\\ \frac{(1-3
i+3 i^2) (8132+15120 i-5700 i^2-20325 i^3+3863 i^4+8670 i^5-1800 i^6-1225 i^7+329 i^8)}{180 (48-48 i-164 i^2+200 i^3+390 i^4-399
i^5-161 i^6+210 i^7-35 i^9+7 i^{10})},\\ \frac{(1+3 i+3 i^2) (4212-15120 i+16560 i^2+1275 i^3-11517 i^4+4350 i^5+1620
i^6-1225 i^7+189 i^8)}{180 (48-48 i-164 i^2+200 i^3+390 i^4-399 i^5-161 i^6+210 i^7-35 i^9+7 i^{10})},\\ -\frac{(7+9
i+3 i^2) (108-240 i-120 i^2+645 i^3-223 i^4-510 i^5+510 i^6-175 i^7+21 i^8)}{180 (48-48 i-164 i^2+200 i^3+390 i^4-399 i^5-161
i^6+210 i^7-35 i^9+7 i^{10})}\bigg)$}
    
    \item Middle (mid$-$value) weights: \newline
    {$S_0^{5M}::\quad(w_{i,0,-2}^{5M},w_{i,0,-1}^{5M},w_{i,0,0}^{5M},w_{i,0,+1}^{5M},w_{i,0,+2}^{5M})=\\ \noindent\bigg(\frac{(19-15 i+3 i^2) (176+128 i-660 i^2-752 i^3+562 i^4+468 i^5-183 i^6-84 i^7+21 i^8)}{1920 (48-48
i-164 i^2+200 i^3+390 i^4-399 i^5-161 i^6+210 i^7-35 i^9+7 i^{10})},\\ -\frac{(7-9 i+3 i^2) (9972+10866 i-30895
i^2-48744 i^3+13939 i^4+22846 i^5-4576 i^6-3248 i^7+812 i^8)}{5760 (48-48 i-164 i^2+200 i^3+390 i^4-399 i^5-161 i^6+210 i^7-35 i^9+7 i^{10})},\\ \frac{(1-3
i+3 i^2) (314028+637134 i-104105 i^2-911256 i^3+83561 i^4+404654 i^5-65174 i^6-59752 i^7+14938 i^8)}{5760 (48-48 i-164 i^2+200
i^3+390 i^4-399 i^5-161 i^6+210 i^7-35 i^9+7 i^{10})},\\ -\frac{(1+3 i+3 i^2) (-29028+70866 i+20855 i^2-75744 i^3+2689
i^4+27346 i^5-4576 i^6-3248 i^7+812 i^8)}{5760 (48-48 i-164 i^2+200 i^3+390 i^4-399 i^5-161 i^6+210 i^7-35 i^9+7 i^{10})},\\ \frac{(7+9
i+3 i^2) (-324+378 i+1215 i^2-752 i^3-1313 i^4+1218 i^5-183 i^6-84 i^7+21 i^8)}{1920 (48-48 i-164 i^2+200 i^3+390 i^4-399 i^5-161
i^6+210 i^7-35 i^9+7 i^{10})}\bigg)
$}     
    
    \item Negative (left) weights: \newline
    {$S_0^{5-}::\quad(w_{i,0,-2}^{5-},w_{i,0,-1}^{5-},w_{i,0,0}^{5-},w_{i,0,+1}^{5-},w_{i,0,+2}^{5-})=\\ \noindent\bigg(-\frac{(19-15 i+3 i^2) (16-16 i-60 i^2+96 i^3+222 i^4-51 i^5-127 i^6+7 i^7+21 i^8)}{180 (48-48
i-164 i^2+200 i^3+390 i^4-399 i^5-161 i^6+210 i^7-35 i^9+7 i^{10})},\\ \frac{(7-9 i+3 i^2) (344-164 i-1350 i^2+1184
i^3+4888 i^4+1071 i^5-1663 i^6-287 i^7+189 i^8)}{180 (48-48 i-164 i^2+200 i^3+390 i^4-399 i^5-161 i^6+210 i^7-35 i^9+7 i^{10})},\\ \frac{(1-3
i+3 i^2) (7064+15196 i-310 i^2-21376 i^3+368 i^4+9431 i^5-1163 i^6-1407 i^7+329 i^8)}{180 (48-48 i-164 i^2+200 i^3+390 i^4-399
i^5-161 i^6+210 i^7-35 i^9+7 i^{10})},\\ -\frac{(1+3 i+3 i^2) (696-3516 i+6850 i^2-1544 i^3-4388 i^4+2329 i^5+543
i^6-553 i^7+91 i^8)}{180 (48-48 i-164 i^2+200 i^3+390 i^4-399 i^5-161 i^6+210 i^7-35 i^9+7 i^{10})},\\ \frac{(7+9
i+3 i^2) (12-42 i+25 i^2+132 i^3-91 i^4-122 i^5+151 i^6-56 i^7+7 i^8)}{90 (48-48 i-164 i^2+200 i^3+390 i^4-399 i^5-161 i^6+210
i^7-35 i^9+7 i^{10})}\bigg)$}
\end{itemize}    
    
\subsubsection{Optimal weights} \label{sphericaloptimalweights}
The analytical values of the optimal weights for spherical$-$radial $r$ coordinates are highly intricate but are grid spacing independent and are given below for the uniform grid, where the index number $i={r_{i+\frac{1}{2}}}/{\Delta{r}}$.
\begin{itemize} 

\item Positive (right) weights:: {$(C_{i,0}^+,C_{i,1}^+,C_{i,2}^+)=\\ \noindent\bigg(\frac{(36-198 i+471 i^2-540 i^3+315 i^4-90 i^5+10 i^6) (16-60 i^2+94 i^4-45 i^6+7 i^8)}{10 (12-48
i+72 i^2-45 i^3+10 i^4) (48-48 i-164 i^2+200 i^3+390 i^4-399 i^5-161 i^6+210 i^7-35 i^9+7 i^{10})},\\ \frac{(4-6
i-9 i^2+20 i^3+15 i^4-30 i^5+10 i^6) (2592-9216 i+1908 i^2+29520 i^3-27762 i^4-36204 i^5+61932 i^6...}{10 (3-9 i^2+10 i^4) (12-48 i+72 i^2-45 i^3+10 i^4) (48-48 i-164 i^2+200 i^3+390 i^4-399 i^5-161
i^6+210 i^7-35 i^9+7 i^{10})}\\ \frac{... -6675 i^7-29126 i^8+12558 i^9+3036 i^{10}-2695
i^{11}+420 i^{12})}{10 (3-9 i^2+10 i^4) (12-48 i+72 i^2-45 i^3+10 i^4) (48-48 i-164 i^2+200 i^3+390 i^4-399 i^5-161
i^6+210 i^7-35 i^9+7 i^{10})},\\ \frac{(4+6 i-9 i^2-20 i^3+15 i^4+30 i^5+10 i^6) (108-240 i-120 i^2+645
i^3-223 i^4-510 i^5+510 i^6-175 i^7+21 i^8)}{10 (3-9 i^2+10 i^4) (48-48 i-164 i^2+200 i^3+390 i^4-399 i^5-161 i^6+210 i^7-35
i^9+7 i^{10})}\bigg)$}

\item Middle (mid$-$value) weights:: {$(C_{i,0}^M,C_{i,1}^M,C_{i,2}^M)= \\ \noindent\bigg(-\frac{3 (36-198 i+471 i^2-540 i^3+315 i^4-90 i^5+10 i^6) (176+128 i-660 i^2-752 i^3+562 i^4+468 i^5-183
i^6-84 i^7+21 i^8)}{80 (-20+58 i-21 i^2-20 i^3+10 i^4) (48-48 i-164 i^2+200 i^3+390 i^4-399 i^5-161 i^6+210 i^7-35 i^9+7 i^{10})},\\ \frac{(4-6
i-9 i^2+20 i^3+15 i^4-30 i^5+10 i^6) (-81696+135168 i+487832 i^2-473176 i^3-1302479 i^4+832366 i^5+1162664 i^6-754472 i^7}{(80 (7+4 i-21 i^2-20 i^3+10 i^4) (-20+58 i-21 i^2-20 i^3+10 i^4)
(48-48 i-164 i^2+200 i^3+390 i^4-399 i^5-161 i^6+210 i^7-35 i^9+7 i^{10}))}\\ \frac{-362767 i^8+292130
i^9+17034 i^{10}-41160 i^{11}+6860 i^{12})}{(80 (7+4 i-21 i^2-20 i^3+10 i^4) (-20+58 i-21 i^2-20 i^3+10 i^4)
(48-48 i-164 i^2+200 i^3+390 i^4-399 i^5-161 i^6+210 i^7-35 i^9+7 i^{10}))},\\ -\frac{3 (4+6 i-9 i^2-20
i^3+15 i^4+30 i^5+10 i^6) (-324+378 i+1215 i^2-752 i^3-1313 i^4+1218 i^5-183 i^6-84 i^7+21 i^8)}{80 (7+4 i-21 i^2-20 i^3+10
i^4) (48-48 i-164 i^2+200 i^3+390 i^4-399 i^5-161 i^6+210 i^7-35 i^9+7 i^{10})}\bigg)
$}

\item Negative (left) weights:: {$(C_{i,0}^-,C_{i,1}^-,C_{i,2}^-)=\\ \noindent\bigg(\frac{(36-198 i+471 i^2-540 i^3+315 i^4-90 i^5+10 i^6) (16-16 i-60 i^2+96 i^3+222 i^4-51 i^5-127 i^6+7 i^7+21
i^8)}{10 (4-22 i+51 i^2-40 i^3+10 i^4) (48-48 i-164 i^2+200 i^3+390 i^4-399 i^5-161 i^6+210 i^7-35 i^9+7 i^{10})},\\ \frac{(4-6
i-9 i^2+20 i^3+15 i^4-30 i^5+10 i^6) (17856-78336 i+24528 i^2+525848 i^3-493806 i^4-1868490 i^5+2594599 i^6+3894831 i^7...}{(10 (4-22 i+51 i^2-40
i^3+10 i^4) (1-i-3 i^2+5 i^3+10 i^4) (69+96 i-63 i^2-90 i^3+50 i^4) (48-48 i-164 i^2+200 i^3+390 i^4-399 i^5-161
i^6+210 i^7-35 i^9+7 i^{10})}\\ \frac{ ... -4959771 i^8 -3980631
i^9+5852829 i^{10}+327519 i^{11}-2477843 i^{12}+642525 i^{13}+299640 i^{14}-163450 i^{15}+21000 i^{16})}{(10 (4-22 i+51 i^2-40
i^3+10 i^4) (1-i-3 i^2+5 i^3+10 i^4) (69+96 i-63 i^2-90 i^3+50 i^4) (48-48 i-164 i^2+200 i^3+390 i^4-399 i^5-161
i^6+210 i^7-35 i^9+7 i^{10})},\\ \frac{(4+6 i-9 i^2-20 i^3+15 i^4+30 i^5+10 i^6) (12-42 i+25 i^2+132
i^3-91 i^4-122 i^5+151 i^6-56 i^7+7 i^8)}{10 (1-i-3 i^2+5 i^3+10 i^4) (48-48 i-164 i^2+200 i^3+390 i^4-399 i^5-161 i^6+210
i^7-35 i^9+7 i^{10})}\bigg)$}
\end{itemize}    

\subsubsection{Weights for interface value integration} \label{sph_int_wt}
In 2D case, the original weights for interpolation might be used according to the situation. In $z$ coordinates, the weights are the same as of Cartesian grids given in \ref{cart_int_wt}. Weights for $\theta-$directional integration are complex and advised to be computed numerically. $r-$directional integration weights are given below, where $r_i$ is the radius of the cell center. 

\begin{itemize}
\item Fifth order quadrature (all middle values)::{$\quad(w_{i,-2}^M,w_{i,-1}^M,w_{i,0}^M,w_{i,+1}^M,w_{i,+2}^M)= \\ \noindent\bigg(\frac{-69 {\Delta r}^2+1904 {\Delta r} r_i-476 r_i^2}{13440 ({\Delta r}^2+12 r_i^2)},\frac{321 {\Delta r}^2-4312 {\Delta r}
r_i+2156 r_i^2}{3360 ({\Delta r}^2+12 r_i^2)},\frac{1835 {\Delta r}^2+24164 r_i^2}{2240 ({\Delta r}^2+12 r_i^2)}, \frac{321 {\Delta r}^2+4312
{\Delta r} r_i+2156 r_i^2}{3360 ({\Delta r}^2+12 r_i^2)}, -\frac{69 {\Delta r}^2+1904 {\Delta r} r_i+476 r_i^2}{13440 ({\Delta r}^2+12 r_i^2)}\bigg)$} 

\item Sixth order quadrature (all interface values)::{$\quad(w_{i,-5/2}^+,w_{i,-3/2}^+,w_{i,-1/2}^+,w_{i,+1/2}^-,w_{i,+3/2}^-,w_{i,+5/2}^-)=\\ \noindent\bigg(\frac{15 {\Delta r}^2-12 {\Delta r} r_i+308 r_i^2}{3360 ({\Delta r}^2+12 r_i^2)},\frac{-129 {\Delta r}^2+172 {\Delta r}
r_i-2604 r_i^2}{3360 ({\Delta r}^2+12 r_i^2)},\frac{897 {\Delta r}^2-3588 {\Delta r} r_i+11228 r_i^2}{1680 ({\Delta r}^2+12 r_i^2)}, \frac{897
{\Delta r}^2+3588 {\Delta r} r_i+11228 r_i^2}{1680 ({\Delta r}^2+12 r_i^2)}, \\ -\frac{129 {\Delta r}^2+172 {\Delta r} r_i+2604 r_i^2}{3360 ({\Delta r}^2+12
r_i^2)}, \frac{15 {\Delta r}^2+12 {\Delta r} r_i+308 r_i^2}{3360 ({\Delta r}^2+12 r_i^2)}\bigg)$}
\end{itemize} 

For 3D cases, one$-$dimensional Jacobian values are given in table \ref{tab:2}. For ($r-\theta$) and ($r-\phi$) planes, the one directional sweeps in $r$ direction can be evaluated from the weights given in \ref{cyl_int_wt} and $\theta-$ or $\phi-$directional integration weights given in \ref{cart_int_wt}. For ($\theta-\phi$) planes, analytical values are complex as one$-$dimensional Jacobians are unity and $sin \xi$. Thus, they require direct numerical procedure.

\subsubsection{Weights for source term integration} \label{sph_source_wt}
The one$-$dimensional Jacobian values for this case are given in table \ref{tab:1}. The original and regularized quadrature values in $\phi$ direction can be computed from \ref{cart_source_wt}, $\theta$ direction by direct numerical operation, and radial ($r$) direction from the weights given below:

\begin{itemize}
\item 3 point Simpson quadrature (2 interface, 1 middle values)::
\begin{enumerate}
\item Original weights: {$\quad(w_{i,-1/2}^+,w_{i,0}^M,w_{i,+1/2}^-)= \noindent\bigg(\frac{3 {\Delta r}^2-20 {\Delta r} r_i+20 r_i^2}{10 {\Delta r}^2+120 r_i^2},\frac{2 ({\Delta r}^2+20 r_i^2)}{5 ({\Delta r}^2+12
r_i^2)},\frac{3 {\Delta r}^2+20 {\Delta r} r_i+20 r_i^2}{10 {\Delta r}^2+120 r_i^2}\bigg)
$}

\item Regularized weights: $\quad(\hat{w}_{i,-1/2}^+,\hat{w}_{i,0}^M,\hat{w}_{i,+1/2}^-)=  \noindent\bigg(-\frac{{\Delta r}-2 r_i}{{\Delta r}^2+12 r_i^2},\frac{8 r_i}{{\Delta r}^2+12 r_i^2},\frac{{\Delta r}+2 r_i}{{\Delta r}^2+12 r_i^2}\bigg)$ 
\end{enumerate}

\item Fifth order quadrature (all middle values):: 
\begin{enumerate}
\item Original weights: Refer to \ref{sph_int_wt}

\item Regularized weights: $\quad(\hat{w}_{i,-2}^M,\hat{w}_{i,-1}^M,\hat{w}_{i,0}^M,\hat{w}_{i,+1}^M,\hat{w}_{i,+2}^M)= \\ \noindent\bigg(\frac{17 (2 {\Delta r}-r_i)}{480 ({\Delta r}^2+12 r_i^2)},-\frac{77 ({\Delta r}-r_i)}{120 ({\Delta r}^2+12 r_i^2)},\frac{863
r_i}{80 ({\Delta r}^2+12 r_i^2)},\frac{77 ({\Delta r}+r_i)}{120 ({\Delta r}^2+12 r_i^2)},-\frac{17 (2 {\Delta r}+r_i)}{480 ({\Delta r}^2+12
r_i^2)}\bigg)$
\end{enumerate}

\item Sixth order quadrature (all interface values):: 
\begin{enumerate}
\item Original weights: Refer to \ref{sph_int_wt}

\item Regularized weights: {$\quad(\hat{w}_{i,-5/2}^+,\hat{w}_{i,-3/2}^+,\hat{w}_{i,-1/2}^+,\hat{w}_{i,+1/2}^-,\hat{w}_{i,+3/2}^-,\hat{w}_{i,+5/2}^-)=\\\noindent\bigg(\frac{-3 {\Delta r}+154 r_i}{1680 ({\Delta r}^2+12 r_i^2)},\frac{43 {\Delta r}-1302 r_i}{1680 ({\Delta r}^2+12 r_i^2)},\frac{-897
{\Delta r}+5614 r_i}{840 ({\Delta r}^2+12 r_i^2)},\frac{897 {\Delta r}+5614 r_i}{840 ({\Delta r}^2+12 r_i^2)}, -\frac{43 {\Delta r}+1302
r_i}{1680 ({\Delta r}^2+12 r_i^2)},\frac{3 {\Delta r}+154 r_i}{1680 ({\Delta r}^2+12 r_i^2)}\bigg)$}
\end{enumerate}
\end{itemize}

\section{Stability analysis of WENO$-$C for hyperbolic conservation laws}\label{Appendix:B}
For WENO$-$C to be practically useful, it is crucial that it enables a stable discretization for hyperbolic conservation laws when coupled with a proper time$-$integration scheme. In this section, we analyze WENO$-$C scheme for model problems involving smooth flow in 1$-$D Cartesian, cylindrical$-$radial, and spherical$-$radial coordinates, based on a modified von Neumann stability analysis \cite{liu2016wls}.

\subsection{Model problem in 1D}
We consider scalar advection equation (\ref{eq:stability_analysis1}) in 1D Cartesian, cylindrical$-$radial, and spherical$-$radial coordinates.

\begin{equation} \label{eq:stability_analysis1}
    \frac{\partial Q}{\partial t}+\frac{1}{(\partial \mathcal{V}/\partial \xi)}\frac{\partial}{\partial \xi}\bigg(\bigg(\frac{\partial \mathcal{V}}{\partial \xi}\bigg)Qv\bigg)=0 \quad \quad \xi\in [0,\infty],\quad t>0
\end{equation}
where $Q$ is the conserved variable, $(\partial \mathcal{V}/\partial \xi)=\xi^m$ is the one$-$dimensional Jacobian where $m=0,1,$ and $2$ in Cartesian, cylindrical$-$radial, and spherical$-$radial coordinates. Boundary conditions are not considered in the present approach to reduce the complexity of the analysis. Assuming a uniform grid $0=\xi_1<\xi_2<...<\xi_i<...<\xi_{\infty}=\infty$ with $\xi_i=i\Delta \xi$ and $\xi_{i+1}-\xi_i =\Delta \xi \quad\forall \quad i$ and $(i \pm 1/2)$ denotes the boundaries of the finite volume $i$. In the finite volume framework, Eq. (\ref{eq:stability_analysis1}) transforms into Eq. (\ref{eq:stability_analysis2}), which can be further approximated by conservative scheme given in Eq. (\ref{eq:stability_analysis5}).

\begin{equation} \label{eq:stability_analysis2}
    \frac{\partial \bar{Q}_i}{\partial t}=-\frac{1}{\Delta \mathcal{V}_i}(F(Q(\xi_{i+1/2},t))-F(Q(\xi_{i-1/2},t)))
\end{equation}

and
\begin{equation} \label{eq:stability_analysis5}
    \frac{\partial \bar{Q}_i}{\partial t}=-\frac{1}{\Delta \mathcal{V}_i}(\hat{F}_{i+1/2}-\hat{F}_{i-1/2})
\end{equation}

where
\begin{equation} \label{eq:stability_analysis3}
\bar{Q}(\xi_i,t)= -\frac{1}{\Delta \mathcal{V}_i}\int_{\xi_{i-1/2}}^{\xi_{i+1/2}} Q(\xi,t)\bigg(\frac{\partial \mathcal{V}(\xi,t)}{\partial \xi}\bigg)d \xi
\end{equation}

and 
\begin{equation} \label{eq:stability_analysis4}
\Delta \mathcal{V}_i= \int_{\xi_{i-1/2}}^{\xi_{i+1/2}} \bigg(\frac{\partial \mathcal{V}(\xi,t)}{\partial \xi}\bigg)d \xi
\end{equation}

The numerical flux $\hat{F}_{i+1/2}$ is replaced by the Lax$-$Friedrichs flux, as given in Eq. (\ref{eq:stability_analysis6}), with $\alpha=$max$_Q|F'(Q)|$. 
\begin{equation} \label{eq:stability_analysis6}
\vec{F}.\vec{n}=\frac{1}{2}\bigg[(\vec{F}(Q^-)+\vec{F}(Q^+)).\vec{n}-\alpha(Q^+-Q^-)\bigg]
\end{equation}
where $+$ and $-$ denote right and left sides of an interface respectively. For this particular problem, let $v=1$ in Eq. (\ref{eq:stability_analysis1}). Therefore, only the values on the left side of the interface are considered, i.e., $\hat{F}_{i+1/2}-\hat{F}_{i-1/2}=[{Q(\partial \mathcal{V}/\partial \xi)]^-}_{i+1/2}-[{Q(\partial \mathcal{V}/\partial \xi)]^-}_{i-1/2}$. For the time integration, we use a TVD Runge$-$Kutta (RK) method. A $n-$stage RK method for the ODE $Q_t=L(Q)$ has the general form as shown in Eq. (\ref{eq:stability_analysis7}).

\begin{equation} \label{eq:stability_analysis7}
\begin{split}
k_0 &=Q(t)\\
k_I &=\sum_{j=0}^{I-1}(\alpha_{Ij}K_j+\beta_{Ij}\Delta tL(k_j)), \quad \quad I=1,...,n
\end{split}
\end{equation}

where $k_I$ denotes the solution after $I^{th}$ stage, and $Q(t+\delta t)=k_n$. An RK method is total variation diminishing (TVD) if all the coefficients $\alpha_{Ij}$ and $\beta_{Ij}$ are nonnegative. The CFL coefficient of such a scheme is given by Eq. (\ref{eq:stability_analysis8}).
\begin{equation} \label{eq:stability_analysis8}
c=\textrm{min}_{I,k}\{\alpha_{Ik}/\beta_{Ik}\}
\end{equation}
For TVD RK order 3 scheme, the CFL coefficient is $c=1$.

\subsection{von Neumann stability analysis}

Based on the von Neumann stability analysis, the semi$-$discrete solution can be expressed as a discrete Fourier series, as given in Eq. (\ref{eq:stability_analysis9}).

\begin{equation} \label{eq:stability_analysis9}
\bar{Q}_i(t)=\sum_{k=-N/2}^{N/2}\hat{Q}_k(t)e^{ji\theta_k}, \quad \quad \omega_k \in R
\end{equation}
where $j=\sqrt{-1}$.
By the superposition principle, only one term in the series can be used for analysis, as illustrated in Eq. (\ref{eq:stability_analysis10}).

\begin{equation} \label{eq:stability_analysis10}
\bar{Q}_i(t)=\hat{Q}_k(t)e^{ji\theta_k}, \quad \quad \theta_k=\omega_k \Delta \xi
\end{equation}

By substituting Eq. (\ref{eq:stability_analysis10}) in Eq. (\ref{eq:stability_analysis5}), we can separate the spatial operator $L$, as given in Eq. (\ref{eq:stability_analysis11}).

\begin{equation} \label{eq:stability_analysis11}
    L=-\frac{(\hat{F}_{i+1/2}-\hat{F}_{i-1/2})}{\Delta \mathcal{V}_i}=-\frac{[Q(\partial \mathcal{V}/\partial \xi)]^-_{i+1/2}-[Q(\partial \mathcal{V}/\partial \xi)]^-_{i-1/2}}{\Delta \mathcal{V}_i}=-\frac{z(\theta_k)\bar{Q}_i}{\Delta \xi}
\end{equation}

where the complex function $z(\theta_k)$ is the Fourier symbol. By substituting the values of $Q^-_{i-1/2}$ and $Q^-_{i+1/2}$ using fifth order positive weights of cells $(i-1)$ and $i$ respectively for a smooth solution, the value of $z(\theta_k)$ can be evaluated using Eq. (\ref{eq:stability_analysisimportant}). 

\begin{equation}\label{eq:stability_analysisimportant}
z(\theta_k)=\frac{m+1}{{i}^{(m+1)}-({i}-1)^{(m+1)}}{\sum_{l=-2}^{+2}\bigg[w^{5+}_{i,0,l}{i}^me^{jl\theta_k}-w^{5+}_{(i-1),0,l}({i}-1)^me^{j(l-1)\theta_k}}\bigg]
\end{equation}
where index number $i=\xi_{i+1/2}/\Delta \xi$, $(i-1)=\xi_{i-1/2}/\Delta \xi$ and $m=0,1,$ and $2$ represents Cartesian, cylindrical$-$radial, and spherical$-$radial coordinates. Let $\bar{Q}_i^n=\bar{Q}_i(t^n)$ be the numerical solution at time $t^n=n \Delta t$. We define the amplification factor $g$ in Eq. (\ref{eq:stability_analysis12}) by substituting (\ref{eq:stability_analysis10}) into the fully$-$discrete system.

\begin{equation} \label{eq:stability_analysis12}
\bar{Q}_i^{n+1}=g(\hat{z}_k)\bar{Q}_i^n, \quad \hat{z}_k=-\sigma z (\theta_k),\quad \quad k=-N/2,...,N/2
\end{equation}
where $\sigma=\Delta t/\Delta \xi$. Therefore, the linear stability domain of an explicit time-stepping scheme is $S_t=\{\hat{z} : |g (\hat{z})| \leq 1\}$. Also, we define the spectrum $S$ of a spatial discretization scheme  in Eq. (\ref{eq:stability_analysis13}) \cite{liu2016wls}.

\begin{equation} \label{eq:stability_analysis13}
S=\{-z (\theta_k):\quad \theta_k\in 0,\Delta \theta, 2\Delta \theta,2\pi\}, \quad \quad \Delta \theta =2\pi \Delta \xi
\end{equation}

The stability limit is thus the largest CFL number $\Tilde{\sigma}$ such that the rescaled
spectrum $\Tilde{\sigma}S$ lies inside the stability domain $S_t$. 

\begin{equation} \label{eq:stability_analysis14}
\Tilde{\sigma}S\in S_t
\end{equation}

For the third$-$order Runge$-$Kutta scheme, the amplification factor $g$ is given in Eq. (\ref{eq:stability_analysis15}).

\begin{equation} \label{eq:stability_analysis15}
g(\Tilde{z})=1+\Tilde{z}+\frac{1}{2}\Tilde{z}^2+\frac{1}{6}\Tilde{z}^3
\end{equation}

Boundaries of the stability domain $\partial S_t=\{\Tilde{z}:|g(\Tilde{z})|=1\}$ is found by setting $g(\Tilde{z})=e^{j\phi}$ and solving Eq. (\ref{eq:stability_analysis16}).

\begin{equation} \label{eq:stability_analysis16}
\Tilde{z}^3+3\Tilde{z}^2+6\Tilde{z}+6+6e^{j\phi}=0
\end{equation}

As for the figures in this section, the stable and unstable regions are shown as off$-$white and blue regions respectively for TVD RK order 3. The stability domain depends on temporal discretization and is thus fixed irrespective of the spatial discretization scheme.

Given the spectrum $S$ and the stability domain $S_t$, the maximum stable CFL number of this scheme can be computed by finding the largest rescaling parameter $\Tilde{\sigma}$, so that the rescaled spectrum still lies in the stability domain. Using interval bisection, we find the CFL number of the proposed WENO$-$C scheme with TVD RK order 3 time marching.

\begin{figure}[] 
\centering
    \includegraphics[width=0.35\linewidth]{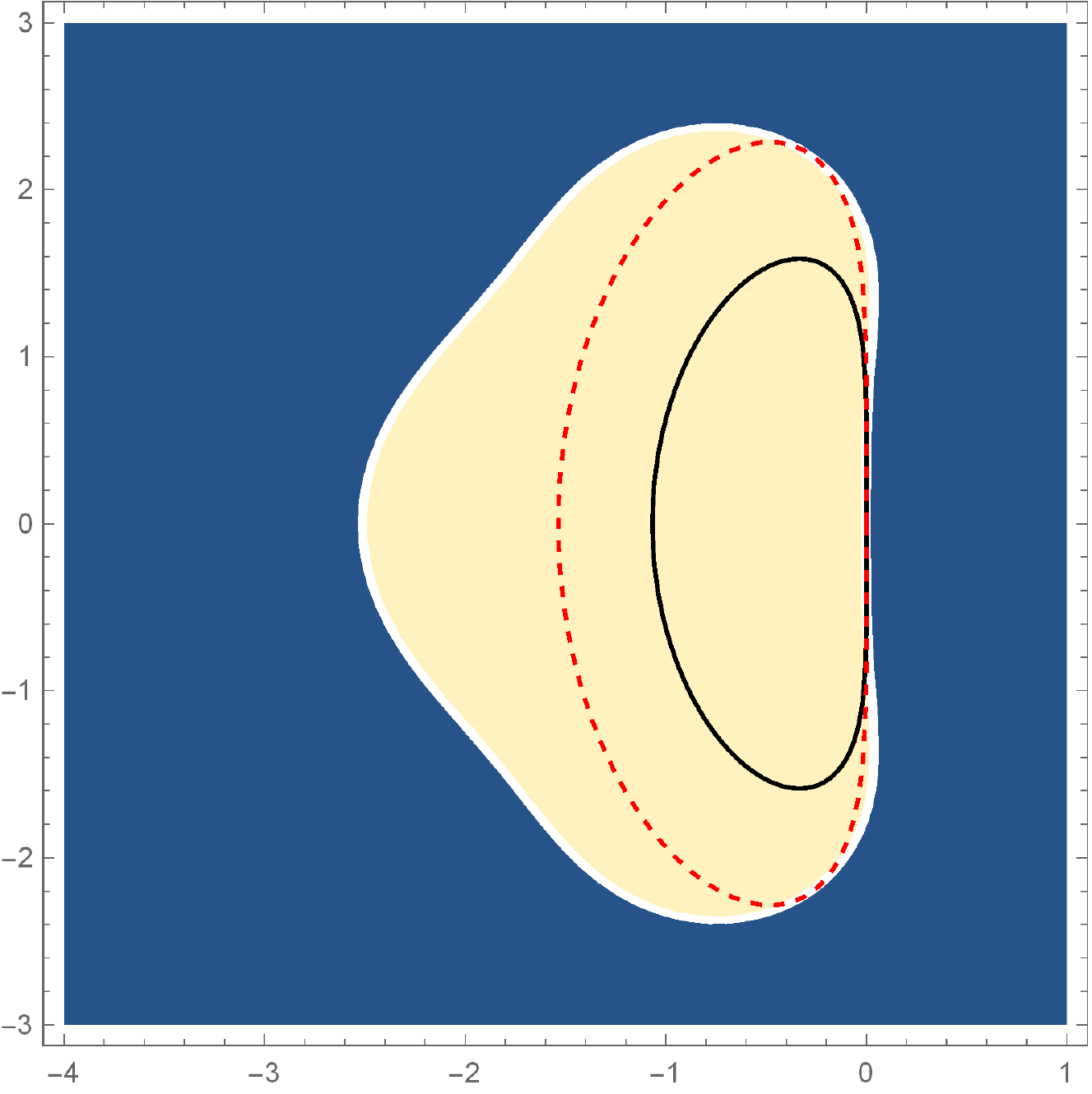} 
        \includegraphics[width=2cm,height=3cm,keepaspectratio]{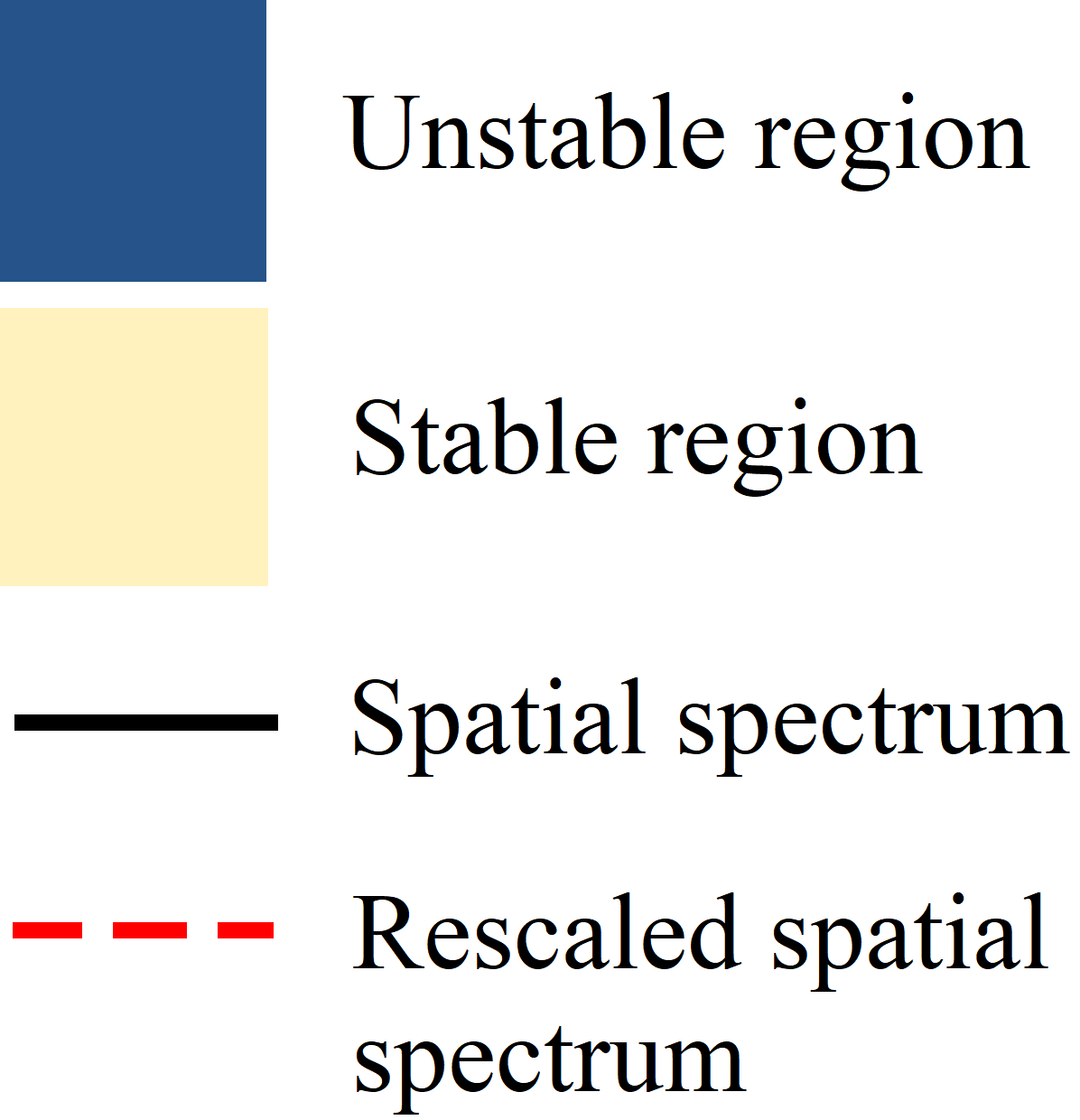}
      \caption{Rescaled spectrum (with maximum stable CFL number $\tilde{\sigma}=1.44$) and stability domains of fifth$-$order WENO$-$C in Cartesian coordinates ($m=0$) in a complex plane}
  \label{stab:Cart} 
  
  \end{figure}

\begin{figure}[] 
\vspace{-1.5in}
    \begin{subfigure}[b]{0.35\linewidth}
  \centering
    \includegraphics[width=2cm,height=3cm,keepaspectratio]{legend.PNG} 
       \caption{Legend}
    \end{subfigure}
    \begin{subfigure}[b]{0.35\linewidth}
    \includegraphics[width=\linewidth]{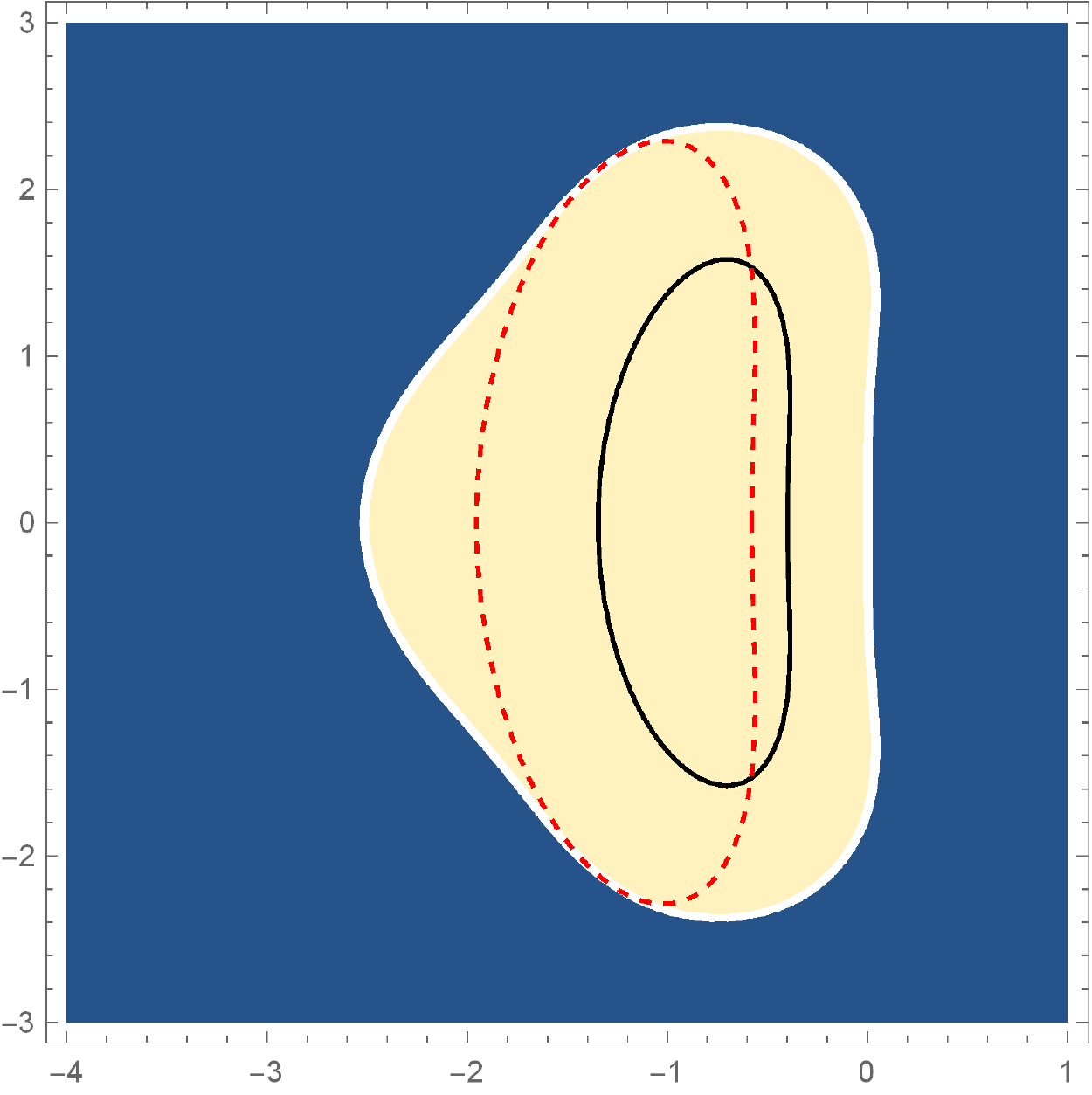}
   \caption{$i=3,\Tilde{\sigma}=1.45$}
  \end{subfigure}
    \begin{subfigure}[b]{0.35\linewidth}
    \includegraphics[width=\linewidth]{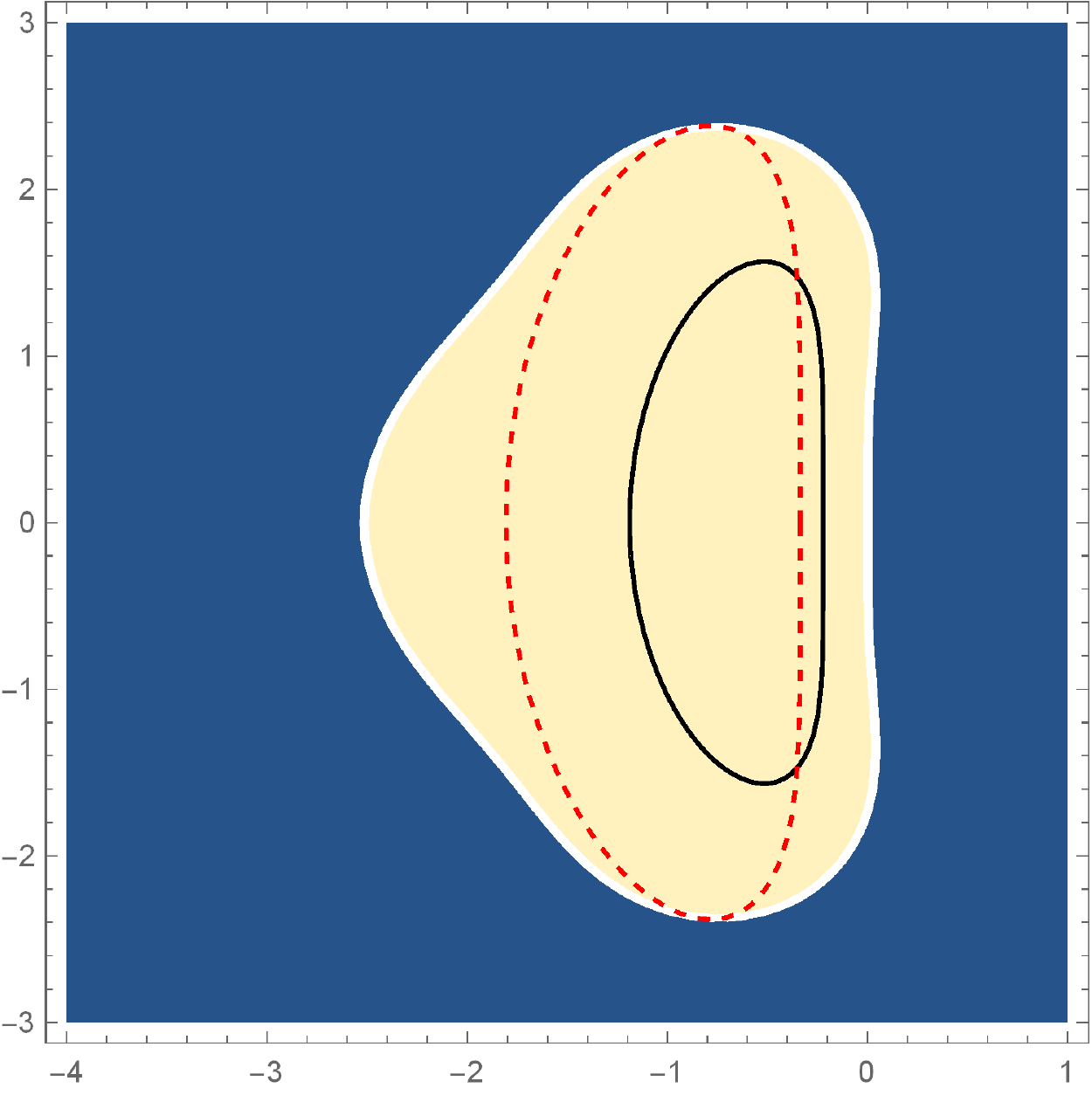} 
       \caption{$i=5,\Tilde{\sigma}=1.52$}
  \end{subfigure} 
  \begin{subfigure}[b]{0.35\linewidth}
    \includegraphics[width=\linewidth]{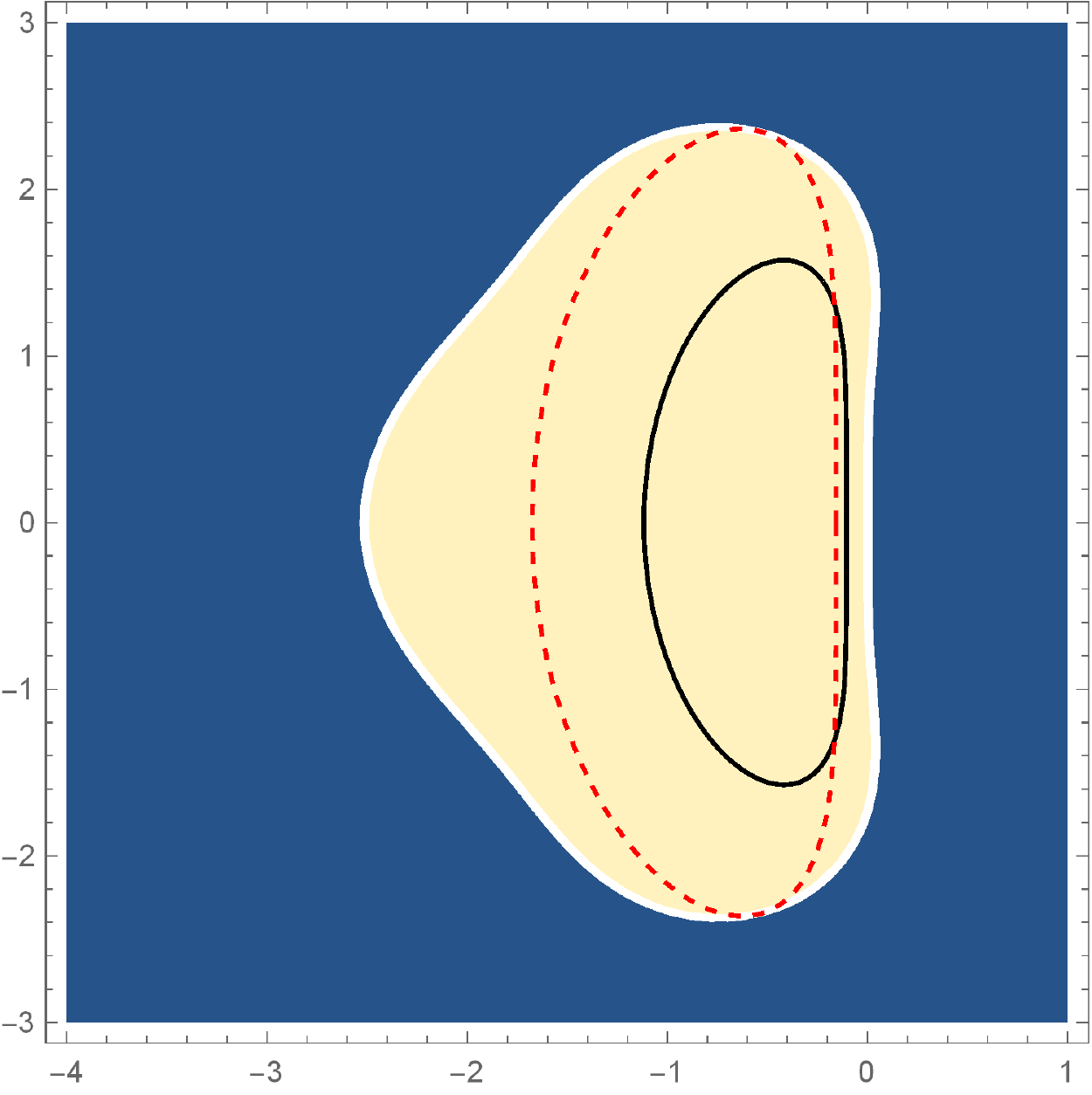}
       \caption{$i=10,\Tilde{\sigma}=1.50$}
  \end{subfigure} 
  \begin{subfigure}[b]{0.35\linewidth}
    \includegraphics[width=\linewidth]{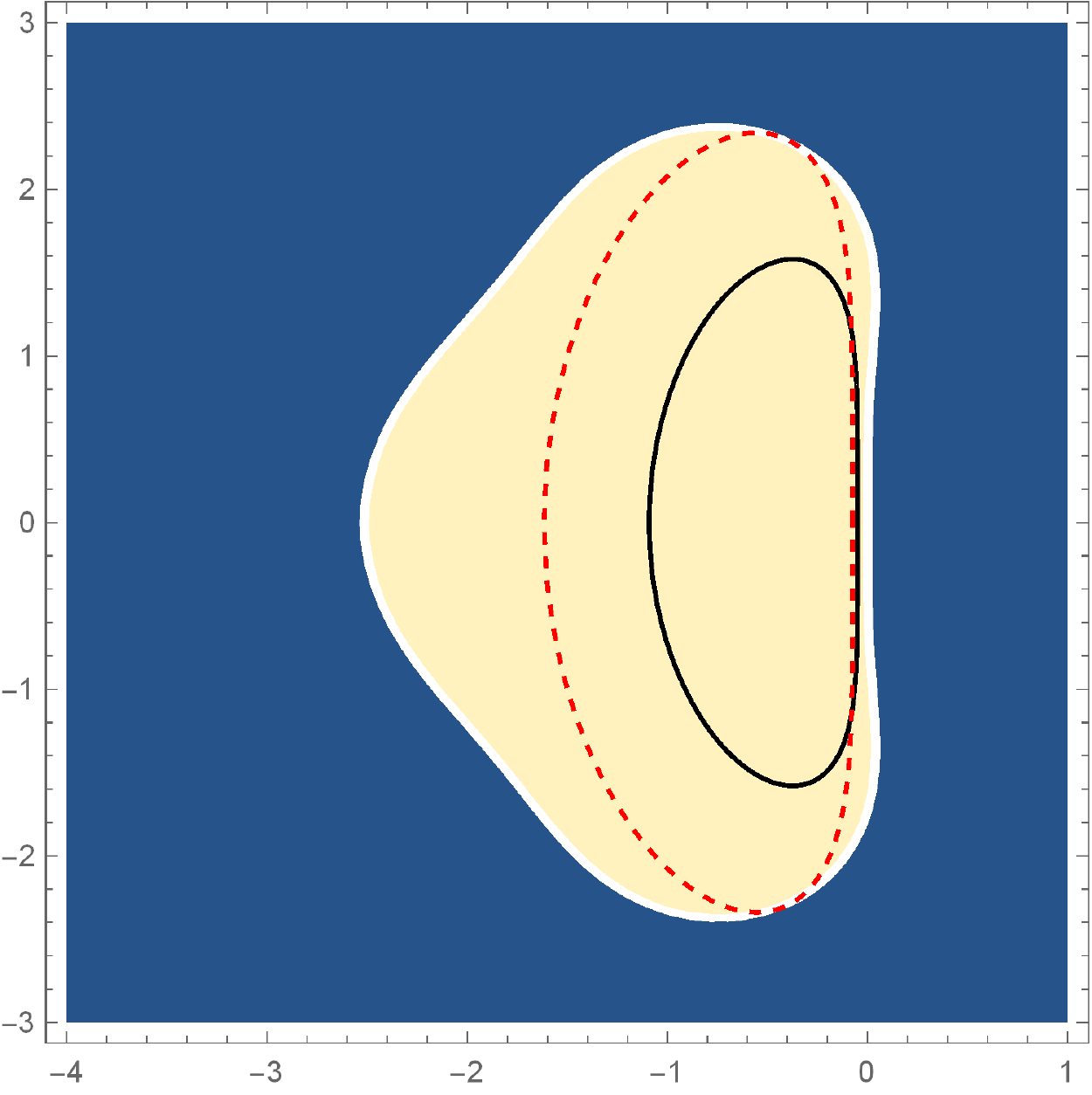} 
       \caption{$i=20,\Tilde{\sigma}=1.48$}
  \end{subfigure}
  \begin{subfigure}[b]{0.35\linewidth}
    \includegraphics[width=\linewidth]{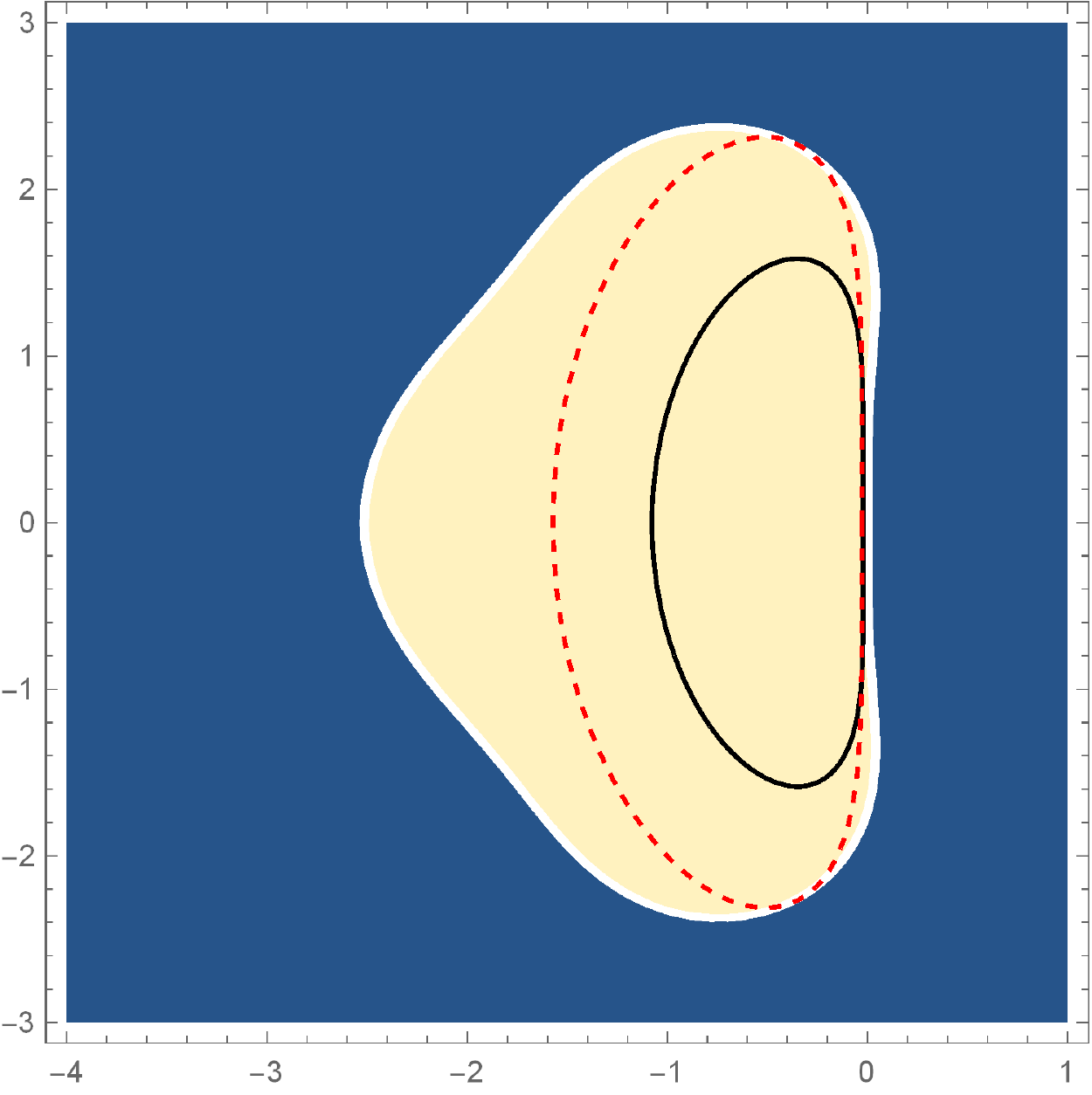} 
       \caption{$i=50,\Tilde{\sigma}=1.46$}
  \end{subfigure} 
    \begin{subfigure}[b]{0.35\linewidth}
    \includegraphics[width=\linewidth]{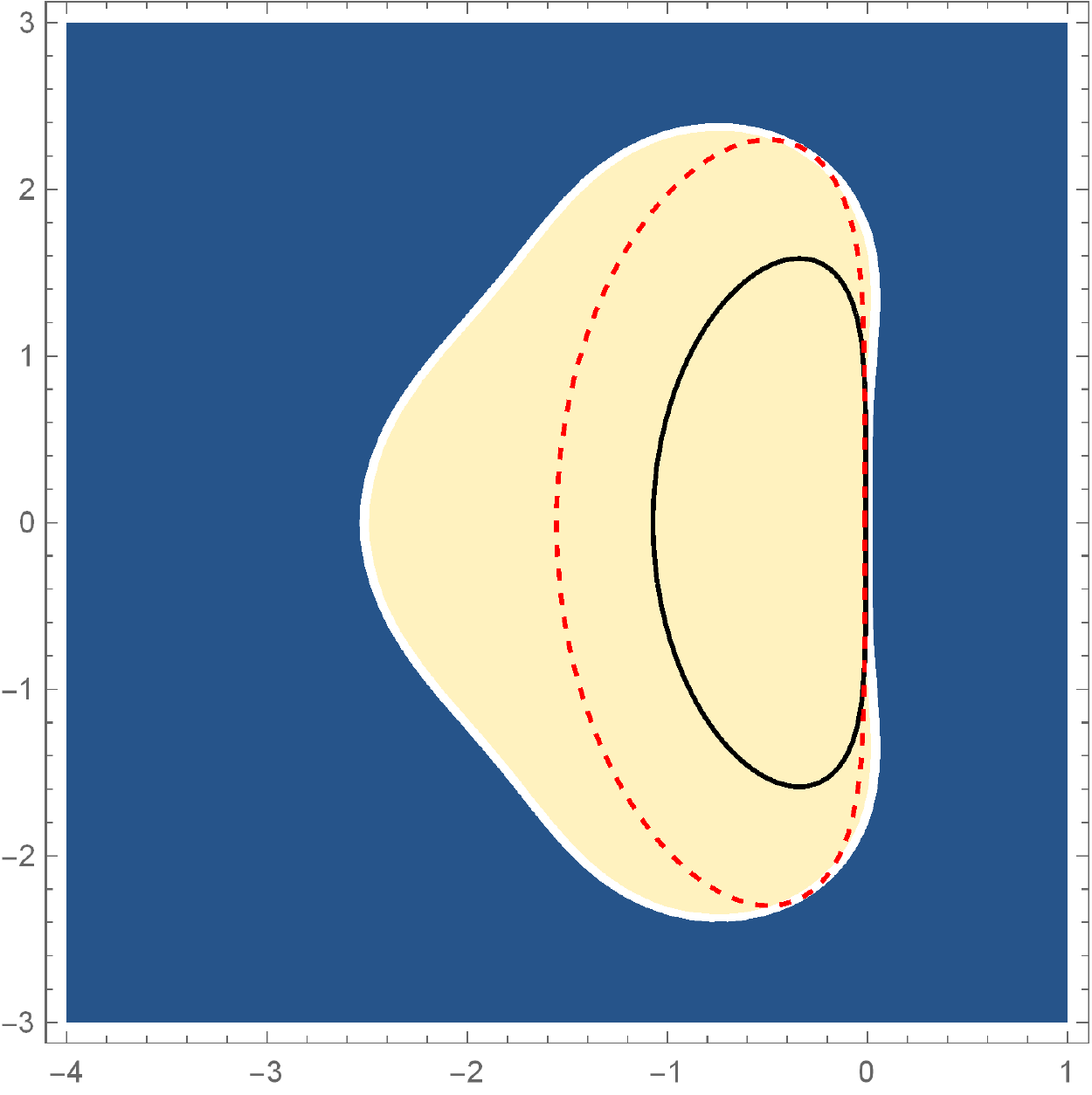} 
       \caption{$i=100,\Tilde{\sigma}=1.45$}
  \end{subfigure} 
  \caption{Rescaled spectrums (with maximum stable CFL number $\Tilde{\sigma}$) and stability domains of fifth$-$order WENO$-$C in cylindrical coordinates ($m=1$) in a complex plane for different index numbers $i$}
  \label{stab:cyl} 
\end{figure}

\begin{figure}[] 
\vspace{-1.5in}
    \begin{subfigure}[b]{0.35\linewidth}
  \centering
    \includegraphics[width=2cm,height=3cm,keepaspectratio]{legend.PNG}
    \caption{Legend}
    \end{subfigure}
    \begin{subfigure}[b]{0.35\linewidth}
    \includegraphics[width=\linewidth]{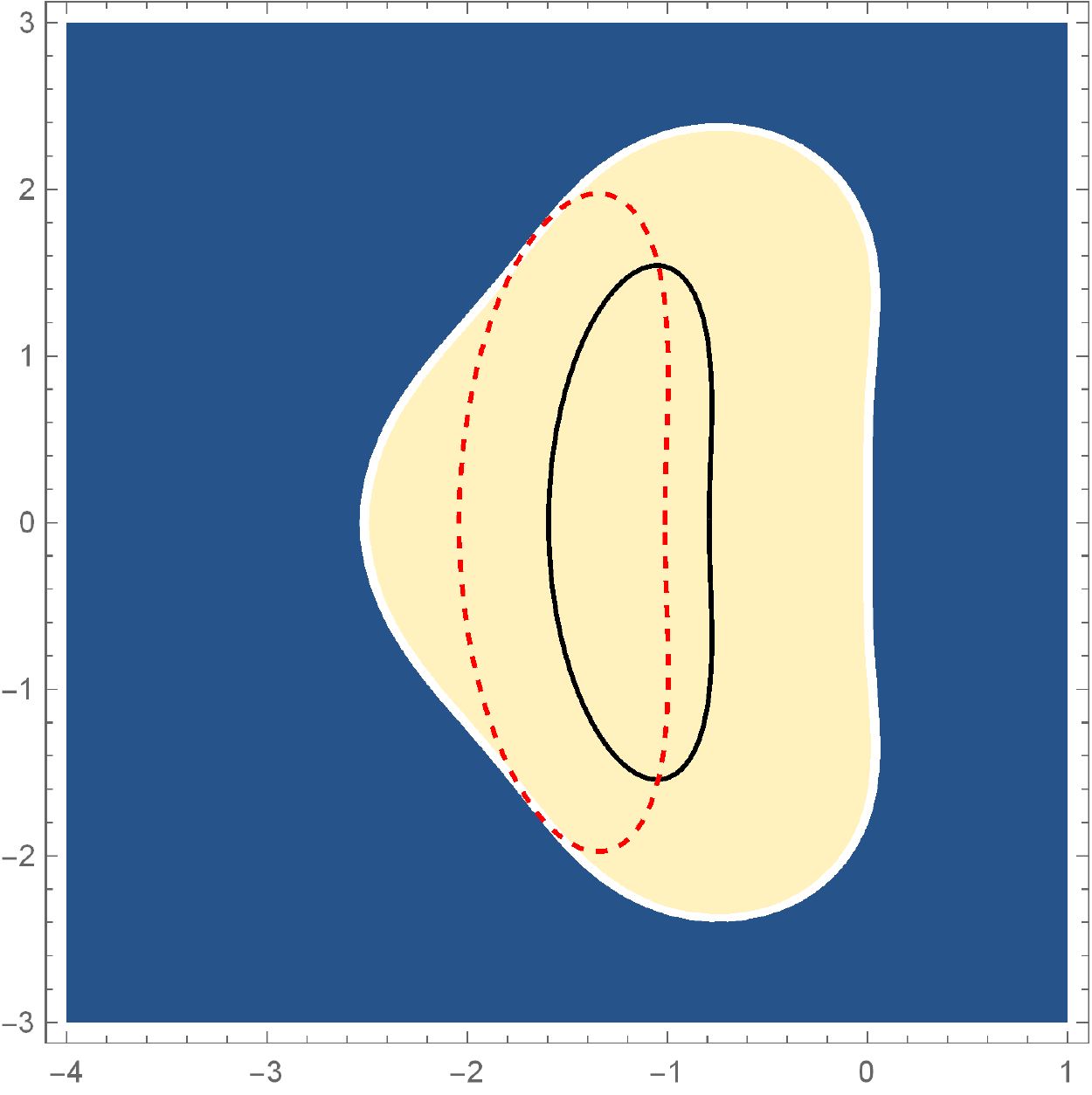}
   \caption{$i=3,\Tilde{\sigma}=1.28$}
  \end{subfigure}
    \begin{subfigure}[b]{0.35\linewidth}
    \includegraphics[width=\linewidth]{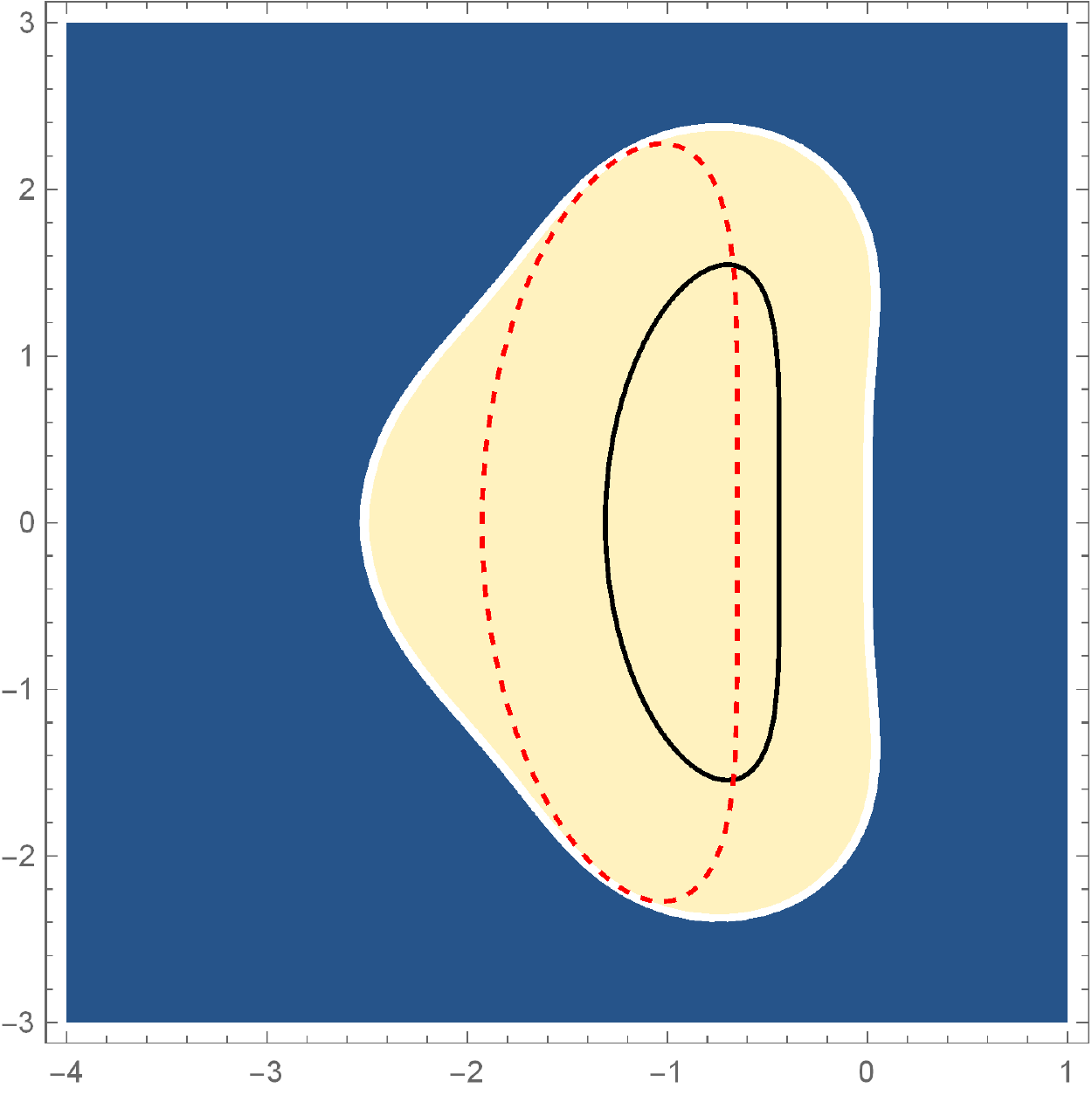} 
       \caption{$i=5,\Tilde{\sigma}=1.47$}
  \end{subfigure} 
  \begin{subfigure}[b]{0.35\linewidth}
    \includegraphics[width=\linewidth]{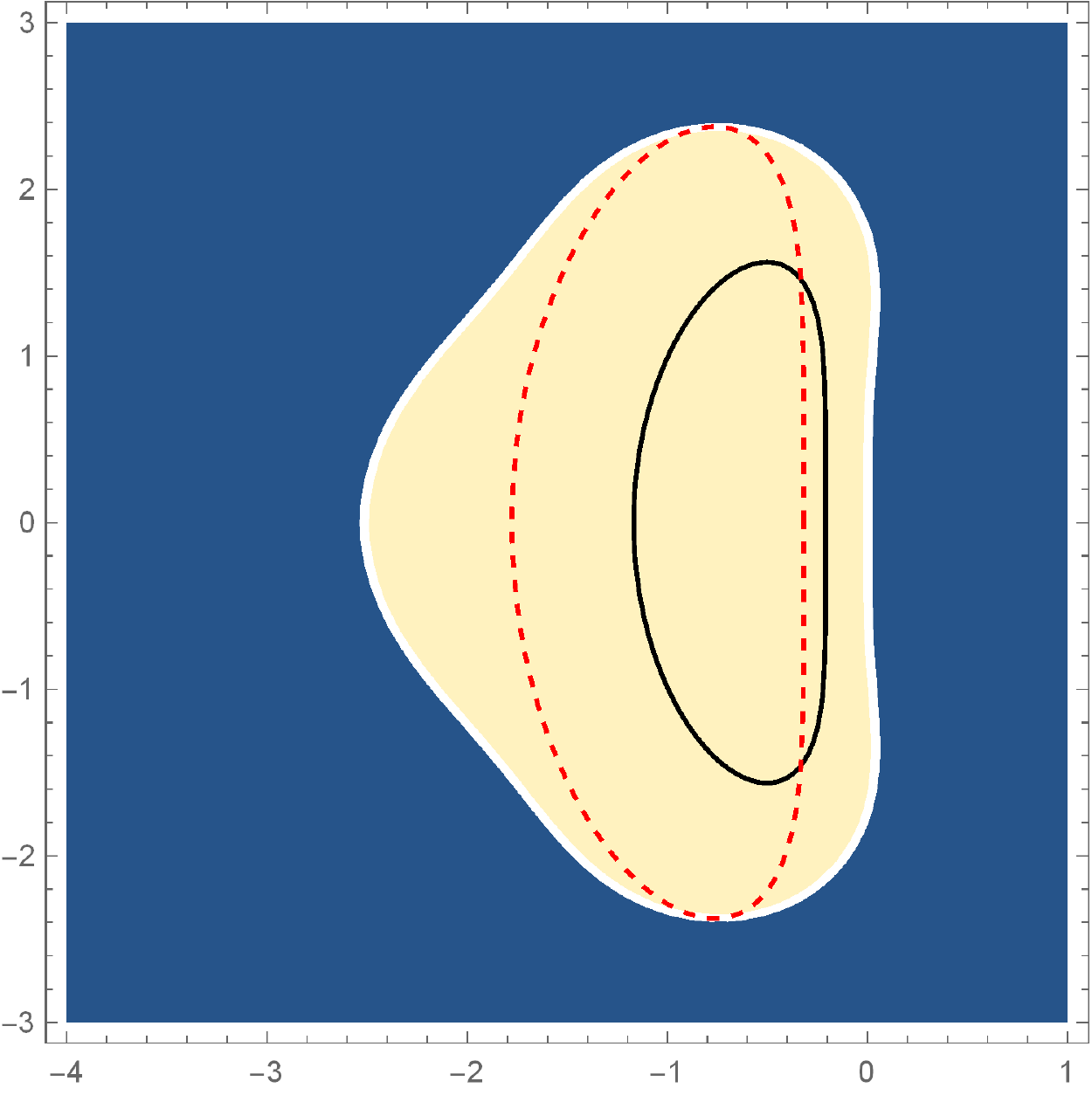}
       \caption{$i=10,\Tilde{\sigma}=1.52$}
  \end{subfigure} 
  \begin{subfigure}[b]{0.35\linewidth}
    \includegraphics[width=\linewidth]{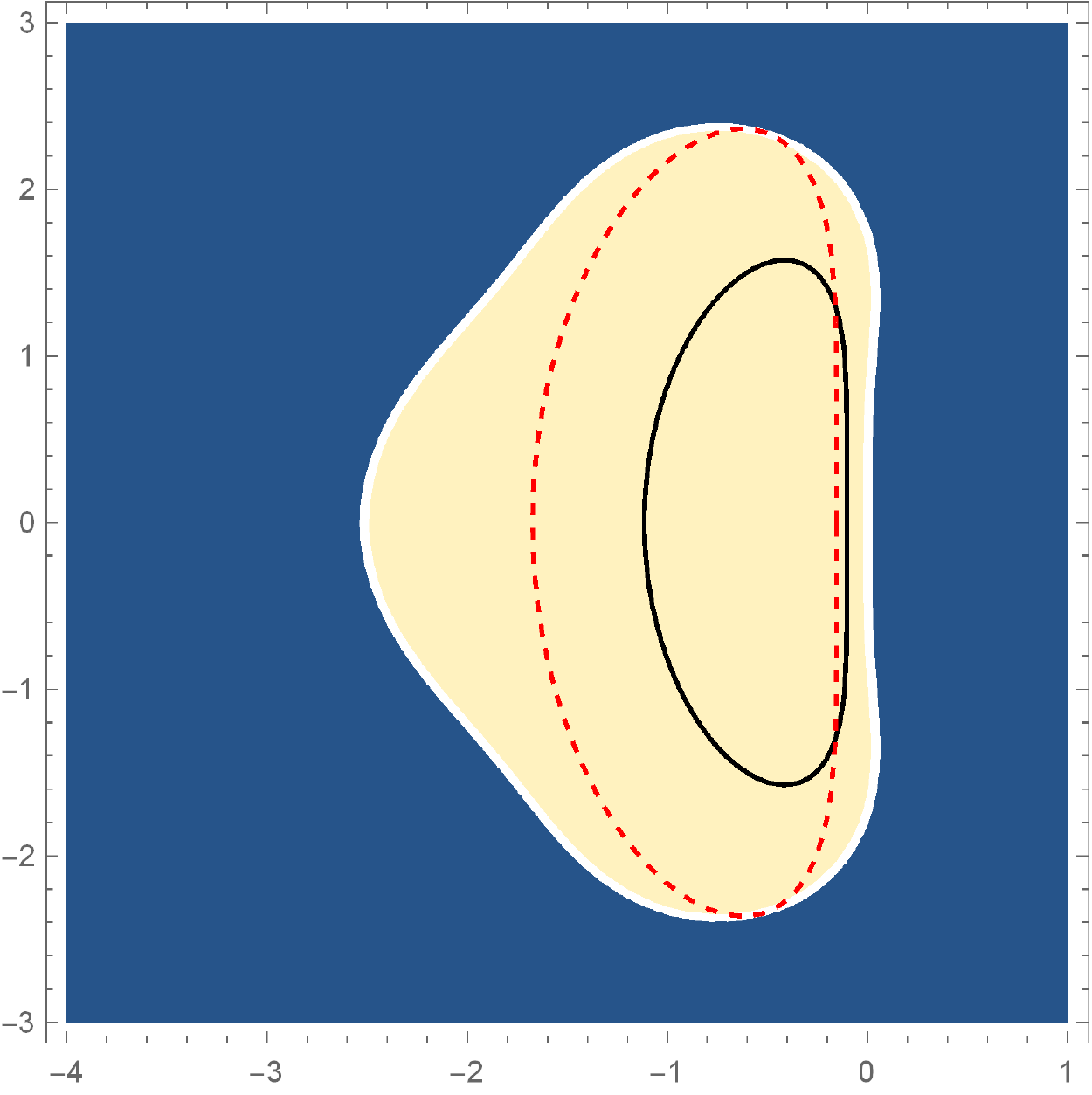} 
       \caption{$i=20,\Tilde{\sigma}=1.50$}
  \end{subfigure}
  \hfill
  \begin{subfigure}[b]{0.35\linewidth}
    \includegraphics[width=\linewidth]{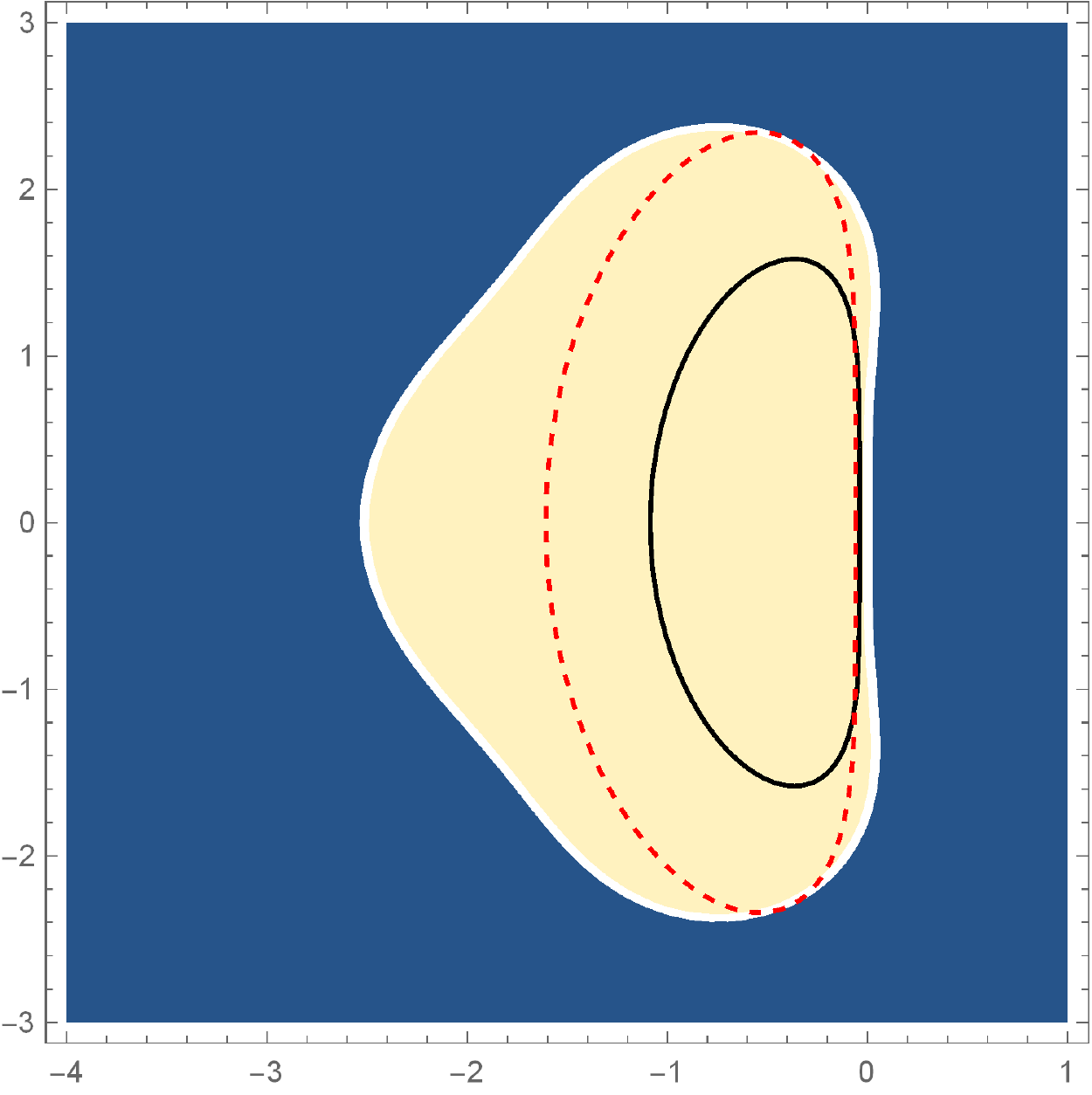} 
       \caption{$i=50,\Tilde{\sigma}=1.48$}
  \end{subfigure} 
    \begin{subfigure}[b]{0.35\linewidth}
    \includegraphics[width=\linewidth]{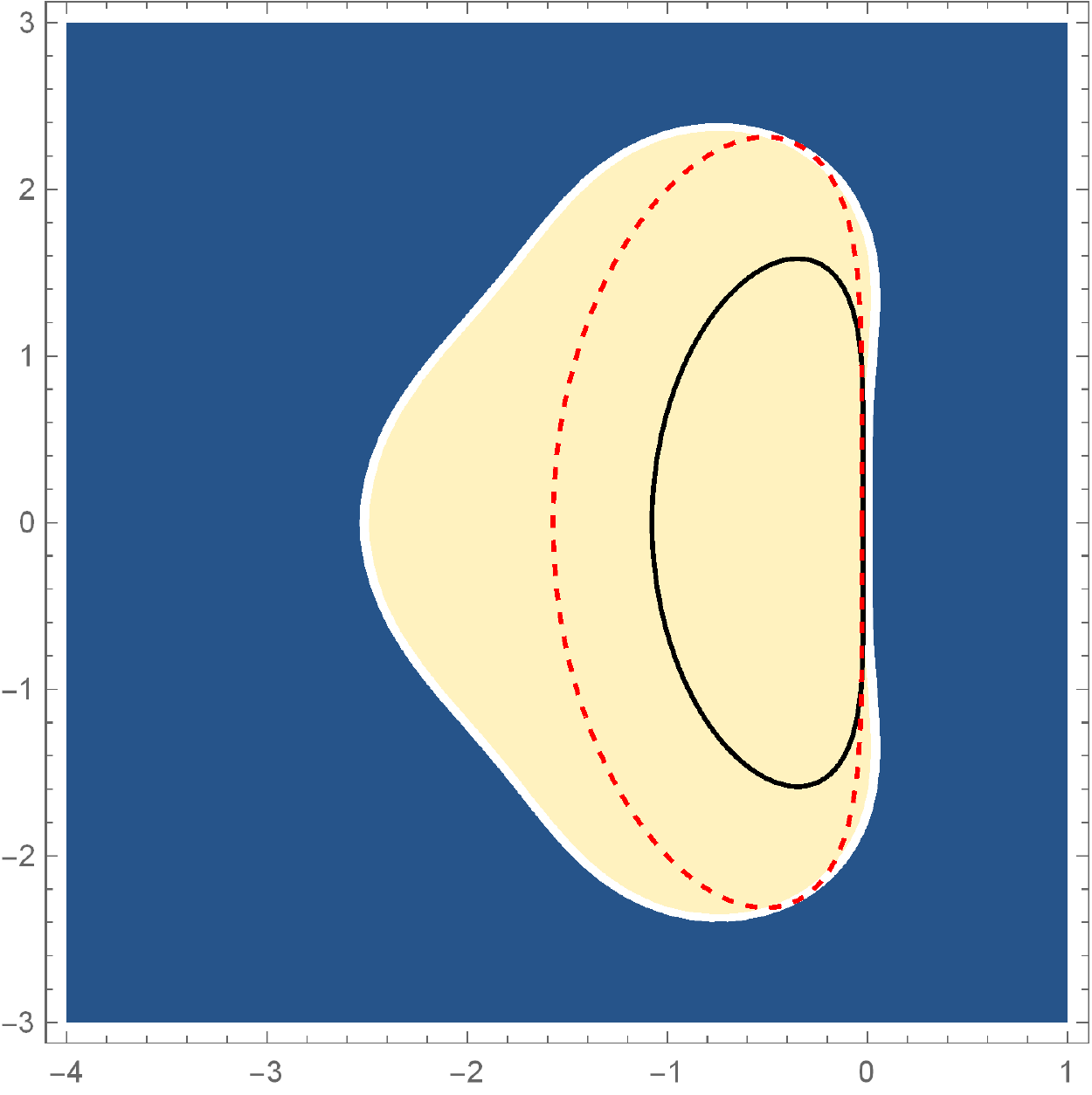} 
       \caption{$i=100,\Tilde{\sigma}=1.46$}
  \end{subfigure} 
  \caption{Rescaled spectrums (with maximum stable CFL number $\Tilde{\sigma}$) and stability domains of fifth$-$order WENO$-$C in spherical coordinates ($m=2$) in a complex plane for different index numbers $i$}
  \label{stab:sph} 
\end{figure}

For the Cartesian case as shown in Fig. \ref{stab:Cart}, the maximum CFL number value obtained is 1.44, similar to a previous study \cite{liu2016wls}. It can be observed respectively from Figs. \ref{stab:cyl} and \ref{stab:sph} for cylindrical$-$radial and spherical$-$radial coordinates that the spatial spectrums $S$ differs with the index numbers $i$ due to the geometrical variation of the finite volume. Some regions $(i=1,2)$ require boundary conditions and thus, are not considered in the present analysis. The values of CFL number for cylindrical$-$radial and spherical$-$radial coordinates lie in between 1.45 to 1.52 and 1.25 to 1.52 respectively. As a final remark, it can be concluded that the proposed scheme will be stable with third or higher order of RK method with an appropriate value of CFL number.

 %\section*{References}

\bibliography{mybibfile}

\end{document}